\documentclass[prd,twocolumn,nofootinbib,showpacs]{revtex4-1}

\usepackage{amsfonts}
\usepackage{yfonts}

\expandafter\let\csname equation*\endcsname\relax
\expandafter\let\csname endequation*\endcsname\relax
\usepackage{amsmath}

\usepackage{amssymb}
\usepackage{bm}
\usepackage{dcolumn}
\usepackage{graphicx}
\usepackage{graphics}
\usepackage[latin1]{inputenc}
\usepackage{latexsym}
\usepackage{rotating}
\usepackage{hyperref}
\usepackage[all]{hypcap} 
\usepackage{xspace} 
\usepackage[usenames,dvipsnames]{color}
\usepackage{mathrsfs}
\usepackage{subfigure}

\usepackage{ulem}
\usepackage{outlines}
\usepackage{enumitem}
\setenumerate[1]{label=\Roman*.}
\setenumerate[2]{label=\Alph*.}
\setenumerate[3]{label=\roman*.}
\setenumerate[4]{label=\alph*.}

\definecolor {darkgreen}{rgb}{0.2,0.7,0.2}

\usepackage{etoolbox}

\makeatletter
\def\@mkboth#1#2{}
\newlength\appendixwidth
\preto\appendix{\addtocontents{toc}{\protect\patchl@section}}
\newcommand{\patchl@section}{%
  \settowidth{\appendixwidth}{\textbf{Appendix }}%
  \addtolength{\appendixwidth}{1.5em}%
  \patchcmd{\l@section}{1.5em}{\appendixwidth}{}{\ddt}%
}
\makeatother

\newcommand\be{\begin{equation}}
\newcommand\ba{\begin{eqnarray}}
\newcommand\ee{\end{equation}}
\newcommand\ea{\end{eqnarray}}

\newcommand\bw{\begin{widetext}}
\newcommand\ew{\end{widetext}}

\newcommand{\nn}{\nonumber}

\newcommand{\orb}{{\mbox{\tiny orb}}}

\newcommand{\Ch}{\text{ch}}
\newcommand{\Sh}{\text{sh}}

\begin{document}
\title{Analytic Waveforms for Eccentric Gravitational Wave Bursts}

\author{Nicholas Loutrel}
\address{Department of Physics, Princeton University, Princeton, NJ, 08544, USA}

\date{\today}

\begin{abstract} 
We here present the first analytic effective fly-by (EFB) waveforms designed to accurately capture the burst of gravitational radiation from the closest approach of highly eccentric compact binaries. The waveforms are constructed by performing a re-summation procedure on the well-known Fourier series representation of the two-body problem at leading post-Newtonian order. This procedure results in two models: one in the time-domain, and one in the Fourier domain, which makes use of the stationary phase approximation. We discuss the computational efficiency of these models, and find that the time-domain model is roughly twice as fast as a numerical quadrupole waveform. We compare the time-domain model to both numerical, leading post-Newtonian order, quadrupole waveforms and numerical relativity fly-by waveforms using the match statistic. While the match is typically $>0.97$ when compared to the quadrupole waveforms, it is much lower when comparing to the numerical relativity fly-by waveforms, due to neglecting relativistic effects within the model. We further show how to use these individual waveforms to detect a repeated burst source.
\end{abstract}



\maketitle


\newpage
\section{Introduction}

A tantalizing problem currently exists in the field of gravitational wave (GW) modeling and data analysis: how best to model and detect binary systems with eccentricity close to unity. Such systems may be formed in the cores of dense stellar environments, where dynamical friction forces black holes into the gravitational center of the system, increasing the probability of two-, three-, and four-body interactions~\cite{2013ApJ...773..187N, 2012PhRvD..85l3005K, 2009MNRAS.395.2127O, Samsing:2017xmd, Samsing:2013kua, Samsing:2019dtb, Kremer:2018cir, Zevin:2018kzq, Samsing:2018ykz, Samsing:2017oij, Leigh:2017wff, Rodriguez:2017pec, Antonini:2015zsa}. These mechanisms are capable of creating black hole (BH) binaries with a wide range of eccentricities, but a subset are formed with high eccentricity, close to the unbound limit. In globular clusters, resonant interactions force these systems into three distinct categories, namely: ejected inspirals, in-cluster mergers, and GW captures. Ejected inspirals and in-cluster mergers generally form with GW frequencies $f \lesssim 10^{-1}$~\cite{Samsing:2018isx, DOrazio:2018jnv, Samsing:2018nxk}, making them possible candidates for detection by the Laser Interferometer Space Antenna (LISA)~\cite{Prince:2003aa} and the Decihertz Interferometer Gravitational wave Observatory (DECIGO)~\cite{Kawamura:2011zz}. On the other hand, GW captures generally form binaries with $f \ge 10^{-1}$ Hz, a subset of which form within the Laser Interferometer Gravitational-wave Observatory (LIGO)~\cite{Abramovici:1992ah, Harry:2010zz, TheLIGOScientific:2014jea} detection band with high ($e\sim1$) eccentricity~\cite{Samsing:2019dtb}, and with estimated event rates of 1-2 yr$^{-1}$ Gpc$^{-3}$ in the local universe~\cite{Rodriguez:2018pss}.

Highly eccentric sources for ground-based detectors constitute some of the most relativistic signals possible, with pericenter velocities reaching greater than ten percent the speed of light. These systems, thus, present themselves as unique laboratories for studying gravitational physics and astrophysics in the so-called dynamical, strong field regime of gravity, where the spacetime curvature is large and rapidly varying~\cite{PPE}. The accuracy of general relativity (GR) within this regime has only been tested with the currently detected quasi-circular LIGO sources~\cite{LIGOScientific:2019fpa}, where the velocities of the component objects are only large in the late inspiral and merger. Gravitational wave bursts from highly eccentric systems would allow us to probe this regime during each closest approach of the binary in the inspiral phase of the coalescence~\cite{Loutrel:2014vja}. If matter is present in the binary components, finite size and tidal effects will become important in each pericenter passage~\cite{Yang:2019kmf, Yang:2018bzx}. Further, if one of the binary components is a neutron star (NS), f-modes on the NS surface and other oscillation modes can be excited, which would generate an observable GW signature and allow for better constraints on the NS equation of state~\cite{Stephens:2011as, Chaurasia:2018zhg, Papenfort:2018bjk, Chirenti:2016xys, East:2015vix, Choptuik:2015mma, Rosofsky:2018vyg}.

However, the large pericenter velocities present a tough problem in terms of modeling, as typical post-Newtonain (PN)~\cite{PW, Blanchet:2013haa} treatments of the two-body problem may not be sufficiently accurate to model such systems.
Ideally, one would want to start by considering the full numerical solution of the Einstein field equations for such systems, but even this presents computation difficulties. The timescale associated with closest approach can be several orders of magnitude smaller than the orbital timescale. Any numerical relativity (NR) simulations would have to resolve these disparate time scales, which is currently too computationally expensive to produce accurate simulations of more than a few orbits~\cite{Stephens:2011as, 2012PhRvD..85l4009E}. Kludge waveforms like those in~\cite{2013PhRvD..87d3004E} have been shown to be more accurate than PN waveforms for single bursts, but without having full NR simulations to compare to, there is no way of knowing whether this model is accurate enough to describe the full inspiral-merger-ringdown signal.

The lack of accurate models poses significant problems if one wants to detect such signals, regardless of the detector being considered~\cite{Salemi:2019owp}. One method is to search for regions of excess power in time-frequency space, and add up the power over multiple bursts, with the total SNR scaling as $N^{1/4}$ where $N$ is the number of bursts~\cite{Tai:2014bfa}. The biggest concern with such a detection strategy is that ground-based detectors often pick up regions of anomalous excess power known as glitches~\cite{Cabero:2019orq, TheLIGOScientific:2016zmo, LIGOScientific:2018mvr}, some of which resemble the GW bursts from highly eccentric binaries. Fortunately, if one knows where an initial burst occurs in time-frequency space and its morphology, one can predict the location and morphology of all subsequent bursts in time-frequency space if given a radiation reaction model. These burst models provide a prior on the bursts that would help us distinguish them from glitches in the detector. Such a model currently only exists within the PN approximation~\cite{Loutrel:2014vja, Loutrel:2017fgu}.

An alternative approach, that has proven to be both fast and robust to modeling error, would be to use neural networks to detect such signals~\cite{Wei:2019zlc, Rebei:2018lzh, George:2017pmj, George:2017vlv, Shen:2019ohi, Luo:2019hvt, Kim:2019ktw}. In this method, deep learning networks are trained on sample data, with an injected waveform model(s). The trained networks can then be run on detector data in real time, with detection sensitivity comparable to, but still less than, that of matched filtering~\cite{George:2016hay, Gabbard:2017lja} and parameter estimation results comparable to Bayesian inference~\cite{Shen:2019vep}. This method has the added benefit of being fast, allowing for rapid follow-up for electromagnetic counterparts~\cite{Chang:2019edd, Allen:2019dkq}. Although promising, limitations of neural networks hinder their ability to make statistically significant detections at the moment, but are still a powerful tool for generating triggers~\cite{Gebhard:2019ldz}. A study of the detectability of highly eccentric binaries using this method is currently being considered~\cite{Huerta:2019ecc}.

Despite the strengths of these search strategies, the gold standard for detecting GWs is, at this point in time, matched filtering, whereby an accurate waveform template is used to extract the GW signal from detector noise~\cite{PhysRevD.43.2470, 2005dgw..book.....B}. The trouble with this method is the need for accurate templates, since small phase errors can result in the search missing the signal entirely~\cite{Martel:1999tm}. Yet, this method has shown extreme success in detecting quasi-circular, spin-aligned binaries~\cite{LIGOScientific:2018mvr, Usman:2015kfa, Sachdev:2019vvd}, due to extensive modeling efforts~\cite{Blanchet:2013haa, Taracchini:2013rva}. The biggest challenge to applying this search strategy to highly eccentric binaries is the requirements of phase accuracy across multiple bursts. For lower mass systems that form on the edge of the LIGO band, there can be hundreds of pericenter passages during the inspiral, and any model would have to be phase accurate over all of these to achieve detection. 

In this article, we take the first steps toward achieving a matched filtering search for highly eccentric binaries. We construct the first analytic waveforms specifically designed to accurately capture the burst of radiation from an eccentric binary. We work to leading order within the PN formalism~\cite{PW}, where the conservative dynamics of the binary are described by the Kepler problem, and the dissipative dynamics are described by the quadrupole approximation, i.e. Newtonian gravity plus quadrupole-order radiation. Working within the PN formalism allows us to write all relevant quantities in a Fourier series on harmonics of the orbital period, which follow directly from Kepler's equation.

The problem with this Fourier decomposition is that, when the eccentricity of the binary approaches unity, the series become badly convergent. We address this issue using the re-summation procedure in~\cite{Loutrel:2016cdw, Forseth:2015oua}, where the Bessel functions appearing in the Fourier series are asymptotically expanded. The series are then re-summed through integration, which results in closed form, analytic expressions of Keplerian quantities for highly eccentric binaries. This procedure destroys the periodic behavior the trajectory, effectively describing the dynamics of the binary as a single fly-by.

We develop two waveform models using this procedure. In the first, which we refer to as the EFB-T model, we perform the re-summation in the time domain, and employ a Taylor series radiation reaction model. The waveform polarizations are given by Eqs.~\eqref{eq:hpc-time}, with the necessary functions given in Appendix~\ref{h-t-coeffs}. The EFB-T model has the benefit of being faster to sample than full numerical, leading PN order waveforms. The second model, referred to as the EFB-F model, utilizes the stationary phase approximation to compute the analytic Fourier transform of the Newtonian order waveforms for eccentric binaries, and is constructed by performing the re-summation in the Fourier domain. The waveform polarizations are given by Eqs.~\eqref{eq:hpc-f} and expressions in Appendix~\ref{h-f-coeffs}, and depends on hypergeometric functions. This makes sampling the EFB-F model computationally expensive. We validate the EFB-T model against both numerical leading PN order waveforms, and NR waveforms, using the match statistic. We show that the EFB-T waveform is highly accurate to leading PN waveforms, with matches $>0.98$ for a $(10,10)M_{\odot}$ binary.

While the EFB waveforms are a faithful representation of the numerical Newtonian order waveforms, it is important to quantify their robustness to modeling error. If the pericenter distance becomes sufficiently small, non-linear strong field effects will play an important role in the dynamics of the binary, which are not modeled in the EFB waveforms. To study how important these extreme gravity effects are, we compute the match between the EFB-T model and NR fly-by waveforms, with the results given in Sec.~\ref{nr-comp}. For the binary studied, the match is $>0.92$ for pericenter distances $r_{p} > 8.75 M$. The match does, however, drop off sharply as $r_{p}$ decreases, specifically, the match is $0.75$ for $r_{p} = 8.125M$. Thus, the modeling of extreme gravity effects will be critical for developing waveforms that cover small pericenter distances.

Finally, while the EFB waveforms only cover a single burst, we show how multiple EFB waveforms can be combined to create a multi-burst model. Each EFB waveform is characterized by three parameters that evolve in time, namely the semi-latus rectum, orbital eccentricity, and time of pericenter passage. While the latter of these effectively describes an overall time-shift of the waveform, this is not independent of the previous time of pericenter passage and orbital parameters when the burst is part of a full inspiral sequence. Using the methods described in~\cite{Loutrel:2017fgu}, we develop a timing model that accurately tracks the evolution of these quantities, applying a Newtonian plus quadrupole radiation approximation. We compute the match between a multi-burst EFB waveform combined with the timing model, and a numerical Newtonian order waveform which covers ten pericenter passages. We show that the multi-burst EFB model achieves a match $>0.97$ for the correct number of waveforms used, i.e. one EFB-T waveform for each pericenter passage. However, we further show that small errors, due to relativistic effects for example, in the timing model can throw off detection.

This paper is organized as follows. Sec.~\ref{bin} reviews the Newtonian order two-body problem, and the necessary ingredients for the remainder of the analysis. Sec.~\ref{time-domain} presents the re-summation procedure in the time-domain and the construction of the EFB-T waveform model, while Sec.~\ref{freq-domain} presents the SPA, re-summation procedure in the Fourier domain, and the EFB-F model. In Sec.~\ref{valid}, we discuss the computation efficiency of the waveform models, and validate the EFB-T model against numerical waveforms. Finally, in Sec.~\ref{discuss}, we discuss the future prospects of such models. Throughout this work, we use units where $G = c = 1$.

\section{Binary Inspirals at Newtonian Order}
\label{bin}

We shall begin with a brief review of the dynamics of binary systems at leading PN order, and some of the difficulties that arise in the problem.

\subsection{The Kepler Problem}
\label{kep}

To begin, we review the common parameterizations of the two-body problem in Newtonian gravity, commonly called the Kepler problem. More specifically, the Kepler problem reduces to solving the effective one-body equations of motion $\vec{a} = - (M/r^{2}) \vec{n}$, where $\vec{a}$ is the relative acceleration of the binary, $M = m_{1} + m_{2}$ is the total mass of the binary with $m_{1,2}$ the component masses, $r$ is the relative radial separation of the two bodies, and $\vec{n} = \vec{x}/r$ with $\vec{x}$ the trajectory. Restricting the motion of the binary to the $xy$-plane, and writing $\vec{n} = [\cos\phi, \sin\phi, 0]$, the equation of motion can be split into two equations, 
\begin{align}
\label{eq:r-eom}
\ddot{r} + r \dot{\phi}^{2} &= - \frac{M}{r^{2}}\,,
\\
\label{eq:phi-eom}
\frac{d}{dt} \left(r^{2} \dot{\phi}\right) &= 0\,,
\end{align}
governing the radial and azimuthal motion, respectively, and where the overdot corresponds to derivatives with respect to time. The latter of these admits one constant of motion, specifically the reduced orbital angular momentum $h = r^{2} \dot{\phi}$. Using this to replace $\dot{\phi}$ in Eq.~\eqref{eq:r-eom}, this equation can be directly integrated by multiplying by $\dot{r}$ to obtain
\begin{equation}
\label{eq:en-def}
\frac{1}{2} \dot{r}^{2} = \varepsilon - \frac{h^{2}}{2 r^{2}} + \frac{M}{r}\,,
\end{equation}
with $\varepsilon$ the reduced orbital energy. The existence of these two constants of motion has now reduced the problem to quadratures.

We may obtain a more explicit solution to the Kepler problem by returning to Eq.~\eqref{eq:r-eom}. Writing $w = 1/r$ and changing variables from $t$ to $\phi$, we obtain the equation
\begin{equation}
\frac{d^{2} w}{d\phi^{2}} + w = \frac{M}{h^{2}}\,.
\end{equation}
This equation may be directly solved to obtain
\begin{equation}
\label{eq:r-kep}
r = \frac{p}{1 + e \cos(\phi-\omega)}
\end{equation}
where $p = h^{2}/m$ is the semi-latus rectum of the orbit, $e$ is the orbital eccentricity, and $\omega$ is an integration constant known as the longitude of pericenter. To obtain the evolution of $\phi$, we use the definition of $h$, which gives
\begin{equation}
\label{eq:phi-kep}
\dot{\phi} = \left(\frac{M}{p^{3}}\right)^{1/2} \left[1 + e \cos(\phi - \omega)\right]^{2}\,.
\end{equation}
The trajectory of the binary is now uniquely parameterized by the azimuthal angle $\phi$, or alternatively, by the true anomaly $V = \phi - \omega$. The evolution equation for $V$ is still given by Eq.~\eqref{eq:phi-kep} since $\omega$ is a constant\footnote{In general, this is not true when considering perturbations of the Kepler problem.}. As a final point, Eq.~\eqref{eq:r-kep} can be inserted into Eq.~\eqref{eq:en-def} to obtain the mapping between orbital energy and eccentricity, specifically
\begin{equation}
e = \left(1 - \frac{2 \epsilon h^{2}}{M^{2}}\right)^{1/2}\,.
\end{equation}

While the parameterization in terms of the true anomaly $V$ is complete, there is the problem of obtaining $V(t)$, which involves integrating Eq.~\eqref{eq:phi-kep}. Technically, the integral can be done to obtain $t(V)$, but the resulting function is too complicated to be inverted analytically. Instead, we rely on a new parameterization in terms of the eccentric anomaly $u$, with the mapping between $V$ and $u$ given by
\begin{equation}
\label{eq:V-to-u}
\cos V = \frac{\cos u - e}{1 - e \cos u}\,, \qquad \sin V = \frac{(1-e^{2})^{1/2} \sin u}{1 - e \cos u}\,.
\end{equation}
The time evolution of $u$ may be obtained by applying these equations to Eq.~\eqref{eq:phi-kep}. After integrating, we obtain
\begin{equation}
\label{eq:kep-eq}
\ell = m^{1/2} \left(\frac{1-e^{2}}{p}\right)^{3/2} \left(t - t_{p}\right) = u - e \sin u\,,
\end{equation}
where $\ell$ is the mean anomaly, and $t_{p}$ is an integration constant. This equation, known as Kepler's equation, does not admit a closed-form solution for $u(t)$ due to it being transcendental.

Generally, one must solve Kepler's equation numerically. However, it does lay the ground work for obtaining a Fourier series representation for the solution to the Kepler problem. To see this, consider $\cos V$ given by Eq.~\eqref{eq:V-to-u}, which may be written as a Fourier series of the form
\begin{equation}
\cos V = \sum_{k=-\infty}^{\infty} c_{k} e^{i k \ell}\,, \qquad c_{k} = \int_{-\pi}^{\pi} \frac{d\ell}{2\pi} \; \frac{\cos u - e}{1 - e \cos u} e^{-i k \ell}\,.
\end{equation}
The coefficients $c_{k}$ can be found explicitly by changing variables in the integral from $\ell$ to $u$, and making use of the integral definition of Bessel functions of the first kind
\begin{equation}
J_{n}(x) = \frac{1}{2\pi}\int_{-\pi}^{\pi} du \; e^{i (n u - x \sin u)}\,,
\end{equation}
with the end result being
\begin{equation}
\label{eq:cosv}
\cos V = - e + \frac{2}{e} \left(1-e^{2}\right) \sum_{k=1}^{\infty} J_{k}(k e) \cos(k \ell)\,.
\end{equation}
Similarly
\begin{equation}
\label{eq:sinv}
\sin V = 2 (1 - e^{2})^{1/2} \sum_{k=1}^{\infty} J_{k}'(k e) \sin(k \ell)\,,
\end{equation}
where the prime corresponds to differentiation with respect to the argument. The entire Kepler problem is now determined as a function of time.

While the solution to the Kepler problem is now complete, the Fourier series representation does have a drawback. As the eccentricity increases, more terms are needed in the sums of Eqs.~\eqref{eq:cosv}-\eqref{eq:sinv} to obtain sufficient phase accuracy~\cite{PhysRevD.80.084001}. For high eccentricities $(e \sim 1)$, the series converge slowly, and one may need to keep several hundred terms in the sums~\cite{Loutrel:2016cdw}. This presents a problem for developing waveform models using these Fourier series, namely, while they may be fast and accurate for moderate eccentricities, they need not necessarily be so for highly eccentric systems. In the next section, we present a method for developing analytic waveform models for eccentric gravitational wave bursts that circumvents this issue.

\subsection{Gravitational Waves}

The discussion of the preceding section deals with only the conservative dynamics of the binary. The binary also inspirals due to the emission of GWs, which to leading PN order are described by the quadrupole approximation~\cite{PW},
\begin{equation}
h_{ij} = \frac{2}{D_{L}} \ddot{I}_{ij}\,,
\end{equation}
where $h_{ij}$ is the metric perturbation, $D_{L}$ is the luminosity distance,
\begin{equation}
I_{ij} = \mu x_{<i} x_{j>}
\end{equation}
is the quadrupole momentum of the binary, with $x_{i}$ the binary's trajectory, $\mu$ the reduced mass of the binary, and $< >$ corresponds to the symmetric trace free part of the tensor.

To find the observable part of the metric perturbation, we project into the transverse traceless gauge. We define the line of sight vector
\begin{equation}
\vec{N} = \left[\sin\iota \cos\beta', \sin\iota \sin\beta', \cos\iota\right]
\end{equation}
where $\iota$ is the inclination angle of the binary (the angle between $\vec{N}$ and the binary's orbital angular momentum), and $\beta$ is an arbitrary polarization angle. We further define two vectors,
\begin{align}
\vec{\Theta} &= \left[\cos\iota \cos\beta', \cos\iota \sin\beta', -\sin\iota \right]
\\
\vec{\Phi} &= \left[-\sin\beta', \cos\beta', 0 \right]
\end{align}
which define the transverse sub-space orthogonal to ${\vec{N}}$. The GW polarizations are then defined by the projections
\begin{align}
h_{+} &= \frac{1}{2} \left(\Theta^{i} \Theta^{j} - \Phi^{i} \Phi^{j}\right) h_{ij}\,,
\\
h_{\times} &= \frac{1}{2} \left(\Theta^{i} \Phi^{j} + \Phi^{i} \Theta^{j}\right) h_{ij}\,.
\end{align}
Using the results of the previous sections, these reduce to~\cite{Martel:1999tm}
\begin{align}
\label{eq:hp-V}
h_{+} &= -\frac{M^{2} \eta}{p D_{L}}\left\{ \Big[ 2\cos(2V-2\beta) + \frac{5}{2} e \cos(V-2\beta) 
\right.
\nn \\
&\left.
+ \frac{e}{2} \cos(3 V - 2 \beta) + e^{2} \cos(2\beta)\Big] \left(1 + \cos^{2}\iota\right) 
\right.
\nn \\
&\left.
+ \left(e \cos V + e^{2}\right)\sin^{2}\iota\right\}\,,
\\
\label{eq:hc-V}
h_{\times} &= - \frac{M^{2} \eta}{p D_{L}} \cos\iota \left[4 \sin(2V-2\beta) + 5 e \sin(V-2\beta) 
\right.
\nn \\
&\left.
+ e \sin(3V-2\beta) - 2 e^{2} \sin(2\beta)\right]\,,
\end{align}
where $\beta = \beta' - \omega$. Finally, these waveform polarizations can, alternatively, be written in a Fourier series on harmonics of the orbital period~\cite{10.1093/mnras/274.1.115},
\begin{align}
\label{eq:h-harm}
h_{+,\times} &= - \frac{m^{2} \eta}{p D_{L}} (1 - e^{2}) \sum_{k=1}^{\infty} \left[C_{+,\times}^{(k)} \cos(k\ell) + S_{+,\times}^{(k)} \sin(k\ell)\right]\,.
\end{align}
where $[C_{+,\times}^{(k)}, S_{+,\times}^{(k)}]$ are given by Eqs.~(9a)-(9d) in~\cite{Moore:2018kvz}. This completes our review.
\section{Time Domain Waveforms}
\label{time-domain}

Now that we have reviewed the parameterizations of the Kepler problem, we may begin to consider re-summing these parameterizations to obtain analytic waveforms for eccentric binaries. We will begin by focusing on the re-summation of quantities in the time domain, where re-summation can be directly applied to $\cos V$ and $\sin V$. We will also consider the construction of a simplified radiation reaction model, to capture the inspiraling nature of the binary under the emission of GWs.

\subsection{Re-summations of the Kepler Problem \& the Post-Parabolic Approximation}
\label{resum-t}

We here present a method for re-summing the series appearing in Eqs.~\eqref{eq:cosv}-\eqref{eq:sinv}. The method was originally developed in~\cite{Loutrel:2016cdw} for re-summing similar series expressions appearing in the GW tails fluxes. The method generally follows three steps: 1) replace any instance of $J_{k}(ke)$ and $J_{k}'(ke)$ with their uniform asymptotic expansions, 2) replace summations on the harmonic index $k$ with integrals, and 3) expand the functions obtained after integration about high eccentricity, namely $\epsilon = 1-e^{2} \ll 1$. This method produced highly accurate representations of the tail enhancement factors to the current limit of the PN expansion for eccentric binaries. We will here adapt it for the Fourier series representation of the Kepler problem.

A detailed description of the uniform asymptotic expansion may be found in~\cite{AS}. For our purposes, it suffices to consider the first two terms in this expansion, specifically
\allowdisplaybreaks[4]
\begin{align}
\label{eq:J-asym}
J_{k}(k e) &\sim \left(\frac{\zeta}{1-e^{2}}\right)^{1/4} \left\{\frac{1}{\pi} \sqrt{\frac{2 \zeta}{3}} K_{1/3}\left(\frac{2}{3} \zeta^{3/2} k\right) 
\right.
\nn \\
&\left.
+ \frac{1}{24 \pi} \sqrt{\frac{\zeta}{6}} \left[\frac{5}{\zeta^{3/2}} - \frac{2 (2 + 3 e^{2})}{(1-e^{2})^{3/2}} \right] \frac{1}{k}  K_{2/3}\left(\frac{2}{3} \zeta^{3/2} k\right)\right\}\,,
\\
\label{eq:Jp-asym}
J_{k}'(k e) &\sim \left(\frac{1-e^{2}}{\zeta}\right)^{1/4} \left\{ \frac{\zeta}{e \pi} \sqrt{\frac{2}{3}} K_{2/3}\left(\frac{2}{3} \zeta^{3/2} k\right) 
\right.
\nn \\
&\left.
- \frac{\zeta}{24 \pi e \sqrt{6}} \left[\frac{7}{\zeta^{3/2}} + \frac{4 - 18 e^{2}}{(1-e^{2})^{3/2}} \right] \frac{1}{k} K_{1/3}\left(\frac{2}{3} \zeta^{3/2} k\right)\right\}\,,
\end{align}
with 
\begin{equation}
\label{eq:zeta}
\zeta = \left\{\frac{3}{2}\left[\ln\left(\frac{1 + \sqrt{1-e^{2}}}{e}\right) - \sqrt{1-e^{2}}\right]\right\}^{2/3}\,.
\end{equation}
The re-summation procedure we will use is as follows: (1) extend the summations to include $k=0$, (2) replace the Bessel functions with their uniform asymptotic expansions, (3) convert the sum to an integral over $k$ and evaluate, (4) match the solution to the exact answer at pericenter, (5) define a ``phase" variable $\psi = (3/2) \ell/\zeta^{3/2}$ and make the replacement $\ell \rightarrow \psi$, and (6) define $\epsilon = 1 - e^{2}$ and expand in $\epsilon \ll 1$, holding $\psi$ fixed. 

There are a few extra steps that we have added here compared to the procedure for re-summing the tail enhancement factors in~\cite{Loutrel:2016cdw}. First, extending the sums to include $k=0$ is necessary when arriving at step (3). Generally, this procedure requires one to evaluate integrals of the form
\begin{equation}
\int_{k_{\rm min}}^{\infty} dk \; K_{n}\left(\frac{2}{3} \zeta^{3/2} k\right) {\cos(k\ell) \choose \sin(k \ell)}\,.
\end{equation}
If $k_{\min} = 1$, the above integrals do not admit a closed-form solution, but do if $k_{\min} = 0$. Hence, we extend the lower limit of the sums to address this. Second is the matching to the exact answer at pericenter. The procedure produces asymptotic series that resemble the exact solution, but are often offset by a constant value. This may be fixed by matching to the known value of the exact solution at pericenter, which can be found by taking $u=0$ in Eq.~\eqref{eq:V-to-u}. Finally, the definition of $\psi$, and the action of holding it fixed when expanding in $\epsilon \ll 1$, helps to ensure that the resulting expressions remain phase accurate compared to an exact answer. 

Consider the Fourier series representation of $\cos V$ given in Eq.~\eqref{eq:cosv}. The summation that we must re-sum is
\begin{equation}
\sum_{k=1}^{\infty} J_{k}(k e) \cos(k \ell) = -1 + \sum_{k=0}^{\infty} J_{k}(k e) \cos(k \ell)\,,
\end{equation}
where we have extended the sum to include $k=0$ in the equality. After replacing $J_{k}(k e)$ with its asymptotic expansion in Eq.~\eqref{eq:J-asym}, and replacing the summations with integrals, we are left with the problem of evaluating
\begin{align}
\label{eq:I-int}
I_{1} &= \int_{0}^{\infty} K_{1/3}\left(\frac{2}{3} \zeta^{3/2} k\right) \cos(k\ell)\,, 
\\
I_{2} &= \int_{0}^{\infty} \frac{dk}{k} K_{2/3}\left(\frac{2}{3} \zeta^{3/2} k\right)\cos(k\ell)\,.
\end{align}
The first of these can be directly evaluated as is, with the end result being
\begin{equation}
I_{1} = \frac{3^{1/2} \pi\; \Ch(\psi,1/3)}{2 \zeta^{3/2} \sqrt{1 + \psi^{2}}}\,,
\end{equation}
where $\text{ch}(\psi,n) = \cosh[n \text{arcsinh}(\psi)]$.\footnote{It is worth noting that this function can also be written as $\Ch(x,n) = (1/2) (x + \sqrt{1+x^{2}})^{n} + (n\rightarrow -n)$ , by the properties of hyperbolic functions.} On the other hand, $I_{2}$ appears to be divergent when $k=0$. We may circumvent this by realizing that $\cos(k\ell)/k = - \int d\ell \; \sin(k \ell)$. Applying this, integrating over $k$, and then integrating over $\ell$ results in
\begin{equation}
I_{2} = - \frac{3^{1/2}}{2} \pi \; \Ch(\psi, 2/3)\,.
\end{equation}
The practical problem of evaluating the integrals is now solved.

Applying the remainder of the procedure, we obtain
\begin{align}
\label{eq:cos-asym}
\cos V &\sim -1 + \frac{2 \; \Ch(\psi, 1/3)}{\sqrt{1+\psi^{2}}} + \frac{2}{5} \epsilon \left[-1 + \frac{\Ch(\psi,1/3)}{\sqrt{1+\psi^{2}}}\right] 
\nn \\
&+ \epsilon^{2} \left[-\frac{107}{350} + \frac{51}{175} \frac{\Ch(\psi, 1/3)}{\sqrt{1+\psi^{2}}} + \frac{1}{70} \Ch(\psi,2/3)\right] 
\nn \\
&+ {\cal{O}}(\epsilon^{3})\,.
\end{align}
Similarly, for $\sin V$ we obtain
\begin{align}
\label{eq:sin-asym}
\sin V &\sim \frac{2 \; \Sh(\psi ,2/3)}{\sqrt{1 + \psi^{2}}} + \frac{2}{5} \epsilon \left[-\Sh(\psi, 1/3) + \frac{\Sh(\psi, 2/3)}{\sqrt{1+\psi^{2}}} \right] 
\nn \\
&+ \epsilon^{2} \left[-\frac{7}{25} \Sh(\psi, 1/3) + \frac{51}{175} \frac{\Sh(\psi, 2/3)}{\sqrt{1+\psi^{2}}}\right] + {\cal{O}}(\epsilon^{3})
\end{align}
where $\Sh(\psi,n) = \sinh[n \text{arcsinh}(\psi)]$. Note that we have truncated the series expansion at second order in $\epsilon$. There is no mathematical difficulty causing us to stop at this order. We are only interested in the high eccentricity regime ($\epsilon \ll 1$). If one desired more accuracy at moderate eccentricities, one could simply carry the series to higher order.

An important note about these expressions is that they are actually non-oscillatory, which runs contrary to the notion of a closed elliptical orbit. Mathematically, this results from the fact that the integrand in Eq.~\eqref{eq:I-int} is effectively an exponentially damped sinusoid in $k$, whose integral is not an oscillatory function. From a more physical perspective, it can be shown that these expressions reproduce a parabolic trajectory when $\epsilon = 0 (e=1)$. We show this explicitly in Appendix~\ref{par}. In this way, the procedure actually acts to model an elliptical system as a deformation of a parabola, rather than a deformation of a circle as is done in the post-circular formalism~\cite{PhysRevD.80.084001}. As such, these expressions are only valid over one orbit, with $\ell \in [-\pi, \pi]$, and constitute a \textit{post-parabolic formalism}. Waveforms generated in this formalism will, thus, only be accurate for one pericenter passage, and we shall refer to them as \textit{effective fly-by (EFB) waveforms}.

\begin{figure*}[ht]
\includegraphics[clip=true,scale=0.27]{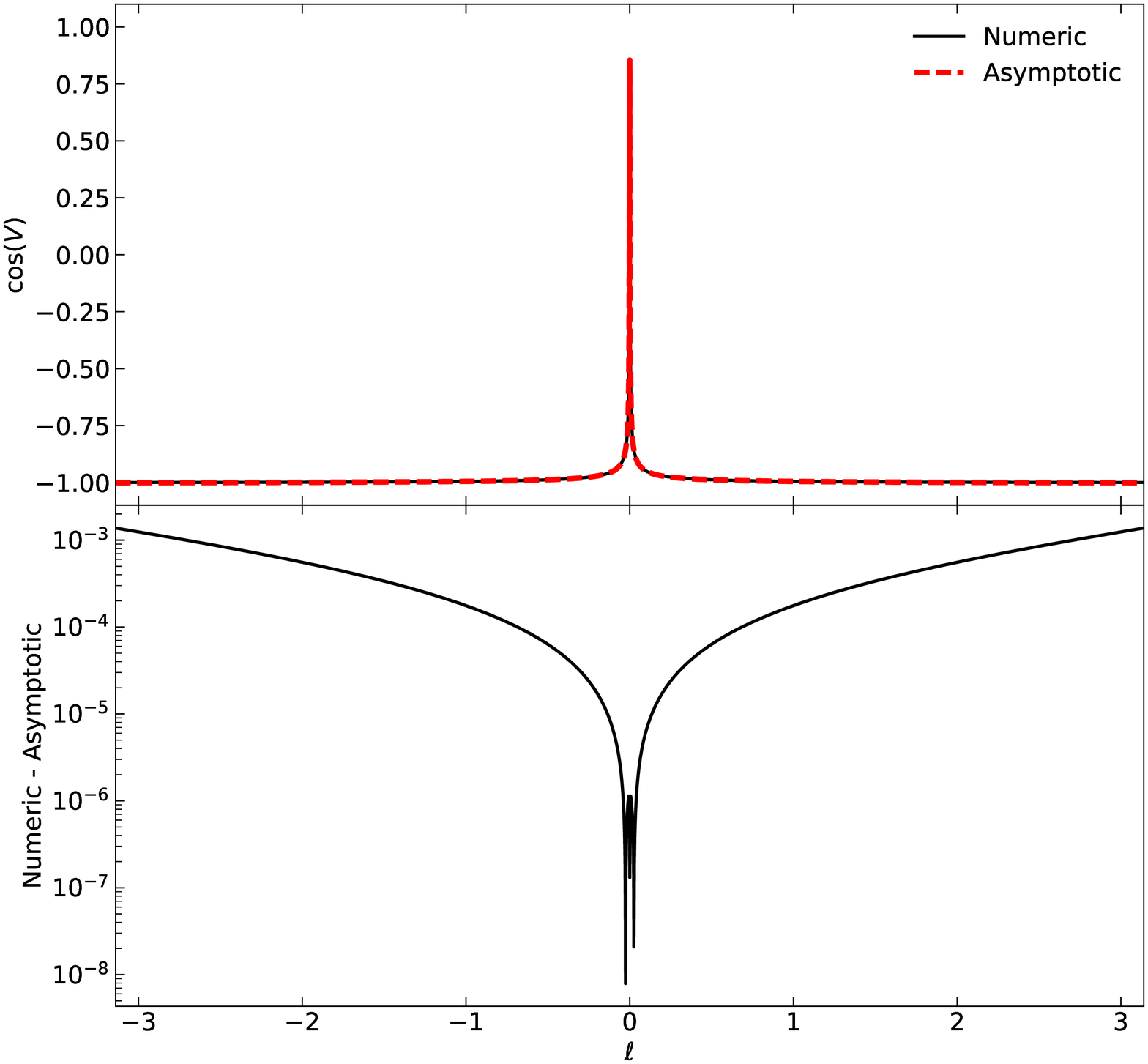}
\includegraphics[clip=true,scale=0.27]{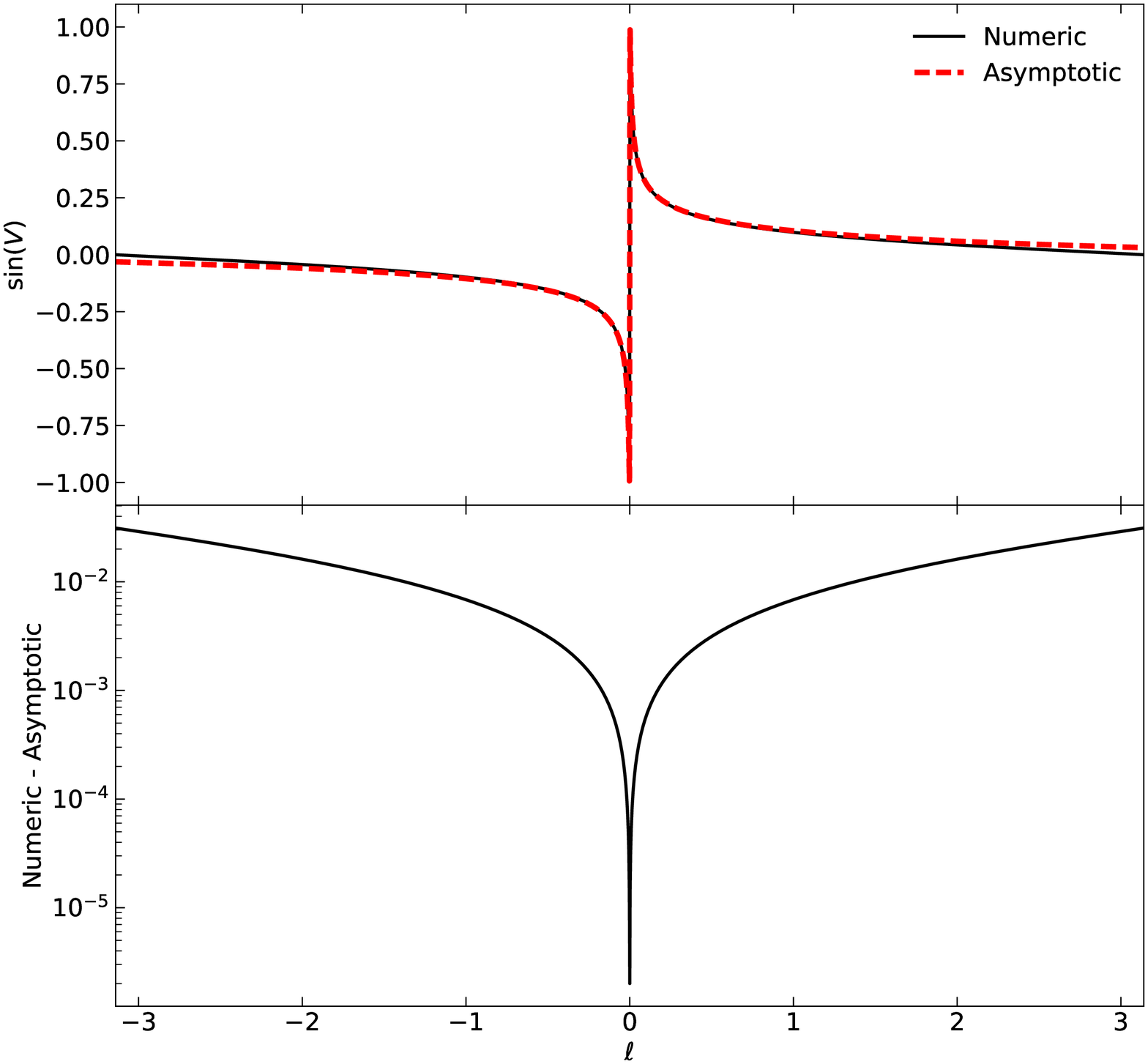}
\caption{\label{comp} Comparison of the asymptotic representation of $\cos V$ (left) and $\sin V$ (right) to their numerical evolution found by solving Eq.~\eqref{eq:phi-kep}. The analytic asymptotic expressions are displayed as dashed lines and are given by Eqs.~\eqref{eq:cos-asym}-\eqref{eq:sin-asym}, while the numerical solutions are represented by solid lines. The bottom panels of each plot shows the difference between the numerical solutions and the asymptotic expressions.}
\end{figure*}

In Fig.~\ref{comp}, we provide comparisons of these analytic expressions to numerical solutions of Eq.~\eqref{eq:phi-kep}, for a system with $e=0.99$. The analytic expressions above are given by the dashed lines in the top panel of each plot, while the numerical solutions are given by the solid lines. The bottom panels display the difference between the numerical and analytical solutions. From this we see that the analytic solutions are highly accurate near pericenter ($\ell = 0$), while the difference is $\sim 10^{-3}$ for $\cos V$ and $\sim 2\times 10^{-2}$ for $\sin V$ at apocenter ($\ell = \pm \pi$).

As a final point, we note that the waveform for an eccentric binary doesn't just contain power in the second harmonic of $V$, but also the first and third harmonics. In theory, one can obtain Fourier series representations for $[\cos(2V), \cos(3V), \sin(2V), \sin(3V)]$ and perform the re-summation procedure presented here to obtain asymptotic representations of these harmonics. Indeed, one could even go one step further and write the full waveform in a Fourier series of orbital harmonics as is done in Eq.~\eqref{eq:h-harm}, and perform the re-summation procedure directly on the detector response $h(t)$. For simplicity, we do not consider this here, and instead rely on the usual relationship among trigonometric functions to obtain the asymptotic expressions of $[\cos(2V), \cos(3V), \sin(2V), \sin(3V)]$ from Eqs.~\eqref{eq:cos-asym}-\eqref{eq:sin-asym}. For example, 
\begin{widetext}
\begin{align}
\cos(2V) &= 2 \cos^{2}V - 1
\nn \\
&\sim 1 - \frac{8 \; \Ch(\psi,1/3)}{\sqrt{1 + \psi^{2}}} + \frac{8 \; \Ch^{2}(\psi,1/3)}{1 + \psi^{2}} + \frac{8}{5} \epsilon \left[1 - \frac{3 \; \Ch(\psi,1/3)}{\sqrt{1+\psi^{2}}} + \frac{2 \; \Ch^{2}(\psi,1/3)}{1+\psi^{2}}\right] 
\nn \\
&+ \frac{2 \epsilon^{2}}{175 (1 + \psi^{2})} \left[4 \left(64 + 35 \psi^{2}\right) - 367 \sqrt{1 + \psi^{2}} \; \Ch(\psi,1/3) + \left(111 - 5 \psi^{2}\right) \Ch(\psi,2/3)\right] + {\cal{O}}(\epsilon^{3})\,.
\end{align}
\end{widetext}
As we will show in Sec.~\ref{faith}, the waveforms obtained via this shortcut are still highly accurate compared to numerical waveforms, and thus, we do not consider further re-summations of time domain quantities.

\subsection{Radiation Reaction Model}
\label{rr}

In the previous section, we held the Keplerian eccentricity $e$ fixed when performing our re-summation procedure. However, if the binary system is inspiraling due to the emission of GWs, then this parameter will evolve in time. What we now seek is an analytic model for this evolution. This problem was first considered by Peters \& Mathews~\cite{PetersMathews}, who found that in the quadrupole approximation, the secular evolution of the Keplerian eccentricity $e$ and semi-latus rectum $p$ of the orbit evolve according to
\begin{align}
\label{eq:dedt}
\frac{de}{dt} &= - \frac{304}{15} \frac{e \eta}{M} \left(\frac{M}{p}\right)^{4} \left(1 - e^{2}\right)^{3/2} \left(1 + \frac{121}{304} e^{2}\right)\,,
\\
\label{eq:dpdt}
\frac{dp}{dt} &= - \frac{64}{5} \eta \left(\frac{M}{p}\right)^{3} \left(1 - e^{2}\right)^{3/2} \left(1 + \frac{7}{8} e^{2}\right)\,.
\end{align}
These evolution equations constitute the adiabatic approximation, where secular changes are small over any given orbit. Thus, if we are only considering the evolution of a binary system over one orbit, we can approximate the evolution of $(e,p)$ by a simple Taylor series, specifically
\allowdisplaybreaks[4]
\begin{align}
\label{eq:e-of-l}
e(t) &= e(\ell = 0) + \left(\frac{de}{d\ell}\right)_{\ell=0} \ell(t) + {\cal{O}}(\ell^{2})
\nn \\
&= e_{0} - \frac{304}{15} \frac{\eta e_{0}}{\bar{p}_{0}^{5/2}} \left(1 + \frac{121}{304} e_{0}^{2}\right) \ell(t) + {\cal{O}}(\ell^{2})\,,
\\
\label{eq:p-of-l}
\bar{p}(t) &= \bar{p}(\ell = 0) + \left(\frac{d\bar{p}}{d\ell}\right)_{\ell=0} \ell(t) + {\cal{O}}(\ell^{2})
\nn \\
&= \bar{p}_{0} \left[1 - \frac{64}{5} \frac{\eta}{\bar{p}_{0}^{5/2}} \left(1 + \frac{7}{8} e_{0}^{2}\right) \ell(t) + {\cal{O}}(\ell^{2})\right]\,,
\end{align}
where $\bar{p} = p/M$, $\bar{p}(\ell = 0) = \bar{p}_{0}$ and $e(\ell = 0) = e_{0}$ are the values at pericenter. Note that the terms proportional to $\ell$ are actually 2.5PN corrections, i.e. they are suppressed by $v^{5} \sim (m/p_{0})^{5/2}$. Thus, the Taylor series expansion also constitutes a PN expansion.

Why are the above Taylor series in terms of $\ell$ and not $t$? Naively, one might expect there to be a linear mapping $\ell(t)$ given by Eq.~\eqref{eq:kep-eq}. However, once radiation reaction is included, this mapping no longer holds and we must consider the more general mapping specified by
\begin{equation}
\label{eq:dldt}
\frac{d\ell}{dt} = M^{-1} \left(\frac{1-e^{2}}{\bar{p}}\right)^{3/2} \equiv n\,.
\end{equation}
To solve, this we insert Eqs.~\eqref{eq:e-of-l}-\eqref{eq:p-of-l} and perform a PN expansion to obtain
\begin{equation}
\frac{d\ell}{dt} = n_{0} + 2 \pi F_{\rm rr} \ell + {\cal{O}}(\ell^{2})\,,
\end{equation}
where
\begin{align}
n_{0} &= M^{-1} \left(\frac{1 - e_{0}^{2}}{\bar{p}_{0}}\right)^{3/2}\,,
\\
\label{eq:Frr}
F_{\rm rr} &= \frac{96}{10 \pi} \frac{\eta}{M \bar{p}_{0}^{4}} \left(1 - e_{0}^{2}\right)^{1/2} \left(1 + \frac{73}{24} e_{0}^{2} + \frac{37}{96} e_{0}^{4}\right)\,.
\end{align}
The above expression constitutes a differential equation for $\ell(t)$, which can be immediately solved with the requirement that $\ell(t = t_{p}) = 0$ to obtain
\begin{equation}
\label{eq:l-of-t}
\ell(t) = \frac{n_{0}}{2\pi F_{\rm rr}} \left\{\exp\left[2\pi F_{\rm rr} (t-t_{p})\right] - 1\right\}\,.
\end{equation}

\begin{figure*}[ht]
\includegraphics[clip=true,scale=0.27]{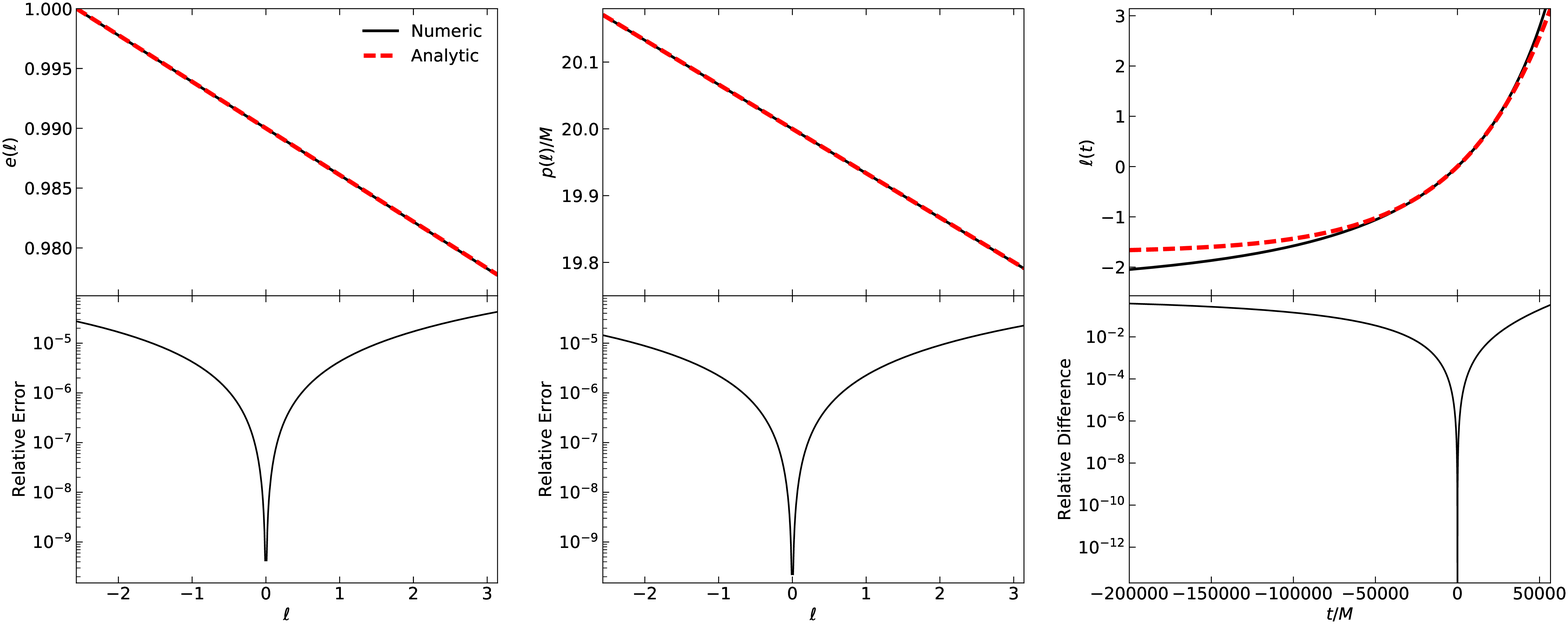}
\caption{\label{rr-comp} (Top) Comparison of the analytic approximations Eqs.~\eqref{eq:e-of-l}-\eqref{eq:p-of-l} and~\eqref{eq:l-of-t} (dashed lines) to numerical evolutions of Eqs.~\eqref{eq:dedt}-\eqref{eq:dpdt} and~\eqref{eq:dldt} (solid lines). (Bottom) The bottom panels of the left and center plots show the error between the numerical evolutions of $e$ and $p$ and their analytic representations. The bottom panel of the right plots shows the de-phasing (difference) between the numerical evolution of $\ell$ and its analytic approximation.}
\end{figure*}

In Fig.~\ref{rr-comp}, we compare the analytic approximations of Eqs.~\eqref{eq:e-of-l}-\eqref{eq:p-of-l} and~\eqref{eq:l-of-t} to the numerical evolutions of Eqs.~\eqref{eq:dedt}-\eqref{eq:dpdt} and~\eqref{eq:dldt}, for a binary system with $t_{p} = 0, e_{0} = 0.99$, and $p_{0} = 20 M$. For this evolution, the binary becomes unbound at finite $\ell \approx - 2.57$, but infinitely far in the past $t = -\infty$. The bottom panels of each plot display the error in the analytic approximation compared to the numerical evolutions. The analytic approximations of $e(\ell)$ and $p(\ell)$ are accurate to $\lesssim 10^{-2} \%$ over the full orbit. Meanwhile, the de-phasing between the analytic $\ell(t)$ and its numerical evolution is typically less than one radian near apocenter, but approaches double precision near pericenter ($t=0$). Thus, the analytic approximation provides an accurate representation of the evolution of the binary over the given orbit.

%

\subsection{Waveform Polarizations}
\label{efb-t}

Now that we have a suitable radiation reaction model, we may combine all of the pieces together to compute the waveform polarizations. In general, the plus and cross polarizations are given by Eqs.~\eqref{eq:hp-V}-\eqref{eq:hc-V} for an eccentric binary. The harmonics of the true anomaly can be replaced with their asymptotic expansions described in Sec.~\ref{resum-t}. After expanding in $\epsilon \ll 1$, we obtain
\begin{widetext}
\begin{align}
\label{eq:hpc-time}
h_{+,\times}(t) &= -\frac{M^{2} \eta}{p[\ell(t)] D_{L}}\sum_{k=0}^{6} \sum_{n=0}^{2} \epsilon^{n} \Big({\cal{C}}_{+,\times}^{(k,n)}[\ell(t); \iota, \beta] \; \Ch\{\psi[\ell(t)], k/3\}
+ {\cal{S}}_{+,\times}^{(k,n)}[\ell(t); \iota, \beta] \; \Sh\{\psi[\ell(t)], k/3\}\Big) + {\cal{O}}(\epsilon^{3})\,,
\end{align}
\end{widetext}
where the functions $[{\cal{C}}(\ell;\iota,\beta), {\cal{S}}(\ell; \iota,\beta)]$ are listed in Appendix~\ref{h-t-coeffs}, and $\ell(t)$ is given by Eq.~\eqref{eq:l-of-t}. The dependence of these functions on $\ell$ comes from the now time evolving eccentricity $e(\ell)$ given by Eq.~\eqref{eq:e-of-l}. Further, the ``phase" variable $\psi$ no longer has a linear mapping to $\ell$ for the same reason. More specifically,
\begin{equation}
\label{eq:psi-of-l}
\psi[\ell(t)] = \frac{\ell(t)}{\ln\left(\frac{1 + \sqrt{1 - e[\ell(t)]^{2}}}{e[\ell(t)]}\right) - \sqrt{1 - e[\ell(t)]^{2}}}\,.
\end{equation}
We shall refer to this model as the time-domain EFB (EFB-T) model.

\begin{figure}[ht]
\includegraphics[clip=true,scale=0.27]{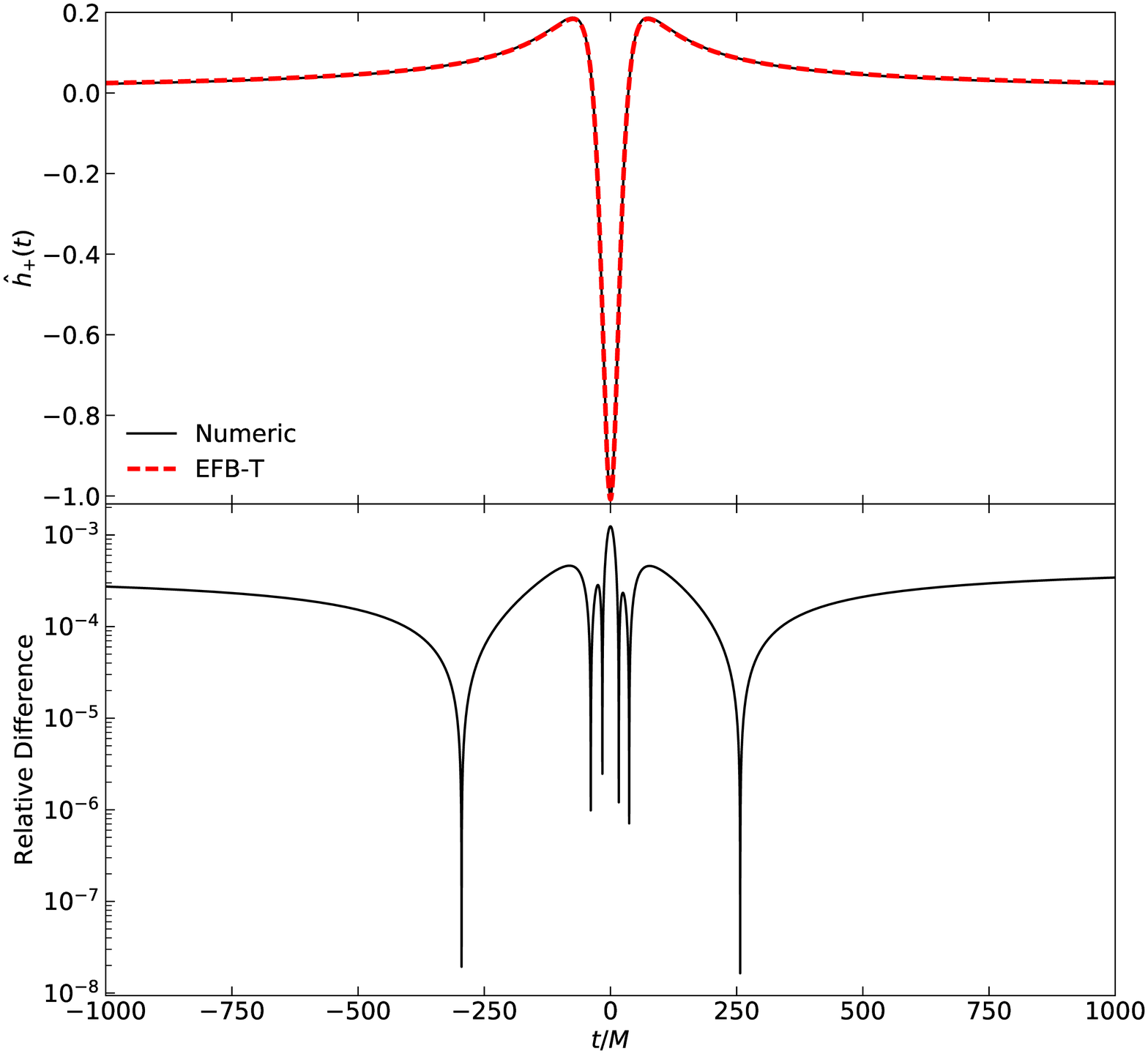}
\caption{\label{hpt} (Top) Comparison of the plus polarization of the EFB-T model (dashed line) to a numerically generated waveform (solid line) with $\bar{p}_{0} = 20$ and $e_{0} = 0.99$. (Bottom) Relative difference between the analytic and numerical waveforms.}
\end{figure}

In Fig.~\ref{hpt}, we compare the EFB-T model to a numerically generated, leading PN order waveform. The waveform is obtained by numerically integrating Eqs.~\eqref{eq:phi-kep},\eqref{eq:dedt}-\eqref{eq:dpdt}, and~\eqref{eq:dldt}, with $\bar{p}_{0} = 20$ and $e_{0} = 0.99$. The numerical solution is then combined with Eq.~\eqref{eq:hp-V}-\eqref{eq:hc-V} to generate the waveform. For simplicity, we take $\iota = 0 = \beta$, and we only plot the plus polarization, where we have normalized the waveforms such that $\hat{h}_{+}(0) = -1$ for the numerical waveform. The bottom panel of of Fig.~\ref{hpt} displays the relative difference between the two waveforms, which is $\lesssim 10^{-3}$ is the region around pericenter passage.

%

\section{Frequency Domain Waveforms}
\label{freq-domain}

So far, our efforts toward creating analytic waveforms have focused on the time domain. The EFB-T model has one particular drawback, namely, there does not seem to be a straightforward way of analytically calculating its Fourier transform. The typical method of using the stationary phase approximation (SPA) does not seem to work in this case due to the complexity of the waveforms, as well as the lack of a readily identifiable waveform phase. We here present an alternative model which does allow for the Fourier transform to be computed analytically. We shall refer to this new model as the EFB-F model.

Just like the EFB-T model, we will also follow a re-summation procedure for the Fourier domain waveform presented here. The procedure is as follow: (1) starting from Eq.~\eqref{eq:h-harm} with the radiation reaction model of Sec.~\ref{rr}, evaluate the Fourier transform of $h_{+,\times}$ using the SPA, (2) replace the Bessel functions appearing in $[C_{+,\times}^{(k)}, S_{+,\times}^{(k)}]$ with their uniform asymptotic expansions in Eqs.~\eqref{eq:J-asym}-\eqref{eq:Jp-asym}, (3) replace the summations on $k$ with an integral and evaluate. We will explain the reasoning and some of the difficulties that arise from this procedure in the following sections.

\subsection{Stationary Phase Approximation}

We desire the frequency domain waveform polarizations $\tilde{h}_{+,\times}(f)$. To do so, we consider the waveform polarizations given by Eq.~\eqref{eq:h-harm}. The Fourier transform is then, schematically
\begin{align}
\tilde{h}_{+,\times}(f) &= h_{0} \sum_{k} \int_{-\infty}^{\infty} dt \left[\frac{1-e(t)^{2}}{p(t)}\right] \exp(2\pi i f t) 
\nn \\
&\times
\left\{E_{+,\times}^{(k)}(t) \exp[i k \ell(t)] + \text{c.c.}\right\}\,,
\end{align}
where $h_{0} = M^{2}\eta/(2 D_{L})$, $E_{+,\times}^{(k)} = C_{+,\times}^{(k)} - i S_{+,\times}^{(k)}$, and c.c. stands for complex conjugate. Here, $(p, e, \ell)$ are still time dependent through Eqs.~\eqref{eq:e-of-l}-\eqref{eq:p-of-l} and~\eqref{eq:l-of-t}, and $E_{+,\times}^{(j)}$ depend on time through the eccentricity $e$. The problem of calculating the Fourier transform now reduces to solving an integral of the form
\begin{equation}
I(f) = \int_{-\infty}^{\infty} dt A(t) \exp[i \Psi_{\pm}(t,f)]\,,
\end{equation}
where $\Psi_{\pm}(t,f) = 2\pi f t \pm k \ell(t)$, and for which the SPA is applicable. The stationary point is found by requiring that $d\Psi_{\pm}/dt = 0$, which results in
\begin{equation}
t^{*}_{j,\pm} = t_{p} + \frac{1}{2\pi F_{\rm rr}} \ln\left(\mp \frac{2 \pi f}{k n_{0}}\right)\,.
\end{equation}
Note that the stationary point of $\Psi_{+}$ is only real valued for negative values of the frequency, while for $\Psi_{-}$, this occurs at positive frequencies. Since we are only interested in the signals observed by GW detectors, we drop the contribution to the Fourier transform from $\Psi_{+}$ since it is only dominant for negative frequencies. The remainder of the SPA procedure may be carried out to obtain
\begin{align}
\tilde{h}_{+,\times}(f) &= h_{0} \sum_{k} \frac{[1-e(t^{*}_{k,-})^{2}]}{p(t^{*}_{k,-})} E_{+,\times}^{(k)\dagger}(t^{*}_{k,-}) \left(\frac{k \chi_{\rm orb}}{\chi}\right)^{-i\chi} 
\nn \\
&\times \frac{\exp[i(k \chi_{\rm orb} - \chi - \pi/4 + 2 \pi f t_{p})]}{F_{\rm rr} \sqrt{2\pi\chi}}\,,
\end{align}
where $\chi = f/F_{\rm rr}$, $\chi_{\rm orb} = n_{0}/(2\pi F_{\rm rr})$, and $\dagger$ corresponds to complex conjugation.
\subsection{Re-summations in the Fourier Domain and Waveform Polarizations}

After applying the SPA, we are still left with a waveform that involves an infinite summation over harmonics. The question now is whether a similar re-summation procedure to the time domain waveforms can be carried out here. The functions $E_{+,\times}^{(j)}$ involve the exact same Bessel functions, so we may replace them with their asymptotic expansions given in Eqs.~\eqref{eq:J-asym}-\eqref{eq:Jp-asym}. Further, there is nothing preventing us from replacing the infinite summations with integrals. The only practical problem is whether these integrals can be evaluated in closed form. The integrals generally take the form
\begin{equation}
J_{a} = \int_{0}^{\infty} dk \; k^{-i\chi + a} K_{b}\left[\frac{2}{3}k \zeta^{3/2}(t^{*}_{k,-})\right] \exp[i k \chi_{\orb}]\,,
\end{equation}
where $a \le 1$ and $b\in\{1/3,2/3\}$. There are two problems associated with trying to evaluate this. The first arises from the stationary point dependence in $\zeta$. This depends on the stationary point through $e(t)$ given by Eq.~\eqref{eq:e-of-l}, which after evaluating produces
\begin{equation}
e^{*}_{k,-} = e(t^{*}_{k,-}) = e_{0} + \frac{304}{15} \frac{e_{0} \eta}{\bar{p}_{0}^{5/2}} \left(1 + \frac{121}{304} e_{0}^{2}\right) \left(\chi_{\rm orb} - \frac{\chi}{k}\right)\,.
\end{equation}
The dependence on $k$ in the modified Bessel function is thus complicated, and in general, the integral does not have a closed form solution. Fortunately, we may realize that the $k$ dependence in $e^{*}_{k,-}$ is suppressed by $\bar{p}_{0}^{5/2}$, and is thus 2.5PN order. As a result, we perform a PN expansion of any quantities that depend on the stationary point $t^{*}_{k,-}$. For example,
\begin{equation}
K_{b}\left[\frac{2}{3}k \zeta^{3/2}(t^{*}_{k,-})\right] = K_{b} \left(\frac{2}{3} k \zeta_{0}^{3/2}\right) + {\cal{O}}\left(\bar{p}_{0}^{-5/2}\right)\,,
\end{equation}
where $\zeta_{0} = \zeta(e_{0})$.

The second issue arises when $a < 0$. Similar to the time domain re-summation, these integrals appear to be divergent when $k \rightarrow 0$. However, this can be circumvented by realizing that $k^{-1} \exp(i k \chi_{\rm orb}) = \int d\chi_{\rm orb} \exp(i k \chi_{\rm orb})$. Utilizing this, we can reverse the order of integration, first integrating over $k$ and then over $\chi_{\rm orb}$, to evaluate $J_{a}$ when $a < 0$. We find that these terms are actually subdominant compared to the $a=1$ and $a=0$ terms in the Fourier domain waveform, so we safely neglect them here.

%
%
%

After applying the re-summation procedure, we obtain
\begin{widetext}
\begin{align}
\label{eq:hpc-f}
\tilde{h}_{+,\times}(f) = \frac{M^{2} \eta}{p_{0} D_{L}} \frac{(1 - e_{0}^{2})}{e_{0}^{2} F_{\rm rr}} \left(\frac{\chi}{\chi_{\rm orb}}\right)^{i\chi} \frac{\exp(2\pi i f t_{p}-i\chi)}{\chi^{1/2}} \sum_{(l_{1}, l_{2}) \in L} \sum_s {\cal{A}}_{l_{1}, l_{2}, s}^{+,\times}(f) \; {}_{2} F_{1}\left(\frac{l_{1}}{6} - i \frac{\chi}{2}, \frac{l_{2}}{6} - i \frac{\chi}{2}; s; - \frac{9}{4} \frac{\chi_{\rm orb}^{2}}{\zeta_{0}^{3}} \right)\,,
\end{align}
\end{widetext}
where ${}_{2}F_{1}$ is the hypergeometric function, the functions ${\cal{A}}_{l_{1}, l_{2}, s}(f)$ are listed in Appendix~\ref{h-f-coeffs}, $(l_{1}, l_{2})$ are integers that belong to the set $L = \{(2,4), (4,8), (1,5), (5,7), (7,11), (10,8) \}$, and $s \in \{-1/2,1/2\}$. Note that we have not expanded this expression about $\epsilon = 1-e_{0}^{2} \ll 1$. In attempting this, we discovered that this results in a severe loss of accuracy compared to numerical waveforms. As a result, we simply leave the above expression un-expanded. 

\begin{figure}[ht]
\includegraphics[clip=true,scale=0.27]{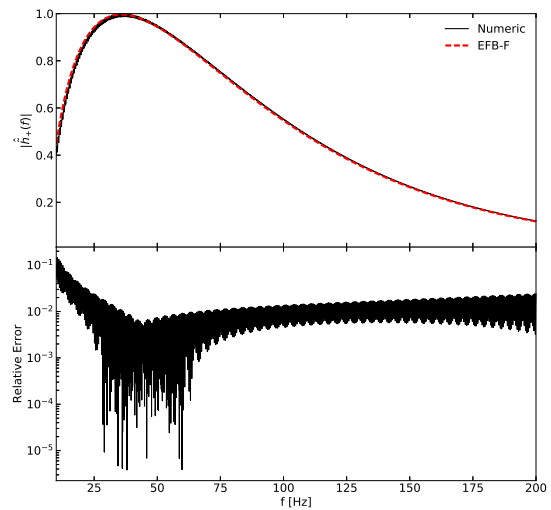}
\caption{\label{hpf} (Top) Comparison of the plus polarization of the EFB-F model in Eq.~\eqref{eq:hpc-f} (dashed line) to a numerically generated waveform (solid line) with $\bar{p}_{0} = 20$ and $e_{0} = 0.99$. (Bottom) Relative error between the analytic and numerical waveforms.}
\end{figure}

We provide a comparison of the EFB-F waveform to a numerically computed one in Fig.~\ref{hpf}. The numerical waveform is generated by taking the discrete Fourier transform (DFT) of the numerical time-domain waveform discussed in Sec.~\ref{efb-t}. We choose the masses for this comparison to be $(m_{1} , m_{2}) = (10,10) M_{\odot}$. The time-domain waveform is sampled at 4096 Hz, and is then padded such that the total length of the waveform contains $2^{20}$ points. The top panel of Fig.~\ref{hpf} displays the plus polarization of both the EFB-F model (dashed line) and the numerical waveform (solid line), normalized to the peak of the numerical waveform. The relative error between these two waveforms is largest at 10 Hz, with the EFB-F waveform being accurate to $\sim 10\%$. At higher frequencies, the EFB-F waveform is more accurate, achieving $\sim 1\%$ accuracy at frequencies above the peak. In principle, this can be improved by considering the next order terms in the asymptotic expansion of Bessel functions, as well as hyperasymptotic techniques~\cite{Boyd} to adjust the low frequency behavior. This completes our discussion of the EFB-F model.

%

\section{Validation}
\label{valid}

Now that we have our waveform models, we seek to validate them against accurate representations of the GW bursts from eccentric systems. We shall also discuss some numerical implementations of these waveforms models, specifically how quickly they may be evaluated.

\subsection{Computational Efficiency}
\label{eff}

While it is appealing to have analytic waveforms from the standpoint of understanding the underlying physics, from a data analysis perspective, it is also necessary for these waveforms to be fast to evaluate. If an analytic waveform is sufficiently complicated, it may take more time to sample the model than it would to simply generate the waveform numerically. In this section, we seek to quantify the efficiency of the EFB-T and EFB-F waveform models presented in Secs.~\ref{time-domain} \&~\ref{freq-domain}, specifically how quickly they can be sampled relative to a numerical waveform. 

The benchmark for this will be a leading PN order waveform given by Eq.~\eqref{eq:hp-V}, and which is generated by evolving the equations of Peters \& Mathews, specifically Eqs.~\eqref{eq:dedt}-\eqref{eq:dpdt}, and Eqs.~\eqref{eq:phi-kep} and~\eqref{eq:dldt}. These equations are solved in the range $[-t_{f}, t_{f}]$, where $t_{f} = \pi/M^{1/2} [p_{0}/(1-e_{0}^2)]^{3/2}$, to ensure that only one pericenter passage is obtained. We choose $\bar{p}_{0} = 20$ and $e_{0} = 0.9$, with masses $m_{1} = 10 M_{\odot} = m_{2}$, and a sampling rate of 4096 Hz. The time domain waveform is padded with zeros until the total length is $2^{20}$ points, before being Fourier transformed. The time domain integration is performed with \texttt{SciPy}'s \texttt{ode} module, while the Fourier transform is computed numerically using the $\texttt{fft}$ module. With these parameter values, it takes approximately 0.36 seconds to generate the waveform.

To generate the EFB-T waveforms, after setting the initial parameters $(p_{0}, e_{0})$ and masses $(m_{1}, m_{2})$, we find the time $t_{\pi}$ associated with $\ell = \pi$ using Eqs.~\eqref{eq:l-of-t}. We then sample the EFB-T waveform over the interval $[-t_{\pi}, t_{\pi}]$ at a rate of 4096 Hz. The waveform is then padded to ensure there are $2^{20}$ total points. After padding, the Fourier transform is then computed used \texttt{SciPy}'s \texttt{fft} routine. For initial parameters $\bar{p}_{0} = 20$ and $e_{0} = 0.9$, and masses $m_{1} = 10 M_{\odot} = m_{2}$, it takes approximately $0.14$ seconds to generate both waveform polarizations in the EFB-T model. This is under half the time to generate the numerical Peters \& Mathews waveforms.

It is worth noting that this evaluation time increases significantly as $e_{0}$ approaches unity for the sampling method described above. For $e_{0} = 0.999$, it takes approximately $0.56$ seconds to generate the EFB-T waveform. For $e_{0} = 0.9999$, the sampled EFB-T waveform has $>2^{20}$ points, simply due to the fact that $t_{\pi}$ can become large (i.e. it takes a long time to get from pericenter to apocenter). For such a case, we pad the EFB-T waveform to have a total of $2^{22}$ points, which takes approximately $3.6$ seconds total to generate the Fourier transform. This can be circumvented by simply choosing a smaller window over which the sampling is performed, i.e. choose $t_{\rm sample} < t_{\pi}$. For such high eccentricities, sampling to apocenter ($\ell = \pi$) is likely unnecessary since there is very little GW emission there. The same issue arises for the numerical waveform that we are comparing to.

While the EFB-T model is relatively fast to evaluate, the same cannot be said of the EFB-F waveform in Eq.~\eqref{eq:hpc-f}. From the numerical and EFB-T waveforms, we obtain a frequency resolution $\delta f$, which we use to sample the EFB-F waveforms from $f_{\rm low} = 10$ Hz to $f_{\rm high} = 2048$ Hz, i.e. the Nyquist frequency. We attempted to sample the EFB-F model in \texttt{Python} using the \texttt{mpmath}~\cite{mpmath} module, but sampling the hypergeometric function proved to be badly convergent for high frequency values, and we were not able to get a full estimate of the time it would take to sample the EFB-F model. The data used to generate Fig.~\ref{hpf} were generated in \texttt{Mathematica}, where it took $\sim 3-4$ hours to complete the sampling, and which doesn't cover the full frequency range.

The reason behind the slow evaluation time of the EFB-F model seems to purely be due to its dependence on the specialized hypergoemetric functions, which are not easily evaluated numerically for large arguments. There are actually two variables that are large in the EFB-F model. The first is $\chi = f /F_{\rm rr}$. For LIGO sources, $F_{\rm rr}$ is typically less than 1 Hz, so $\chi$ can span over several orders of magnitude. The second is the ratio $\chi_{\rm orb}^{2}/\zeta_{0}^{3}$, which is actually a $-2.5$-PN term, i.e. it scales like $v^{-5}$. For small values of the velocity, this ratio is large, and common methods of numerically evaluating the hypergeometric functions are poorly convergent. 

Given these two considerations, it may be possible to produce analytic approximations to the hypergeometric functions appearing in Eq.~\eqref{eq:hpc-f} that would be significantly faster to evaluate. This was attempted in the course of this work, but the resulting approximates were not sufficiently accurate over the full range of frequencies of the LIGO band, so we do not provide the details here. This is not to say that these methods are total failures in speeding up the waveform, only that more work would be necessary to obtain sufficiently accurate waveforms that are also fast to evaluate. We leave this to future work. Due to the excessive computation cost of evaluating the EFB-F model, the remainder of the numerical analysis performed in this section is done only with the EFB-T model.

\subsection{Faithfulness}
\label{faith}

In order to construct the EFB-T model, we were forced to make a few approximations, namely the post-parabolic approximation for the conservative dynamics, and an approximate model based on Taylor expansions of the radiation reaction equations for the dissipative dynamics. In Figs.~\ref{comp} and~\ref{hpt}, we showed the difference between these analytic approximations and numerical calculations. While this is suitable for checking the accuracy of these approximations, it would also be useful to understand how errors in these approximations might affect our ability to detect, and perform parameter estimation on, such signals. To that end, we study the match between the EFB-T model and numerical Peters \& Mathews waveforms

The match between waveforms $h_{A}$ and $h_{B}$ is defined as
\begin{equation}
\label{eq:match}
\text{M} = \underset{t_{p}}{\max} \frac{(h_{A} | h_{B})}{\sqrt{(h_{A} | h_{A}) (h_{B} | h_{B})}}\,,
\end{equation}
where $(h_{A} | h_{B})$ is the noise-weighted inner product defined as
\begin{equation}
\label{eq:in-prod}
(h_{A} | h_{B}) = 4 \text{Re} \int df \; \frac{\tilde{h}_{A}(f) \tilde{h}_{B}^{\dagger}(f)}{S_{n}(f)}\,,
\end{equation}
with $S_{n}(f)$ the noise power spectral density of the detector being considered. The match in Eq.~\eqref{eq:match} is maximized over the time of pericenter passage, which amounts to an arbitrary time shift of the waveform. Physically, the match provides an estimate of how biased, or unbiased, parameter estimation will be if model $h_{B}$ is used to detect signal $h_{A}$. In this case, we are using it to determine how faithful the EFB-T model is compared to the numerical Peters \& Mathews waveforms discussed in the previous section.

\begin{figure*}[ht]
\includegraphics[clip=true,scale=0.27]{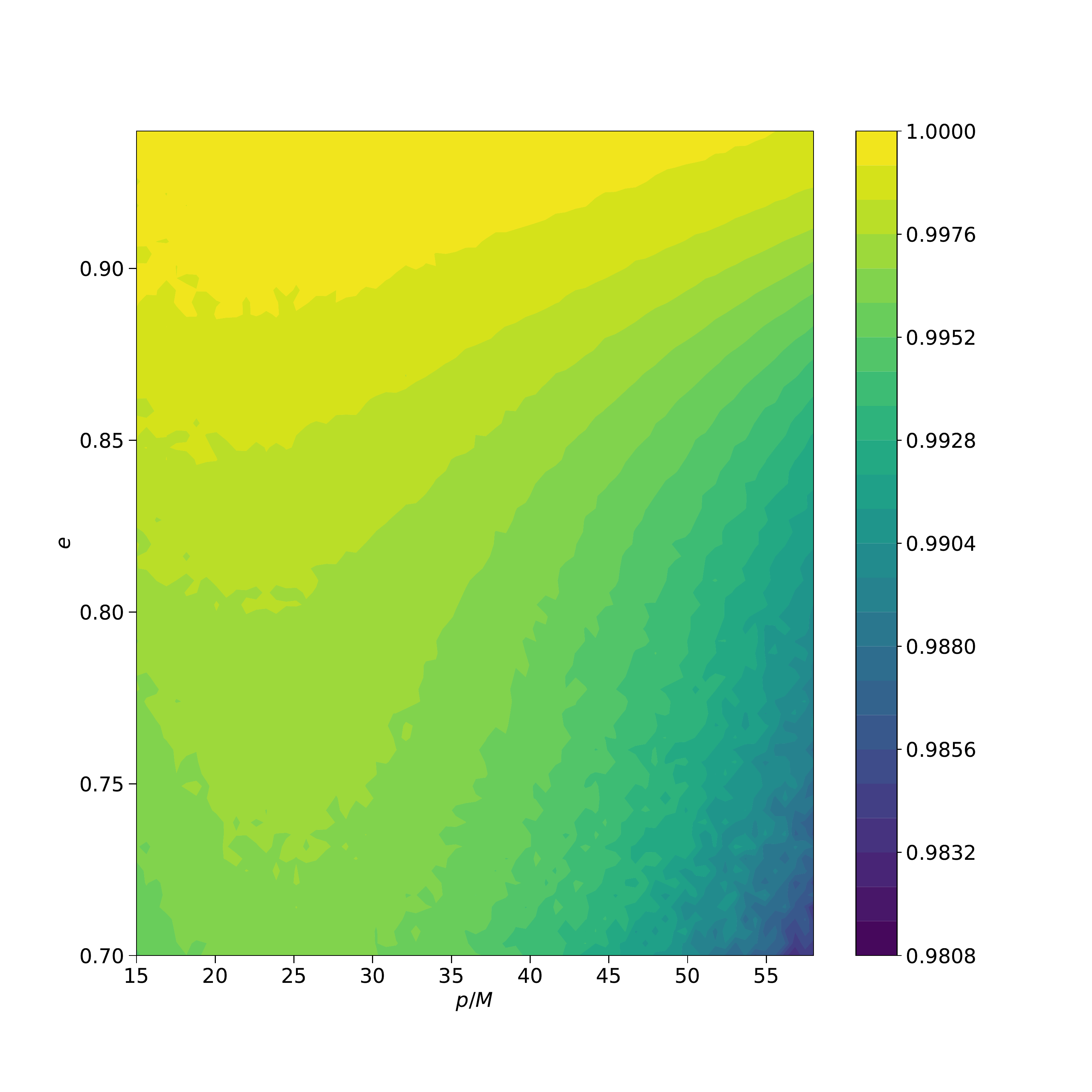}
\includegraphics[clip=true,scale=0.27]{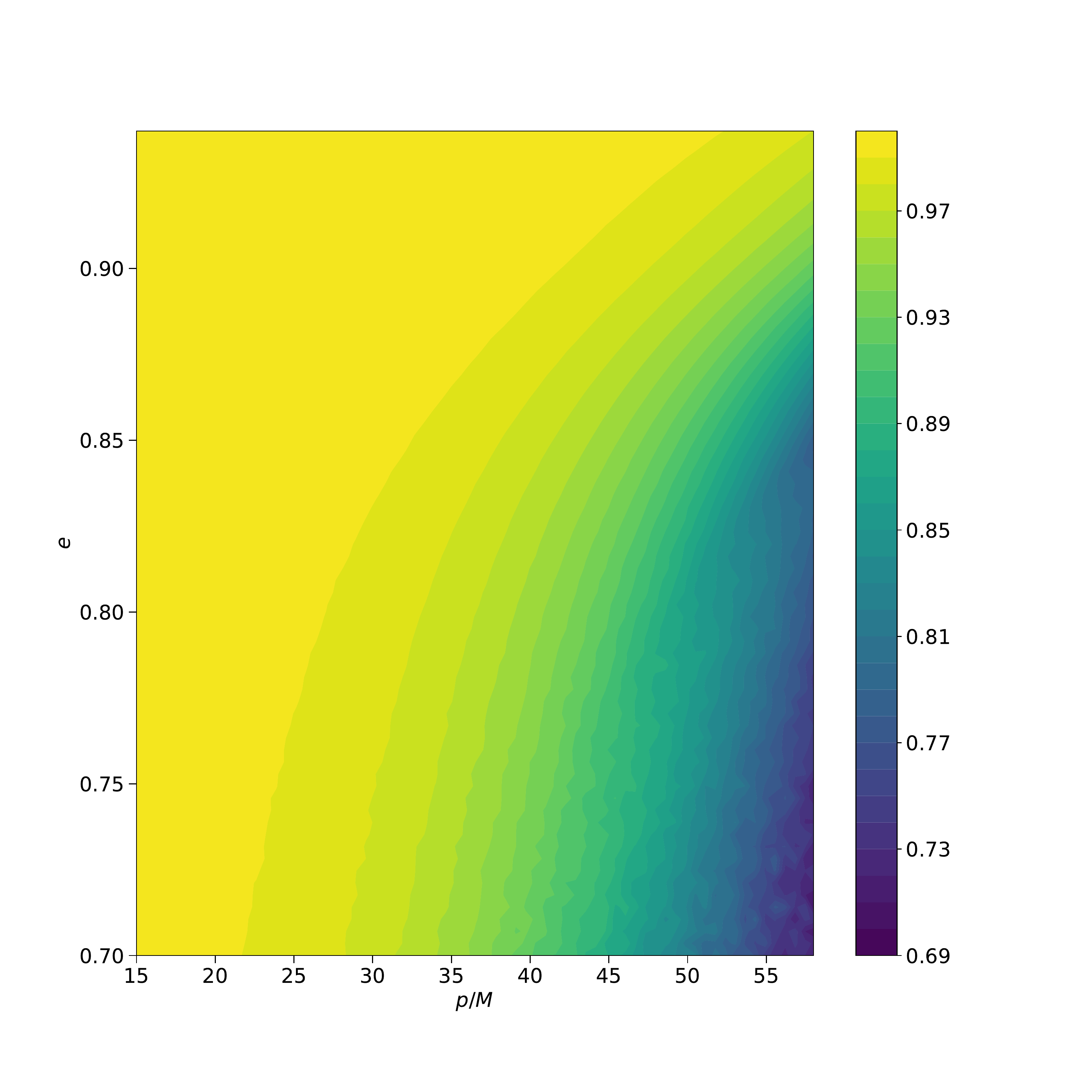}
\caption{\label{pw} (Left) Faithfulness (match) of the EFB-T waveform to numerical Peters \& Mathews waveforms for binaries with total mass $M = 20 M_{\odot}$ and mass ratio $q = 1$. (Right) Same as the left panel, but with $M = 50 M_{\odot}$ and $q = 4$. Note that the color scale is different in each plot.}
\end{figure*}

We consider the match for two LIGO sources, with $m_{1} = 10 M_{\odot} = m_{2}$, and $m_{1} = 10 M_{\odot}$ and $m_{2} = 40 M_{\odot}$. We compute the match as a function of the semi-latus rectum $p$ and eccentricity $e$ of the orbit. For simplicity, we only compute the match between the plus polarizations of the waveforms. The Fourier domain waveforms are computed via the method discussed in Sec.~\ref{eff}. For $S_{n}(f)$, we use the publicly available data for LIGO at design sensitivity~\cite{LIGOsn}. To compute the integral in Eq.~\eqref{eq:in-prod}, we take the limits of integration as $f_{\rm low} = 10$ Hz and $f_{\rm high} = f_{\rm Ny}$. Finally, to maximize over the time shift $t_{p}$, we compute the inverse Fourier transform of the integrand in Eq.~\eqref{eq:in-prod}, and find its maximum. This gives us an approximate value of $t_{p}$ that maximizes Eq.~\eqref{eq:match}. We then perform a grid search around this point to find the true value. The results of this computation are displayed in Fig.~\ref{pw}.

For the $m_{1} = 10M_{\odot} = m_{2}$ case, the match is always above $\sim 0.98$ for all of the cases studied. This is above the threshold of $0.97$ that is commonly used as a requirement\footnote{It is worth noting a more stringent requirement could be obtained by considering the percentage of events lost during a search, which scales as $1-\text{M}^{3}$.} for performing matched filtering searches~\cite{Buonanno:2009zt}. Thus, for this low mass case, the EFB-T waveform is an excellent approximation for the typical bursts that might occur within the LIGO band. On the other hand, the high mass case with $m_{1} = 10M_{\odot}$ and $m_{2} = 40 M_{\odot}$ only achieves such high matches for high values of eccentricity and/or low values of the semi-latus rectum. The reason for this is that the peak frequency of the waveforms is sensitive to the total mass of the binary, the eccentricity, and the semi-latus rectum, which determines how widely separated the binary is. For higher masses and higher semi-latus recta, the peak frequency can become smaller than $10$ Hz, resulting in only the exponential high-frequency tail being within the LIGO band, and lower matches overall. This is not unexpected, a similar effect occurs in the quasi-circular case, i.e. higher mass systems generally merge at lower frequencies, and as a result, spend less time in the LIGO band. 

\subsection{Robustness to Modeling Error}
\label{nr-comp}

The calculation in the previous section provides a useful measure of errors introduced by the approximations necessary to create the EFB-T waveforms, and show that they are a relatively faithful representation of the numerical Peters \& Mathews waveforms. However, the pericenter velocity of binaries emitting GWs in the detection band of ground based detectors need not necessarily be small. One may wonder how accurate the EFB-T waveform is compared to a realistic signal. More specifically, what is the (modeling) error induced by working to leading PN order?

We seek to answer this question by comparing to full NR waveforms of single pericenter passages. We use the waveforms from~\cite{2012PhRvD..85l4009E}, which specifically looked at black hole-neutron star binaries with $q = 4$. We are primarily interested in the case of binary black holes (BBHs), and while these simulations may treat one of the compact objects as a neutron star, the underlying dynamics should be an accurate trace of the BBH case since tidal effects and f-mode oscillations are subdominant in the waveforms~\cite{Yang:2018bzx}. For our analysis, we thus choose the masses to be $m_{1} = 10 M_{\odot}$ and $m_{2} = 40 M_{\odot}$.

The initial data for these simulations is set by choosing velocities corresponding to a Newtonian orbit with parameters $[r_{p}, e]$. The time domain data from the NR simulations is for the Weyl scalar $\Psi_{4} = \ddot{h}_{+} + i \ddot{h}_{\times}$. To perform a match comparison, we require the Fourier domain waveform $\tilde{h}_{+}(f)$, which we compute by using the properties of Fourier transforms to realize that 
\begin{equation}
\tilde{h}_{+}(f) = - \frac{{\cal{F}}\{\text{Re}[\Psi_{4}]\}}{4\pi^{2} f^{2}}\,,
\end{equation}
where ${\cal{F}}[h]$ is shorthand for the Fourier transform of $h$. For simplicity, we once again only consider the match between plus polarizations of the NR waveforms and the EFB-T model.

\begin{figure*}[ht]
\includegraphics[clip=true,scale=0.27]{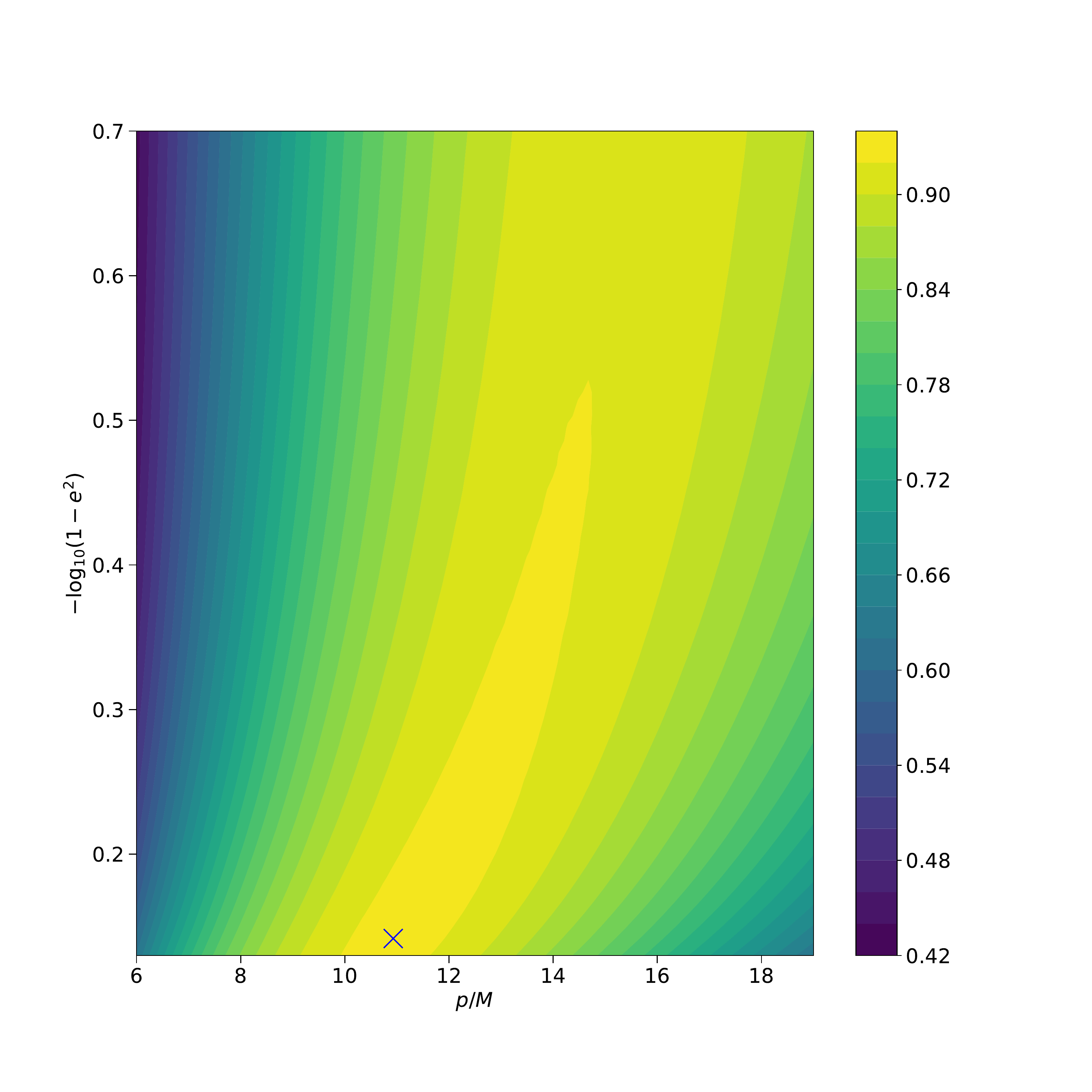}
\includegraphics[clip=true,scale=0.27]{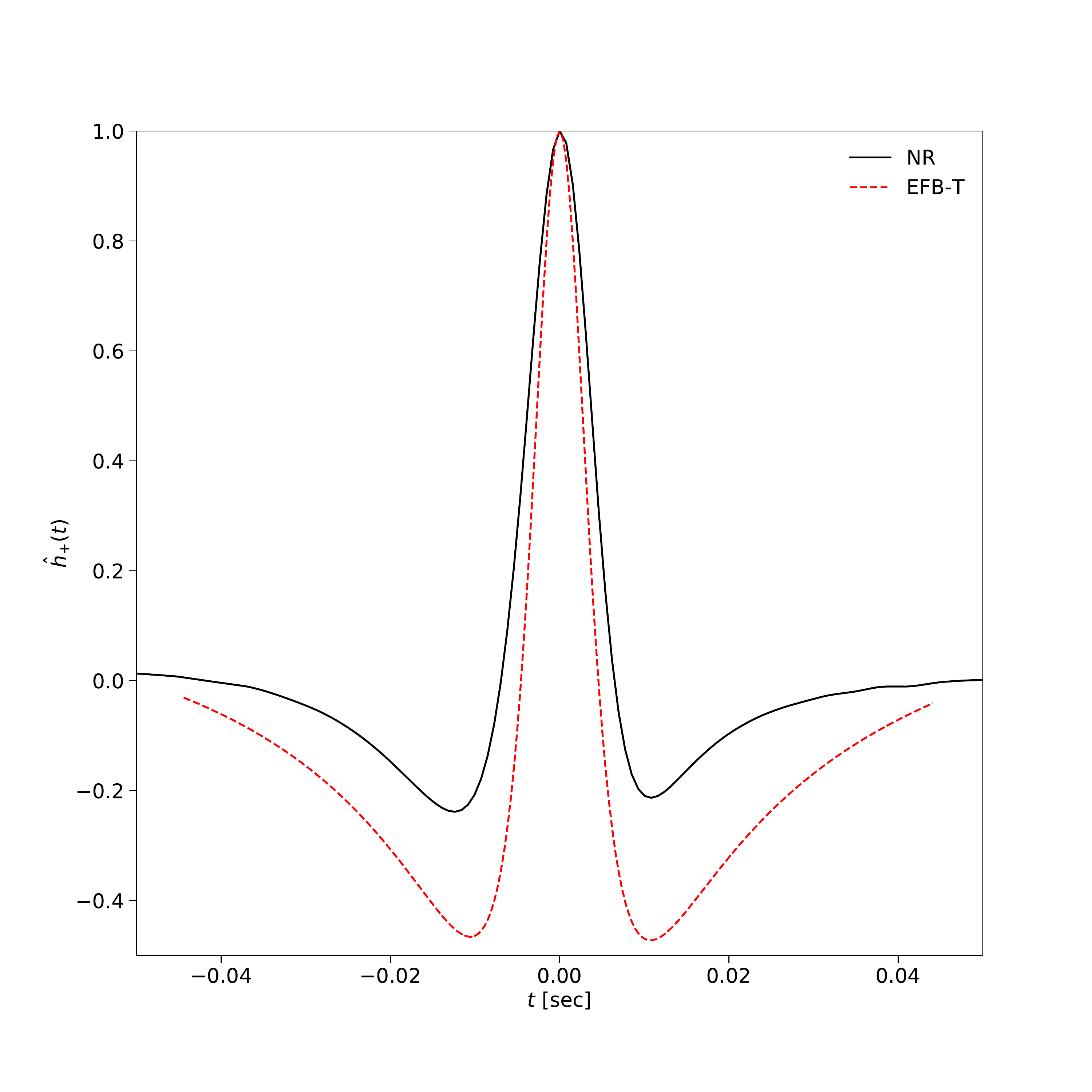}
\caption{\label{match1} (Left) Match (color) between the EFB-T waveform model and an NR waveform with input Newtonian values $r_{p} = 10 M$ and $e_{0} = 0.75$, and masses $m_{1} = 10 M_{\odot}$ and $m_{2} = 40 M_{\odot}$. The maximum match, displayed by the cross symbol, is 0.927, and is achieved at $p_{0} = 10.9 M$ and $e_{0} = 0.528$, which corresponds to $r_{p,0} = 7.13M$. (Right) Comparison of the best fit EFB-T waveform to the NR waveform. Both waveforms have been normalized so that $\hat{h}_{+} = 1$ at the peak amplitude. }
\end{figure*}

\begin{figure*}[ht]
\includegraphics[clip=true,scale=0.27]{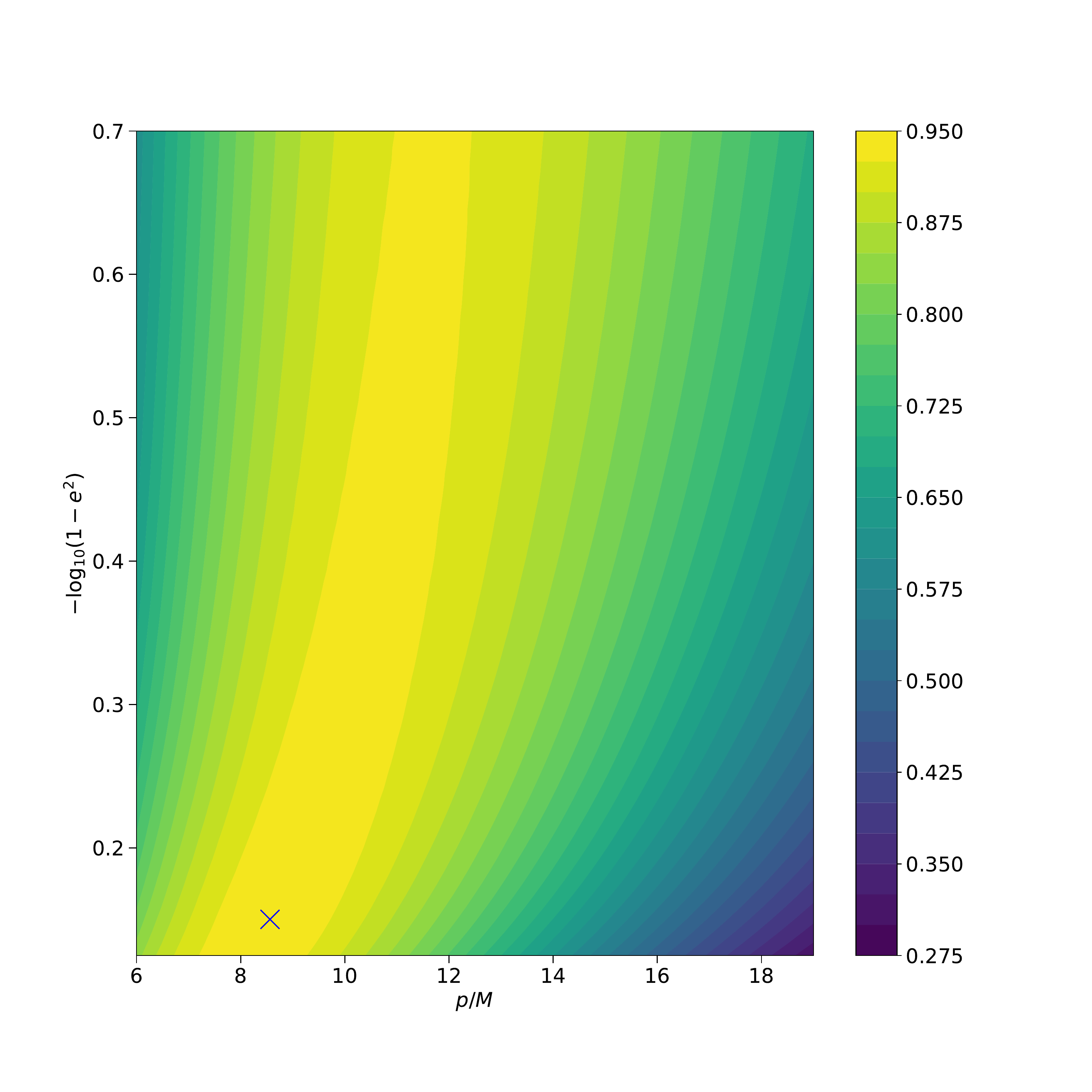}
\includegraphics[clip=true,scale=0.27]{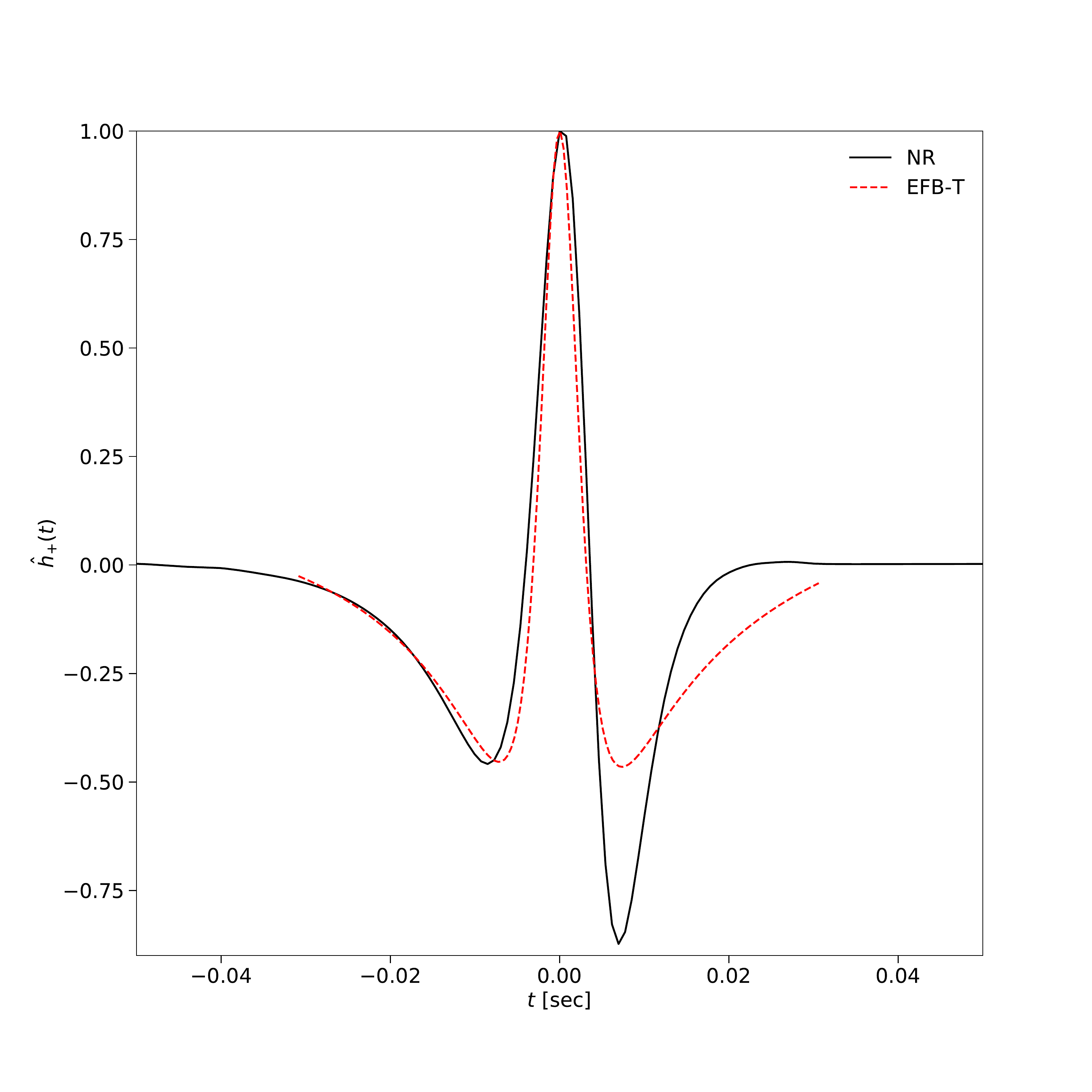}
\caption{\label{match2} The same as Fig.~\ref{match1}, but with an NR waveform with input Newtonian values of $r_{p} = 8.75M$ and $e_{0} = 0.75$. The maximum match is 0.945, and is achieved at $p_{0} = 8.56 M$ and $e_{0} = 0.541$, which corresponds to $r_{p,0} = 5.55M$.}
\end{figure*}

\begin{figure*}[ht]
\includegraphics[clip=true,scale=0.27]{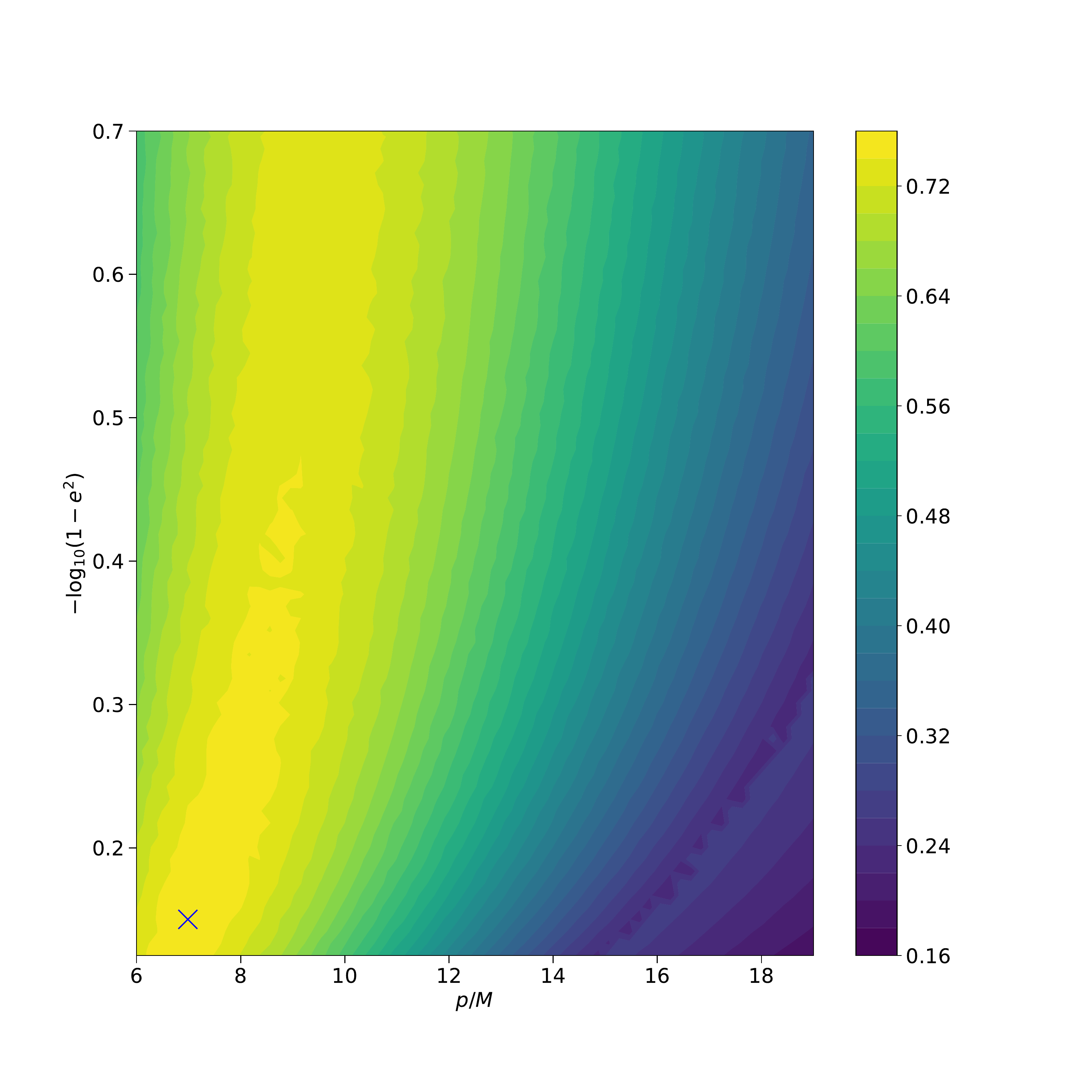}
\includegraphics[clip=true,scale=0.27]{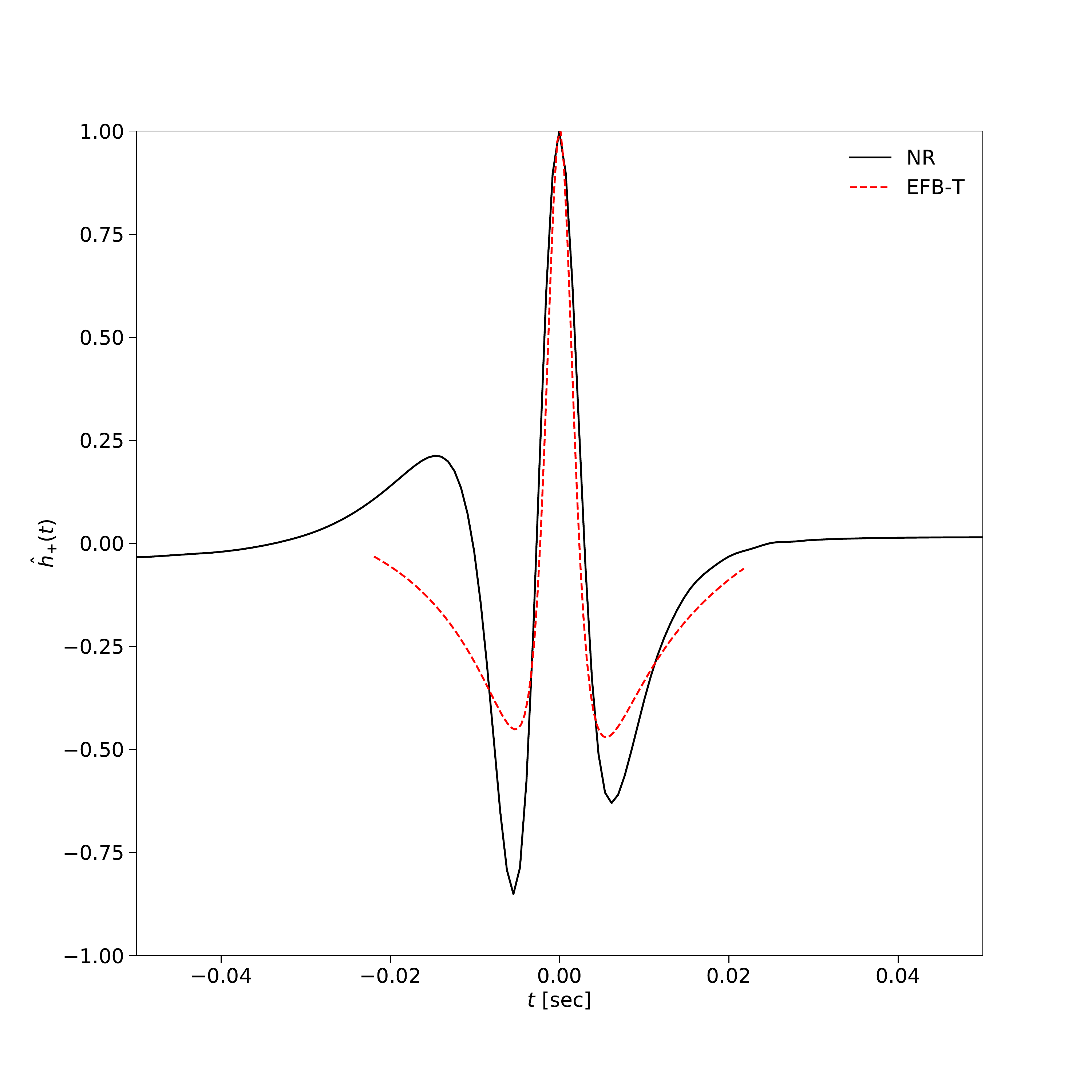}
\caption{\label{match3} The same as Fig.~\ref{match1}, but with an NR waveform with input Newtonian values of $r_{p} = 8.125M$ and $e_{0} = 0.75$. The maximum match is 0.754, and is achieved at $p_{0} = 6.99 M$ and $e_{0} = 0.541$, which corresponds to $r_{p,0} = 4.54M$.}
\end{figure*}

The NR simulations are discretized with a time step of $\delta t = 1.5625 M$, which corresponds to the sampling rate of $2586.34$ Hz for the masses we have chosen. The method for computing the Fourier transform of $\text{Re}[\Psi_{4}]$ follows the same procedure detailed in Sec.~\ref{eff} for computing the Fourier transform of the EFB-T model. For the match comparison, the values of the Newtonian parameters $[r_{p}, e]$ of the NR simulations need not give the best match for the EFB-T waveforms. This is due to the fact that these Newtonian parameters do not correspond to the true pericenter and eccentricity of the orbit, as well as the EFB-T model is not an exact representation of the NR waveform, so its parameters can be biased. We thus vary these parameters in the EFB-T model, or more specifically $[p_{0}, e_{0}]$, to find the highest match possible. The results of this calculation are displayed in Figs.~\ref{match1}-\ref{match3}.

The left panel of Fig.~\ref{match1} shows the results for the NR simulation with $r_{p} = 10M$. The maximum match is $0.927$, and is achieved at values of $p_{0} = 10.9M$ and $e_{0} = 0.528$ for the EFB-T model. The right panel compares the ``best fit" EFB-T waveform to the NR waveform in the time domain. To obtain $h_{+}(t)$ for the NR simulation, we simply compute the inverse Fourier transform of $\tilde{h}_{+}(f)$ using \texttt{SciPy}'s \texttt{ifft} module. Both the NR and EFB-T waveforms are normalized by their peak amplitudes. This comparison shows that the two waveforms have the same morphology, but differ by their amplitudes, which is consistent with what was found in~\cite{Stephens:2011as}.

Fig.~\ref{match2} shows the results of the same comparison, but for an NR simulation with $r_{p} = 8.75M$. In this case, the maximum match is $0.945$, and is achieved at EFB-T parameters $p_{0} = 8.56M$ and $e_{0} = 0.541$. The NR waveform displays an asymmetry around its peak amplitude, due to the binary exhibiting whirl-like behavior around closest approach~\cite{Stephens:2011as}. This effect is not captured by the EFB-T model. Finally, Fig.~\ref{match3} gives the results for the NR simulation with $r_{p} = 8.125M$, with the maximum match of $0.754$ at EFB-T parameters $p_{0} = 6.99M$ and $e_{0} = 0.541$. The NR waveform displays more of the whirl-like behavior than the previous waveform, and as a result, the match is significantly lower.


The results of this analysis show two things. First, the EFB-T waveforms are relatively robust to modeling error, but only to a point. As the pericenter distance becomes smaller, the EFB-T waveform becomes less accurate compared to NR waveforms. This is not unexpected, since the EFB-T model is constructed from leading PN order dynamics, while the Newtonian pericenter velocity is $v_{p} = 0.46 c$ for the case with $r_{p} = 8.125M$, where $c$ is the speed of light. Relativistic effects not captured by the EFB-T model become important at such high velocities. Second, the EFB-T waveform can capture these bursts, but the parameters of the model will be biased relative to the true parameters of the binary. It is difficult to tell how much the parameters are biased in this case, since the true parameters of the binary aren't actually known from the simulations. Both of these considerations necessitate the creation of more accurate waveform models to cover binaries with small pericenter distances.

\subsection{Multi-burst Sequences}
\label{multi-burst}

The previous sections show that the EFB-T model provides an accurate description of the GWs produced during a single pericenter passage, provided the pericenter distance is sufficiently large that relativistic effects can be neglected. However, binary systems will generally go through multiple pericenter passages as they pass through, and ultimately merge in the LIGO band. The usefulness of having a waveform that accurately covers only one pericenter passage seems somewhat limiting. We here show how multiple EFB-T waveforms can be combined to recover a sequence of bursts from an eccentric system. 

To begin, we generate a sequence of bursts by numerically integrating Eqs.~\eqref{eq:dedt}-\eqref{eq:dpdt} with the initial conditions $p(\ell = 0) = 60M$ and $e(\ell = 0) = 0.9$, where $\ell = 0$ corresponds to the first pericenter passage. We further choose the masses to be $m_{1} = 10 M_{\odot} = m_{2}$. We numerically integrate these equations over the range $t\in[-T_{{\rm orb},0}/2, 30 \text{sec}]$, where $T_{{\rm orb},0} = 2\pi/n_{0}$. This results in a sequence of ten pericenter passages. We could extend this to more, but this suffices for our purposes.

To generate a multi-burst EFB-T waveform, we start by generating a single waveform using the method described in Sec.~\ref{eff} with $\bar{p}_{0} = 60$ and $e_{0} = 0.9$. To generate the next burst, we must know what the parameters $[p_{1}, e_{1}]$ at the next pericenter passage will be. Fortunately, such a model was developed in~\cite{Loutrel:2014vja}. We follow a similar procedure here. As an example, the change in the eccentricity $e$ from one pericenter passage to another is given by
\begin{align}
\label{eq:e-map}
e_{I+1} &= e_{I} + \left(\frac{de}{dt}\right)_{p_{I}, e_{I}} T_{{\rm orb}, I}
\nn \\
&= e_{I} - \frac{604 \pi}{15} \frac{\eta e_{I}}{\bar{p}_{I}^{5/2}} \left(1 + \frac{121}{304} e_{I}^{2}\right)\,,
\end{align}
where we have used Eq.~\eqref{eq:dedt} to obtain the second equality. The same procedure can be used for the semi-latus rectum and time of pericenter passage to obtain
\begin{align}
\label{eq:p-map}
p_{I+1} &= p_{I} \left[1 - \frac{128\pi}{5} \frac{\eta}{\bar{p}_{I}^{5/2}} \left(1 + \frac{7}{8} e_{I}^{2}\right)\right]\,,
\\
\label{eq:tp-map}
t_{p,I+1} - t_{p,I} &= 2\pi M \left(\frac{\bar{p}_{I}}{1-e_{I}^{2}}\right)^{3/2} \Bigg[1 
\nn \\
&- \frac{96 \pi}{5} \frac{\eta}{\bar{p}_{I}^{5/2}} \left(\frac{1 + \frac{73}{24} e_{I}^{2} + \frac{37}{96} e_{I}^{4}}{1 - e_{I}^{2}}\right)\Bigg]\,,
\end{align}
respectively. There are two slight differences in this timing model compared to that of~\cite{Loutrel:2014vja}, namely, we do not expand about $e_{I} \ll 1$ and we include the radiation reaction effect on the orbital period in Eq.~\eqref{eq:tp-map}. These are to ensure the model is accurate over a wider range of orbital parameters. 

The above equations constitute a timing model to predict when the subsequent burst will occur, and what the orbital parameters will be during that closest approach. From $[p_{0}, e_{0},t_{p,0}]$, we can obtain all future values. The initial burst is characterized by an EFB-T waveform in the interval $[-t_{\pi,0}, t_{\pi,0}]$, where $t_{\pi,0}$ is described in Sec.~\ref{eff}. To generate the second burst, we calculate $[p_{1}, e_{1}, t_{p,1}]$ from the initial values, and we sample the new EFB-T waveform with these parameters in the range $[t_{\pi,0}, T_{{\rm orb},0} + t_{\pi,1}]$. The change in the sample interval is to ensure that the sampling interval of the total waveform is continuous, and we do not have to perform padding in between each EFB-T waveform. Generating all subsequent EFB-T waveforms follows the same procedure.

\begin{figure}[ht]
\includegraphics[clip=true,scale=0.27]{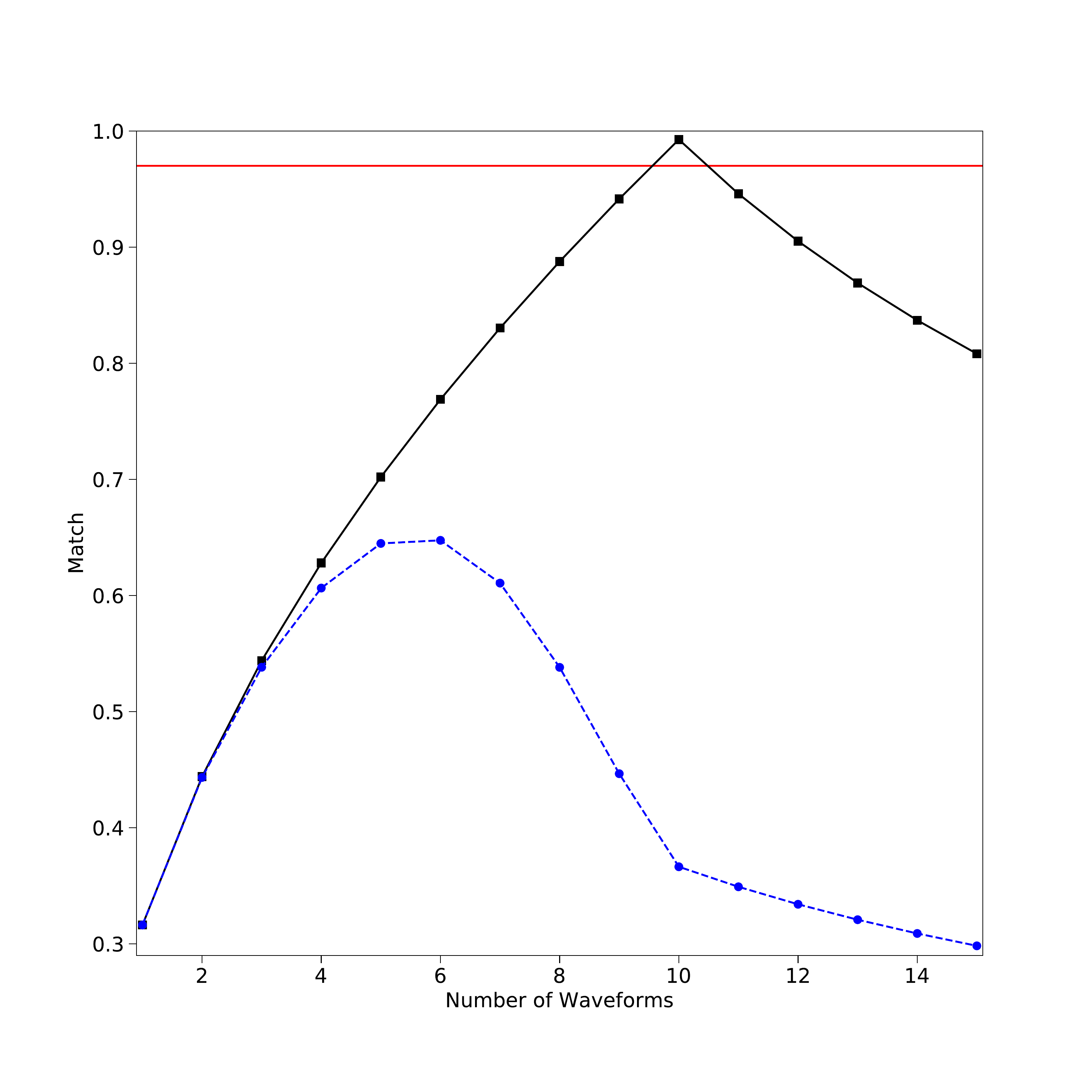}
\caption{\label{multi} Match between a multi-burst EFB-T waveform to a numerical ten burst sequence generated by numerically integrating Eqs.~\eqref{eq:dedt}-\eqref{eq:dpdt} with $p_{0} = 60M$ and $e_{0} = 0.9$, over the time interval $t = [-1.75, 30]$ seconds. We compute the match as a function of the number of EFB-T waveforms used, up to fifteen. The square sequence is the match with the timing model given by Eqs.~\eqref{eq:e-map}-\eqref{eq:tp-map}, while the circles have a 1\% mis-modeling error introduced in the timing model. The match is plotted as a function of the number of EFB-T waveforms, with the horizontal line corresponding to 0.97.}
\end{figure}

We compute the match between the numerical ten burst sequence and the multi-burst EFB-T model in Fig.~\ref{multi} (squares), as a function of the number of EFB-T waveforms used. The match is initially very low, but increases with the number of waveforms and reaches a maximum of $0.993$ at ten bursts. This is well above the value of 0.97 used for matched filtering searches, which is indicated by the horizontal line. Thus, a simple method of combining multiple EFB-T waveforms can be used to capture sequences of bursts from eccentric systems.

There is one caveat to this analysis. We have here implicitly used the results of Peters \& Mathews throughout the analysis, namely to generate the numerical waveform, the EFB-T waveforms, and the timing model. Thus, the timing model of Eqs.~\eqref{eq:e-map}-\eqref{eq:tp-map} will only be accurate for sufficiently widely separated binaries where the PN expansion is valid. For sources of ground-based detectors, this may not be accurate enough to perform a matched filtered search using a multi-burst EFB-T waveform. To show this, we induce an error into the timing model by multiplying the radiation reaction terms, i.e. those proportional to $\bar{p}_{I}^{-5/2}$ in Eqs.~\eqref{eq:e-map}-\eqref{eq:tp-map}, by $1+\epsilon$, where $\epsilon$ allows us to control the level of the mis-modeling error. We repeat the above match calculation with $\epsilon = 0.01$, thus introducing a 1\% error in the radiation reaction effects. The results are given by the circles in Fig.~\ref{multi}. The match no longer peaks at the correct number of bursts, and no longer reaches the threshold for matched filtering. One thus needs a highly accurate timing model to perform match filtering searches using these EFB-T waveforms.

\section{Discussion}
\label{discuss}

We have developed here the first analytic waveforms designed to describe the burst of gravitational radiation from highly eccentric binaries. This was achieved by applying a re-summation procedure to commonly used Fourier series representations of quantities at leading PN order. By comparing to NR waveforms, we showed that the EFB-T model is an accurate representation of the bursts from eccentric systems where the pericenter distance is large enough to neglect relativistic effects. Yet, there are still many open questions from the analyses carried out here.

First, how does one construct a more accurate model compared to NR simulations? The EFB-T model is relatively accurate to the NR waveforms used here, but there is significant room for improvement. The most direct way of improving the model would be to repeat the analysis carried out here to higher PN order. Fourier series representations of the PN two-body problem are currently known to 3PN order~\cite{Boetzel:2017zza}. The re-summation procedure carried out here should still be applicable to higher PN order. A more indirect approach would be to construct an analytic kludge model along the lines of~\cite{2013PhRvD..87d3004E}. The conservative dynamics of such a model are described by geodesic motion on an effective Kerr background. While this might be appealing due to its accuracy, re-summations of the type carried out here may not be possible in this case due to the complicated geodesic motion.

Second, how would one go about using the EFB-T model to search for eccentric binaries? The analysis carried out in Sec.~\ref{multi-burst} shows that one needs a very accurate timing model in order to string multiple EFB-T waveforms together to perform matched filtering searches. Timing models constructed in the PN approximation may not be accurate enough for full inspirals. An alternative strategy would be to search for correlated bursts within the detector. If multiple bursts are emitted by the same system, then the parameters of each burst will not be independent of one another. For example, the sky location, inclination angle, and component masses should all be the same (to within some error) among the bursts. Further, the peak times and frequencies of the bursts should be correlated in the typical chirping fashion. One may be able to use these correlations to search for full eccentric signals without the need for a timing model.

Third, what can we learn from detecting these signals with the models developed here? The most cited application of detecting eccentricity within inspiraling binaries is that it acts as a tracer for formation channels. If the binary has a measurable amount of eccentricity, it would have had to form relatively close to merger, pointing to dynamical formation. Furthermore, eccentric binaries also present themselves as unique systems for placing constraints on the NS equation of state due to the importance of tidal effects and f-mode oscillations. With pericenter velocities potentially being large, these systems may also be unique laboratories for performing tests of GR. The models developed here are a first step toward performing such studies.

Finally, the analyses carried out here have only focused on sources for ground-based detectors. However, population synthesis studies have shown that there are many more eccentric sources in the detection bands of space-based detectors such as LISA and DECIGO, some of which will be highly eccentric. Within our galaxy alone, there are $\sim 150$ globular clusters, which could contain $\sim 20$ sources emitting within the LISA detection band~\cite{Kremer:2018tzm}. If these sources are highly eccentric, the waveform models developed here should be excellent candidates for detecting and characterizing the parameters of these systems.

\acknowledgements

I would like to thank Miriam Cabero, Frans Pretorius, and Nicol\'as Yunes for useful discussions and comments. This work was supported by NSF grant PHY-1607449, the Simons Foundation, and the Canadian Institute For Advanced Research (CIFAR).

\appendix
\section{The Parabolic Limit}
\label{par}

We here show that the re-summation procedure described in Sec.~\ref{resum-t} reproduces a parabolic trajectory in the limit $e\rightarrow1$. We begin by reviewing the Newtonian two body problem for parabolic trajectories. The discussion of Keplerian orbits in Sec.~\ref{kep} is completely general, and valid for all values of $e$. Taking $e=1$ in Eq.~\eqref{eq:r-kep} gives us
\begin{equation}
r = \frac{p}{1+\cos V} = \frac{p}{2}\left[1 + \tan^{2}\left(\frac{V}{2}\right)\right]\,,
\end{equation}
where the second equality follows from trigonometric identities. From the conservation of the orbital angular momentum $h=r^{2} \dot{\phi}$, we obtain
\begin{equation}
\dot{V} = 4 \left(\frac{M}{p^{3}}\right)^{1/2} \left[1 + \tan^{2}\left(\frac{V}{2}\right)\right]^{-2}\,.
\end{equation}
In analogy to Kepler's equation, this can be directly integrated to obtain Barker's equation
\begin{equation}
2\ell_{B} = 3 U + U^{3}\,,
\end{equation}
where $U = \tan(V/2)$ and $\ell_{B} = 3 (M/p^{3})^{1/2}(t-t_{p})$, with $t_{p}$ the time of closest approach. Unlike Kepler's equation, Barker's equation can be solved in closed form by making the replacement $U = z - (1/z)$. This results in the solution
\begin{equation}
\tan\left(\frac{V}{2}\right) = \left(\ell_{B} + \sqrt{\ell_{B}^{2} + 1}\right)^{1/3} - \left(\ell_{B} + \sqrt{\ell_{B}^{2} + 1}\right)^{-1/3}\,.
\end{equation}

The question is now whether the re-summation procedure of Sec.~\ref{resum-t} reproduces this expression. The analytic expressions for $\cos V$ and $\sin V$ are given in Eqs.~\eqref{eq:cos-asym}-\eqref{eq:sin-asym}. Taking the limit $\epsilon = 0$, a simple evaluation reveals
\begin{align}
\tan\left(\frac{V}{2}\right) &= \frac{\sin V}{1 + \cos V}\,,
\nn \\
&\sim \left(\psi + \sqrt{\psi^{2} + 1}\right)^{1/3} - \left(\psi + \sqrt{\psi^{2} + 1}\right)^{-1/3}
\nn \\
&+{\cal{O}}(\epsilon)\,,
\end{align}
where we have expanded $\Ch(\psi,n)$ and $\Sh(\psi,n)$ (see footnote 2). When performing the expansion about $\epsilon \ll 1$ in Sec.~\ref{resum-t}, we held $\psi$ fixed. To obtain the appropriate limit, we must now consider the behavior of $\psi$ when $e\rightarrow1$. Recall that
\begin{equation}
\psi = \frac{3 \ell}{2 \zeta^{3/2}} = \frac{3 n}{2 \zeta^{3/2}} (t - t_{p})
\end{equation}
where $n$ and $\zeta$ are given by Eqs.~\eqref{eq:dldt} and~\eqref{eq:zeta}, respectively. Expanding this expression about $\epsilon \ll 1$ reveals $\psi \sim \ell_{B} + {\cal{O}}(\epsilon)$, and we thus obtain the correct limit.
\begin{widetext}
\section{Time Domain Waveform Functions}
\label{h-t-coeffs}

We here provide explicit forms for the ${\cal{C}}^{(k,n)}$ and ${\cal{S}}^{(k,n)}$ functions appearing in Eq.~\eqref{eq:hpc-time}. The non-zero functions are as follows, where $(s_{\theta}, c_{\theta}) = (\sin\theta, \cos\theta)$, and $\psi$ is given by Eq.~\eqref{eq:psi-of-l}.

\allowdisplaybreaks[4]
\begin{align}
{\cal{C}}_{+}^{(0,0)} &= \frac{8 \psi}{1+\psi^{2}} c_{\beta} s_{\beta} \left(3 + c_{\iota}^{2} - s_{\iota}^{2}\right)
\\
{\cal{C}}_{+}^{(0,1)} &= \frac{1}{20} \left\{9 c_{\beta}^2 (3 + c_{\iota}^2 - s_{\iota}^2) - 9 \left[1 + c_{\iota}^2 (-1 + s_{\beta}^2) + s_{\iota}^2 - s_{\beta}^2 (-3 + s_{\iota}^2)\right] + \frac{16 c_{\beta} s_{\beta} (3 + c_{\iota}^2 - s_{\iota}^2) \psi}{1 + \psi^2}\right\}
\\
{\cal{C}}_{+}^{(0,2)} &= \frac{1}{2800} \left\{-9 c_{\iota}^2 (3 + 265 s_{\beta}^2) + 2385 c_{\beta}^2 (3 + c_{\iota}^2 - s_{\iota}^2) + 2385 s_{\beta}^2 (-3 + s_{\iota}^2) + 27 (1 + s_{\iota}^2) + \frac{3392 c_{\beta} s_{\beta} (3 + c_{\iota}^2 - s_{\iota}^2) \psi}{1 + \psi^2}\right\}
\\
{\cal{C}}_{+}^{(1,0)} &= \frac{1}{(1+\psi^{2})^{3/2}} \left\{1 - 15 s_{\beta}^2 + s_{\iota}^2 + 5 s_{\beta}^2 s_{\iota}^2 + \psi^2 + 3 s_{\beta}^2 \psi^2 + s_{\iota}^2 \psi^2 - s_{\beta}^2 s_{\iota}^2 \psi^2 - c_{\beta}^2 (3 + c_{\iota}^2 - s_{\iota}^2) (-5 + \psi^2) 
\right.
\nn \\
&\left.
+ c_{\iota}^2 \left[-1 - \psi^2 + s_{\beta}^2 (-5 + \psi^2)\right]\right\}
\\
{\cal{C}}_{+}^{(1,1)} &= - \frac{3}{10(1+\psi^{2})^{3/2}} \left\{1 - 15 s_{\beta}^2 + s_{\iota}^2 + 5 s_{\beta}^2 s_{\iota}^2 + \psi^2 - 21 s_{\beta}^2 \psi^2 + s_{\iota}^2 \psi^2 + 7 s_{\beta}^2 s_{\iota}^2 \psi^2 + c_{\beta}^2 (3 + c_{\iota}^2 - s_{\iota}^2) (5 + 7 \psi^2) 
\right.
\nn \\
&\left.
- c_{\iota}^2 \left[1 + \psi^2 + s_{\beta}^2 (5 + 7 \psi^2)\right]\right\}
\\
{\cal{C}}_{+}^{(1,2)} &= \frac{1}{1400(1+\psi^{2})^{3/2}} \left\{-111 + 3585 s_{\beta}^2 - 111 s_{\iota}^2 - 1195 s_{\beta}^2 s_{\iota}^2 - 111 \psi^2 + 6915 s_{\beta}^2 \psi^2 - 111 s_{\iota}^2 \psi^2 - 2305 s_{\beta}^2 s_{\iota}^2 \psi^2 
\right.
\nn \\
&\left.
- 5 c_{\beta}^2 (3 + c_{\iota}^2 - s_{\iota}^2) (239 + 461 \psi^2) + c_{\iota}^2 \left[111 (1 + \psi^2) + 5 s_{\beta}^2 (239 + 461 \psi^2)\right]\right\}
\\
{\cal{C}}_{+}^{(2,0)} &= - \frac{2}{1+\psi^{2}} \left(c_{\beta}^{2} - s_{\beta}^{2}\right) \left(3 + c_{\iota}^{2} - s_{\iota}^{2}\right)
\\
{\cal{C}}_{+}^{(2,1)} &= - \frac{1}{5(1+\psi^{2})} \left(c_{\beta}^{2} - s_{\beta}^{2}\right) \left(3 + c_{\iota}^{2} - s_{\iota}^{2}\right)
\\
{\cal{C}}_{+}^{(2,2)} &= \frac{1}{140 (1+\psi^{2})} \left\{1 + 96 s_{\beta}^2 + s_{\iota}^2 - 32 s_{\beta}^2 s_{\iota}^2 + \psi^2 + 3 s_{\beta}^2 \psi^2 + s_{\iota}^2 \psi^2 - s_{\beta}^2 s_{\iota}^2 \psi^2 - c_{\beta}^2 (3 + c_{\iota}^2 - s_{\iota}^2) (32 + \psi^2) 
\right.
\nn \\
&\left.
+ c_{\iota}^2 (-1 - \psi^2 + s_{\beta}^2 (32 + \psi^2))\right\}
\\
{\cal{C}}_{+}^{(4,2)} &= \frac{3}{70(1+\psi^{2})} \left(c_{\beta}^{2} - s_{\beta}^{2}\right) \left(3 + c_{\iota}^{2} - s_{\iota}^{2}\right)
\\
{\cal{S}}_{+}^{(1,0)} &= \frac{8}{1+\psi^{2}} c_{\beta} s_{\beta} \left(3 + c_{\iota}^{2} - s_{\iota}^{2}\right)
\\
{\cal{S}}_{+}^{(1,1)} &= - \frac{8}{5(1+\psi^{2})} c_{\beta} s_{\beta} \left(3 + c_{\iota}^{2} - s_{\iota}^{2}\right)
\\
{\cal{S}}_{+}^{(1,2)} &= \frac{4}{175(1+\psi^{2})} c_{\beta} s_{\beta} \left(3 + c_{\iota}^{2} - s_{\iota}^{2}\right) \left(-179 + 26 \psi^{2}\right)
\\
{\cal{S}}_{+}^{(2,0)} &= \frac{12}{(1+\psi^{2})^{3/2}} c_{\beta} s_{\beta} \left(3 + c_{\iota}^{2} - s_{\iota}^{2}\right)
\\
{\cal{S}}_{+}^{(2,1)} &= - \frac{2}{5(1+\psi^{2})^{3/2}} c_{\beta} s_{\beta} \left(3 + c_{\iota}^{2} - s_{\iota}^{2}\right) \left(19 + 22 \psi^{2}\right)
\\
{\cal{S}}_{+}^{(2,2)} &= - \frac{1}{350(1+\psi^{2})^{3/2}}  c_{\beta} s_{\beta} \left(3 + c_{\iota}^{2} - s_{\iota}^{2}\right) \left(2251 + 2026 \psi^{2}\right)
\\
{\cal{S}}_{+}^{(4,2)} &= \frac{78}{35 (1 + \psi^{2})^{3/2}}  c_{\beta} s_{\beta} \left(3 + c_{\iota}^{2} - s_{\iota}^{2}\right)
\\
{\cal{S}}_{+}^{(5,1)} &= \frac{12}{5 (1+\psi^{2})}  c_{\beta} s_{\beta} \left(3 + c_{\iota}^{2} - s_{\iota}^{2}\right)
\\
{\cal{S}}_{+}^{(5,2)} &= \frac{36}{25 (1+\psi^{2})}  c_{\beta} s_{\beta} \left(3 + c_{\iota}^{2} - s_{\iota}^{2}\right)
\\
{\cal{S}}_{+}^{(6,0)} &= -\frac{4}{(1+\psi^{2})^{3/2}}  c_{\beta} s_{\beta} \left(3 + c_{\iota}^{2} - s_{\iota}^{2}\right)
\\
{\cal{S}}_{+}^{(6,1)} &= - \frac{2}{5(1+\psi^{2})^{3/2}}  c_{\beta} s_{\beta} \left(3 + c_{\iota}^{2} - s_{\iota}^{2}\right)
\\
{\cal{S}}_{+}^{(6,2)} &= - \frac{37}{70 (1+\psi^{2})^{3/2}}  c_{\beta} s_{\beta} \left(3 + c_{\iota}^{2} - s_{\iota}^{2}\right)
\\
{\cal{C}}_{\times}^{(0,0)} &= \frac{16 \psi}{1+\psi^{2}} c_{\iota} \left(c_{\beta}^{2} - s_{\beta}^{2}\right)
\\
{\cal{C}}_{\times}^{(0,1)} &= - \frac{2}{5 (1+\psi^{2})} c_{\iota} \left[-4 c_{\beta}^2 \psi + 4 s_{\beta}^2 \psi + 9 c_{\beta} s_{\beta} (1 + \psi^2)\right]
\\
{\cal{C}}_{\times}^{(0,2)} &= - \frac{53}{350(1+\psi^{2})} c_{\iota} \left[-16 c_{\beta}^2 \psi + 16 s_{\beta}^2 \psi + 45 c_{\beta} s_{\beta} (1 + \psi^2)\right]
\\
{\cal{C}}_{\times}^{(1,0)} &= \frac{1}{35(1+\psi^{2})^{3/2}} c_{\beta} s_{\beta} c_{\iota} \left(-1400 + 280 \psi^{2}\right)
\\
{\cal{C}}_{\times}^{(1,1)} &= \frac{1}{35(1+\psi^{2})^{3/2}} c_{\beta} s_{\beta} c_{\iota} \left(420 + 588 \psi^{2}\right)
\\
{\cal{C}}_{\times}^{(1,2)} &= \frac{1}{35(1+\psi^{2})^{3/2}} c_{\beta} s_{\beta} c_{\iota} \left(239 + 461 \psi^{2}\right)
\\
{\cal{C}}_{\times}^{(2,0)} &= \frac{16}{1+\psi^{2}} c_{\beta} s_{\beta} c_{\iota}
\\
{\cal{C}}_{\times}^{(2,1)} &= \frac{8}{5(1+\psi^{2})} c_{\beta} s_{\beta} c_{\iota}
\\
{\cal{C}}_{\times}^{(2,2)} &= \frac{2}{35 (1 + \psi^{2})} c_{\beta} s_{\beta} c_{\iota} \left(32 + \psi^{2}\right)
\\
{\cal{C}}_{\times}^{(4,2)} &= -\frac{12}{35(1+\psi^{2})} c_{\beta} s_{\beta} c_{\iota}
\\
{\cal{S}}_{\times}^{(1,0)} &= \frac{16}{1+\psi^{2}} c_{\iota} \left(c_{\beta}^{2} - s_{\beta}^{2}\right)
\\
{\cal{S}}_{\times}^{(1,1)} &= - \frac{16}{5(1+\psi^{2})} c_{\iota} \left(c_{\beta}^{2} - s_{\beta}^{2}\right)
\\
{\cal{S}}_{\times}^{(1,2)} &= \frac{8}{175(1+\psi^{2})} c_{\iota} \left(c_{\beta}^{2} - s_{\beta}^{2}\right) \left(-179 + 26 \psi^{2}\right)
\\
{\cal{S}}_{\times}^{(2,0)} &= \frac{24}{(1+\psi^{2})^{3/2}} c_{\iota} \left(c_{\beta}^{2} - s_{\beta}^{2}\right)
\\
{\cal{S}}_{\times}^{(2,1)} &= - \frac{4}{5(1+\psi^{2})^{3/2}} c_{\iota} \left(c_{\beta}^{2} - s_{\beta}^{2}\right) \left(19 + 22 \psi^{2}\right)
\\
{\cal{S}}_{\times}^{(2,2)} &= - \frac{1}{175 (1+\psi^{2})^{3/2}} c_{\iota} \left(c_{\beta}^{2} - s_{\beta}^{2}\right) \left(2251 + 2026 \psi^{2}\right)
\\
{\cal{S}}_{\times}^{(4,2)} &= \frac{156}{35 (1+\psi^{2})^{3/2}} c_{\iota} \left(c_{\beta}^{2} - s_{\beta}^{2}\right)
\\
{\cal{S}}_{\times}^{(5,1)} &= \frac{24}{5 (1+\psi^{2})} c_{\iota} \left(c_{\beta}^{2} - s_{\beta}^{2}\right)
\\
{\cal{S}}_{\times}^{(5,2)} &= \frac{72}{25 (1+\psi^{2})} c_{\iota} \left(c_{\beta}^{2} - s_{\beta}^{2}\right)
\\
{\cal{S}}_{\times}^{(6,0)} &= - \frac{8}{(1+\psi^{2})^{3/2}} c_{\iota} \left(c_{\beta}^{2} - s_{\beta}^{2}\right)
\\
{\cal{S}}_{\times}^{(6,1)} &= - \frac{4}{5 (1+\psi^{2})^{3/2}} c_{\iota} \left(c_{\beta}^{2} - s_{\beta}^{2}\right)
\\
{\cal{S}}_{\times}^{(6,2)} &= - \frac{37}{35 (1+\psi^{2})^{3/2}} c_{\iota} \left(c_{\beta}^{2} - s_{\beta}^{2}\right)
\end{align}
%
\section{Fourier Domain Waveform Functions}
\label{h-f-coeffs}

We here provide expressions for the functions ${\cal{A}}_{l_{1}, l_{2}, s}$ appearing in Eq.~\eqref{eq:hpc-f}.
The non-zero functions are
\begin{align}
{\cal{A}}^{+}_{10,8,1/2} &= -\sqrt{\frac{2}{\pi}} \frac{(1 + i) 3^{i\chi/2} (1 + \cos^2\iota) (1 - e_{0}^2)^{5/4}  \sin(2\beta) \zeta_{0}^{\frac{3}{4} (-5 + 2 i \chi)} [9 (3 i + \chi) \chi_{\rm orb}^2 + 2 i \zeta_{0}^3] \Gamma(3 - \frac{3 i \chi}{2})}{(5 i + 3 \chi) (7 i + 3 \chi) \chi_{\rm orb} \Gamma(1 - \frac{i \chi}{2})}\,,
\\
{\cal{A}}^{+}_{10,8,-1/2} &= -\frac{(1 + i) 3^{i\chi/2} (1 + \cos^2\iota) (1 - e_0^2)^{5/4} \sin(2\beta) \zeta_{0}^{\frac{3}{4} (-5 + 2 i \chi)} (9 \chi_{\rm orb}^2 + 4 \zeta_{0}^{3}) \Gamma(3 - \frac{3 i \chi}{2})}{\sqrt{2 \pi} (-5 + 3 i \chi) (7 i + 3 \chi) \chi_{\rm orb} \Gamma(1 - \frac{i\chi}{2})}
\\
{\cal{A}}^{+}_{2,4,1/2} &= \frac{(1 - i) 3^{i \chi/2} \sqrt{\frac{2}{\pi}} [\cos(2\beta) (1 + \cos^2\iota) (-2 + e_0^2) + e_0^2 \sin^2\iota] \zeta_{0}^{-\frac{3}{4} + \frac{3i\chi}{2}} \Gamma(3 - \frac{3i\chi}{2})}{(1 - e_0^2)^{1/4} (-8 + 18 i \chi + 9 \chi^2) \Gamma(1 - \frac{i\chi}{2})}
\\
{\cal{A}}^{+}_{4,8,1/2} &= \frac{(\frac{1}{2} + \frac{i}{2}) 3^{\frac{3}{2} - i \chi} (1 + \cos^2\iota) (1 - e_0^2)^{1/4} \zeta_{0}^{-\frac{9}{4} + \frac{3 i\chi}{2}} \Gamma(\frac{2}{3} - \frac{i\chi}{2}) \Gamma(\frac{4}{3} - \frac{i\chi}{2})}{\sqrt{2} \pi^{3/2} (i + 3 \chi) (5 i + 3 \chi) \chi_{\rm orb}} \Big\{\cos(2\beta) (-1 + e_0^2) (5 i + 18 \chi - 9 i \chi^2) \chi_{\rm orb} 
\nn \\
&+ 2 \sqrt{1 - e_0^2} \sin(2\beta) \left[9 (2 i + \chi) \chi_{\rm orb}^2 + 2 i \zeta_{0}^3\right]\Big\}
\\
{\cal{A}}^{+}_{4,8,-1/2} &= \frac{(\frac{1}{2} + \frac{i}{2}) 3^{\frac{3}{2} - i \chi} (1 + \cos^2\iota) (1 - e_0^2)^{3/4} \sin(2\beta) \zeta_{0}^{-\frac{9}{4} + \frac{3 i\chi}{2}} (9 \chi_{\rm orb}^2 + 4 \zeta_{0}^3) \Gamma(\frac{2}{3} - \frac{i\chi}{2}) \Gamma(\frac{4}{3} - \frac{i\chi}{2})}{\sqrt{2} \pi^{3/2} (-5 + 3 i \chi) (i + 3 \chi) \chi_{\rm orb}}
\\
{\cal{A}}^{+}_{1,5,1/2} &= \frac{(\frac{1}{2} + \frac{i}{2}) 3^{\frac{1}{2} - i \chi} (1 + \cos^2\iota) (1 - e_0^2)^{3/4} \sin(2\beta) \zeta_{0}^{-\frac{3}{4} + \frac{3i\chi}{2}} \Gamma(\frac{1}{6} - \frac{i \chi}{2}) \Gamma(\frac{5}{6} - \frac{i\chi}{2})}{\sqrt{2} \pi^{3/2}}
\\
{\cal{A}}^{+}_{5,7,1/2} &= \frac{(\frac{1}{4} + \frac{i}{4}) 3^{\frac{1}{2} - i \chi} (1 - 3 i \chi) \zeta_{0}^{-\frac{9}{4} + \frac{3 i \chi}{2}} \Gamma(\frac{1}{6} - \frac{i\chi}{2}) \Gamma(\frac{5}{6} - \frac{i\chi}{2})}{\sqrt{2} (1 - e_0^2)^{1/4} \pi^{3/2} (2 i + 3 \chi) (4 i + 3 \chi) \chi_{\rm orb}} \Big\{-(1 + \cos^2\iota) (1 - e_0^2)^{3/2} \sin(2\beta) (-8 + 18 i \chi + 9 \chi^2) \chi_{\rm orb} 
\nn \\
&+ i [\cos(2\beta) (1 + \cos^2\iota) (-2 + e_0^2) + e_0^2 \sin^2\iota] [9 (2 i + \chi) \chi_{\rm orb}^2 + 2 i \zeta_{0}^3]\Big\}\,,
\\
{\cal{A}}^{+}_{5,7,-1/2} &= \frac{(\frac{1}{8} + \frac{i}{8}) 3^{\frac{1}{2} - i \chi} [\cos(2\beta) (1 + \cos^2\iota) (-2 + e_0^2) + e_0^2 \sin^2\iota] (1 - 3 i \chi) \zeta_{0}^{-\frac{9}{4} + \frac{3 i \chi}{2}} (9 \chi_{\rm orb}^2 + 4 \zeta_{0}^3) \Gamma(\frac{1}{6} - \frac{i \chi}{2}) \Gamma(\frac{5}{6} - \frac{i\chi}{2})}{\sqrt{2} (1 - e_0^2)^{1/4} \pi^{3/2} (2 i + 3 \chi) (4 i + 3 \chi) \chi_{\rm orb}}\,,
\\
{\cal{A}}^{+}_{7,11,1/2} &=\frac{(\frac{1}{4} - \frac{i}{4}) 3^{\frac{1}{2} - i \chi} \cos(2\beta) (1 + \cos^2\iota) (1 - e_0^2)^{5/4} (-5 + 18 i \chi + 9 \chi^{2}) \zeta^{\frac{3}{4} (-5 + 2 i\chi)} [9 (3 i + \chi) \chi_{\rm orb}^2 + 2 i \zeta_{0}^3]}{\sqrt{2} \pi^{3/2} (4 i + 3 \chi) (8 i + 3 \chi) \chi_{\rm orb}}
\nn \\
&\times \Gamma\left(\frac{1}{6} - \frac{i\chi}{2}\right) \Gamma\left(\frac{5}{6} - \frac{i\chi}{2}\right)
\\
{\cal{A}}^{+}_{7,11,-1/2} &= -\frac{(\frac{1}{8} + \frac{i}{8}) 3^{\frac{1}{2} - i \chi} \cos(2\beta) (1 + \cos^2\iota) (1 - e_0^2)^{5/4} (-5 + 18 i \chi + 9 \chi^2) \zeta_{0}^{\frac{3}{4} (-5 + 2 i \chi)} (9 \chi_{\rm orb}^2 + 4 \zeta_{0}^3) \Gamma(\frac{1}{6} - \frac{i\chi}{2}) \Gamma(\frac{5}{6} - \frac{i\chi}{2})}{\sqrt{2} \pi^{3/2} (4 i + 3 \chi) (8 i + 3 \chi) \chi_{\rm orb}}
\\
{\cal{A}}^{\times}_{10,8,1/2} &= -\frac{(2 + 2 i) 3^{i\chi/2} \cos(2\beta) \cos\iota (1 - e_0^2)^{5/4} \sqrt{\frac{2}{\pi}} \zeta_{0}^{\frac{3}{4} (-5 + 2 i \chi)} [9 (3 i + \chi) \chi_{\rm orb}^2 + 2 i \zeta_{0}^3] \Gamma(3 - \frac{3i\chi}{2})}{(5 i + 3 \chi) (7 i + 3 \chi) \chi_{\rm orb} \Gamma(1 - \frac{i\chi}{2})}
\\
{\cal{A}}^{\times}_{10,8,-1/2} &= -\frac{(1 + i) 3^{i\chi/2} \cos(2\beta) \cos\iota (1 - e_0^2)^{5/4} \sqrt{\frac{2}{\pi}} \zeta_{0}^{\frac{3}{4} (-5 + 2 i \chi)} (9 \chi_{\rm orb}^2 + 4 \zeta_{0}^3) \Gamma(3 - \frac{3i\chi}{2})}{(-5 + 3 i \chi) (7 i + 3 \chi) \chi_{\rm orb} \Gamma(1 - \frac{i\chi}{2})}
\\
{\cal{A}}^{\times}_{2,4,1/2} &= -\sqrt{\frac{2}{\pi}} \frac{(2 - 2 i) 3^{i \chi/2} \cos\iota (-2 + e_0^2) \sin(2\beta) \zeta_{0}^{-\frac{3}{4} + \frac{3 i\chi}{2}} \Gamma(3 - \frac{3i\chi}{2})}{(1 - e_0^2)^{1/4} (-8 + 18 i \chi + 9 \chi^2) \Gamma(1 - \frac{i\chi}{2})}
\\
{\cal{A}}^{\times}_{4,8,1/2} &= \frac{(1 + i) 3^{\frac{3}{2} - i\chi} \cos\iota (1 - e_0^2)^{1/4} \zeta_{0}^{-\frac{9}{4} + \frac{3i\chi}{2}} \Gamma(\frac{2}{3} - \frac{i\chi}{2}) \Gamma(\frac{4}{3} - \frac{i\chi}{2})}{\sqrt{2} \pi^{3/2} (i + 3 \chi) (5 i + 3 \chi) \chi_{\rm orb}} \Big\{i (-1 + e_0^2) \sin(2\beta) (-5 + 18 i \chi + 9 \chi^2) \chi_{\rm orb} 
\nn \\
&+ 2 \cos(2\beta) \sqrt{1 - e_0^2} [9 (2 i + \chi) \chi_{\rm orb}^2 + 2 i \zeta_{0}^3]\Big\}
\\
{\cal{A}}^{\times}_{4,8,-1/2} &= \frac{(1 + i) 3^{\frac{3}{2} - i\chi} \cos(2\beta) \cos\iota (1 - e_0^2)^{3/4} \zeta_{0}^{-\frac{9}{4} + \frac{3i\chi}{2}} (9 \chi_{\rm orb}^2 + 4 \zeta_{0}^3) \Gamma(\frac{2}{3} - \frac{i\chi}{2}) \Gamma(\frac{4}{3} - \frac{i\chi}{2})}{\sqrt{2} \pi^{3/2} (-5 + 3 i \chi) (i + 3 \chi) \chi_{\rm orb}}
\\
{\cal{A}}^{\times}_{1,5,1/2} &= \frac{(1 + i) 3^{\frac{1}{2} - i \chi} \cos(2\beta) \cos\iota (1 - e_0^2)^{3/4} \zeta_{0}^{-\frac{3}{4} + \frac{3i\chi}{2}} \Gamma(\frac{1}{6} - \frac{i\chi}{2}) \Gamma(\frac{5}{6} - \frac{i\chi}{2})}{\sqrt{2} \pi^{3/2}}
\\
{\cal{A}}^{\times}_{5,7,1/2} &= \frac{(\frac{1}{2} + \frac{i}{2}) 3^{\frac{1}{2} - i\chi} \cos\iota (1 - 3 i \chi) \zeta_{0}^{-\frac{9}{4} + \frac{3i\chi}{2}} \Gamma(\frac{1}{6} - \frac{i\chi}{2}) \Gamma(\frac{5}{6} - \frac{i\chi}{2})}{\sqrt{2} (1 - e_0^2)^{1/4} \pi^{3/2} (2 i + 3 \chi) (4 i + 3 \chi) \chi_{\rm orb}} \Big\{-\cos(2\beta) (1 - e_0^2)^{3/2} (-8 + 18 i \chi + 9 \chi^2) \chi_{\rm orb} 
\nn \\
&+ (-2 + e_0^2) \sin(2\beta) [9 (2 - i\chi) \chi_{\rm orb}^2 + 2 \zeta_{0}^3]\Big\}
\\
{\cal{A}}^{\times}_{5,7,-1/2} &= -\frac{(\frac{1}{4} - \frac{i}{4}) 3^{\frac{1}{2} - i\chi} \cos\iota (-2 + e_0^2) \sin(2\beta) (i + 3 \chi) \zeta_{0}^{-\frac{9}{4} + \frac{3i\chi}{2}} (9 \chi_{\rm orb}^2 + 4 \zeta_{0}^3) \Gamma(\frac{1}{6} - \frac{i\chi}{2}) \Gamma(\frac{5}{6} - \frac{i\chi}{2})}{\sqrt{2} (1 - e_0^2)^{1/4} \pi^{3/2} (2 i + 3 \chi) (4 i + 3 \chi) \chi_{\rm orb}}
\\
{\cal{A}}^{\times}_{7,11,1/2} &= -\frac{(\frac{1}{2} + \frac{i}{2}) 3^{\frac{1}{2} - i\chi} \cos\iota (1 - e_0^2)^{5/4} \sin(2\beta) (-5 + 18 i \chi + 9 \chi^2) \zeta^{\frac{3}{4} (-5 + 2 i\chi)} [9 (3 - i \chi) \chi_{\rm orb}^2 + 2 \zeta_{0}^3] \Gamma(\frac{1}{6} - \frac{i\chi}{2}) \Gamma(\frac{5}{6} - \frac{i\chi}{2})}{\sqrt{2} \pi^{3/2} (4 i + 3 \chi) (8 i + 3 \chi) \chi_{\rm orb}}
\\
{\cal{A}}^{\times}_{7,11,-1/2} &= \frac{(\frac{1}{4} + \frac{i}{4}) 3^{\frac{1}{2} - i\chi} \cos\iota (1 - e_0^2)^{5/4} \sin(2\beta) (-5 + 18 i\chi + 9 \chi^2) \zeta_{0}^{\frac{3}{4} (-5 + 2 i \chi)} (9 \chi_{\rm orb}^2 + 4 \zeta_{0}^3) \Gamma(\frac{1}{6} - \frac{i\chi}{2}) \Gamma(\frac{5}{6} - \frac{i\chi}{2})}{\sqrt{2} \pi^{3/2} (4 i + 3 \chi) (8 i + 3 \chi) \chi_{\rm orb}}
\end{align}
\end{widetext}
\bibliography{master}

\begin{thebibliography}{77}%
\makeatletter
\providecommand \@ifxundefined [1]{%
 \@ifx{#1\undefined}
}%
\providecommand \@ifnum [1]{%
 \ifnum #1\expandafter \@firstoftwo
 \else \expandafter \@secondoftwo
 \fi
}%
\providecommand \@ifx [1]{%
 \ifx #1\expandafter \@firstoftwo
 \else \expandafter \@secondoftwo
 \fi
}%
\providecommand \natexlab [1]{#1}%
\providecommand \enquote  [1]{``#1''}%
\providecommand \bibnamefont  [1]{#1}%
\providecommand \bibfnamefont [1]{#1}%
\providecommand \citenamefont [1]{#1}%
\providecommand \href@noop [0]{\@secondoftwo}%
\providecommand \href [0]{\begingroup \@sanitize@url \@href}%
\providecommand \@href[1]{\@@startlink{#1}\@@href}%
\providecommand \@@href[1]{\endgroup#1\@@endlink}%
\providecommand \@sanitize@url [0]{\catcode `\\12\catcode `\$12\catcode
  `\&12\catcode `\#12\catcode `\^12\catcode `\_12\catcode `\%12\relax}%
\providecommand \@@startlink[1]{}%
\providecommand \@@endlink[0]{}%
\providecommand \url  [0]{\begingroup\@sanitize@url \@url }%
\providecommand \@url [1]{\endgroup\@href {#1}{\urlprefix }}%
\providecommand \urlprefix  [0]{URL }%
\providecommand \Eprint [0]{\href }%
\providecommand \doibase [0]{http://dx.doi.org/}%
\providecommand \selectlanguage [0]{\@gobble}%
\providecommand \bibinfo  [0]{\@secondoftwo}%
\providecommand \bibfield  [0]{\@secondoftwo}%
\providecommand \translation [1]{[#1]}%
\providecommand \BibitemOpen [0]{}%
\providecommand \bibitemStop [0]{}%
\providecommand \bibitemNoStop [0]{.\EOS\space}%
\providecommand \EOS [0]{\spacefactor3000\relax}%
\providecommand \BibitemShut  [1]{\csname bibitem#1\endcsname}%
\let\auto@bib@innerbib\@empty
\bibitem [{\citenamefont {{Naoz}}\ \emph {et~al.}(2013)\citenamefont {{Naoz}},
  \citenamefont {{Kocsis}}, \citenamefont {{Loeb}},\ and\ \citenamefont
  {{Yunes}}}]{2013ApJ...773..187N}%
  \BibitemOpen
  \bibfield  {author} {\bibinfo {author} {\bibfnamefont {S.}~\bibnamefont
  {{Naoz}}}, \bibinfo {author} {\bibfnamefont {B.}~\bibnamefont {{Kocsis}}},
  \bibinfo {author} {\bibfnamefont {A.}~\bibnamefont {{Loeb}}}, \ and\ \bibinfo
  {author} {\bibfnamefont {N.}~\bibnamefont {{Yunes}}},\ }\href {\doibase
  10.1088/0004-637X/773/2/187} {\bibfield  {journal} {\bibinfo  {journal} {{The
  Astrophysical Journal}}\ }\textbf {\bibinfo {volume} {773}},\ \bibinfo {eid}
  {187} (\bibinfo {year} {2013})},\ \Eprint {http://arxiv.org/abs/1206.4316}
  {arXiv:1206.4316 [astro-ph.SR]} \BibitemShut {NoStop}%
\bibitem [{\citenamefont {{Kocsis}}\ and\ \citenamefont
  {{Levin}}(2012)}]{2012PhRvD..85l3005K}%
  \BibitemOpen
  \bibfield  {author} {\bibinfo {author} {\bibfnamefont {B.}~\bibnamefont
  {{Kocsis}}}\ and\ \bibinfo {author} {\bibfnamefont {J.}~\bibnamefont
  {{Levin}}},\ }\href {\doibase 10.1103/PhysRevD.85.123005} {\bibfield
  {journal} {\bibinfo  {journal} {{Phys. Rev. D.}}\ }\textbf {\bibinfo {volume}
  {85}},\ \bibinfo {eid} {123005} (\bibinfo {year} {2012})},\ \Eprint
  {http://arxiv.org/abs/1109.4170} {arXiv:1109.4170 [astro-ph.CO]} \BibitemShut
  {NoStop}%
\bibitem [{\citenamefont {{O'Leary}}\ \emph {et~al.}(2009)\citenamefont
  {{O'Leary}}, \citenamefont {{Kocsis}},\ and\ \citenamefont
  {{Loeb}}}]{2009MNRAS.395.2127O}%
  \BibitemOpen
  \bibfield  {author} {\bibinfo {author} {\bibfnamefont {R.~M.}\ \bibnamefont
  {{O'Leary}}}, \bibinfo {author} {\bibfnamefont {B.}~\bibnamefont {{Kocsis}}},
  \ and\ \bibinfo {author} {\bibfnamefont {A.}~\bibnamefont {{Loeb}}},\ }\href
  {\doibase 10.1111/j.1365-2966.2009.14653.x} {\bibfield  {journal} {\bibinfo
  {journal} {Mon. Not. R. Astron. Soc.}\ }\textbf {\bibinfo {volume} {395}},\
  \bibinfo {pages} {2127} (\bibinfo {year} {2009})},\ \Eprint
  {http://arxiv.org/abs/0807.2638} {arXiv:0807.2638} \BibitemShut {NoStop}%
\bibitem [{\citenamefont {Samsing}(2017)}]{Samsing:2017xmd}%
  \BibitemOpen
  \bibfield  {author} {\bibinfo {author} {\bibfnamefont {J.}~\bibnamefont
  {Samsing}},\ }\href@noop {} {\  (\bibinfo {year} {2017})},\ \Eprint
  {http://arxiv.org/abs/1711.07452} {arXiv:1711.07452 [astro-ph.HE]}
  \BibitemShut {NoStop}%
\bibitem [{\citenamefont {Samsing}\ \emph {et~al.}(2013)\citenamefont
  {Samsing}, \citenamefont {MacLeod},\ and\ \citenamefont
  {Ramirez-Ruiz}}]{Samsing:2013kua}%
  \BibitemOpen
  \bibfield  {author} {\bibinfo {author} {\bibfnamefont {J.}~\bibnamefont
  {Samsing}}, \bibinfo {author} {\bibfnamefont {M.}~\bibnamefont {MacLeod}}, \
  and\ \bibinfo {author} {\bibfnamefont {E.}~\bibnamefont {Ramirez-Ruiz}},\
  }\href@noop {} {\  (\bibinfo {year} {2013})},\ \Eprint
  {http://arxiv.org/abs/1308.2964} {arXiv:1308.2964 [astro-ph.HE]} \BibitemShut
  {NoStop}%
\bibitem [{\citenamefont {Samsing}\ \emph {et~al.}(2019)\citenamefont
  {Samsing}, \citenamefont {D'Orazio}, \citenamefont {Kremer}, \citenamefont
  {Rodriguez},\ and\ \citenamefont {Askar}}]{Samsing:2019dtb}%
  \BibitemOpen
  \bibfield  {author} {\bibinfo {author} {\bibfnamefont {J.}~\bibnamefont
  {Samsing}}, \bibinfo {author} {\bibfnamefont {D.~J.}\ \bibnamefont
  {D'Orazio}}, \bibinfo {author} {\bibfnamefont {K.}~\bibnamefont {Kremer}},
  \bibinfo {author} {\bibfnamefont {C.~L.}\ \bibnamefont {Rodriguez}}, \ and\
  \bibinfo {author} {\bibfnamefont {A.}~\bibnamefont {Askar}},\ }\href@noop {}
  {\  (\bibinfo {year} {2019})},\ \Eprint {http://arxiv.org/abs/1907.11231}
  {arXiv:1907.11231 [astro-ph.HE]} \BibitemShut {NoStop}%
\bibitem [{\citenamefont {Kremer}\ \emph {et~al.}(2019)\citenamefont {Kremer}
  \emph {et~al.}}]{Kremer:2018cir}%
  \BibitemOpen
  \bibfield  {author} {\bibinfo {author} {\bibfnamefont {K.}~\bibnamefont
  {Kremer}} \emph {et~al.},\ }\href {\doibase 10.1103/PhysRevD.99.063003}
  {\bibfield  {journal} {\bibinfo  {journal} {Phys. Rev.}\ }\textbf {\bibinfo
  {volume} {D99}},\ \bibinfo {pages} {063003} (\bibinfo {year} {2019})},\
  \Eprint {http://arxiv.org/abs/1811.11812} {arXiv:1811.11812 [astro-ph.HE]}
  \BibitemShut {NoStop}%
\bibitem [{\citenamefont {Zevin}\ \emph {et~al.}(2019)\citenamefont {Zevin},
  \citenamefont {Samsing}, \citenamefont {Rodriguez}, \citenamefont {Haster},\
  and\ \citenamefont {Ramirez-Ruiz}}]{Zevin:2018kzq}%
  \BibitemOpen
  \bibfield  {author} {\bibinfo {author} {\bibfnamefont {M.}~\bibnamefont
  {Zevin}}, \bibinfo {author} {\bibfnamefont {J.}~\bibnamefont {Samsing}},
  \bibinfo {author} {\bibfnamefont {C.}~\bibnamefont {Rodriguez}}, \bibinfo
  {author} {\bibfnamefont {C.-J.}\ \bibnamefont {Haster}}, \ and\ \bibinfo
  {author} {\bibfnamefont {E.}~\bibnamefont {Ramirez-Ruiz}},\ }\href {\doibase
  10.3847/1538-4357/aaf6ec} {\bibfield  {journal} {\bibinfo  {journal}
  {Astrophys. J.}\ }\textbf {\bibinfo {volume} {871}},\ \bibinfo {pages} {91}
  (\bibinfo {year} {2019})},\ \Eprint {http://arxiv.org/abs/1810.00901}
  {arXiv:1810.00901 [astro-ph.HE]} \BibitemShut {NoStop}%
\bibitem [{\citenamefont {Samsing}\ \emph
  {et~al.}(2018{\natexlab{a}})\citenamefont {Samsing}, \citenamefont
  {D'Orazio}, \citenamefont {Askar},\ and\ \citenamefont
  {Giersz}}]{Samsing:2018ykz}%
  \BibitemOpen
  \bibfield  {author} {\bibinfo {author} {\bibfnamefont {J.}~\bibnamefont
  {Samsing}}, \bibinfo {author} {\bibfnamefont {D.~J.}\ \bibnamefont
  {D'Orazio}}, \bibinfo {author} {\bibfnamefont {A.}~\bibnamefont {Askar}}, \
  and\ \bibinfo {author} {\bibfnamefont {M.}~\bibnamefont {Giersz}},\
  }\href@noop {} {\  (\bibinfo {year} {2018}{\natexlab{a}})},\ \Eprint
  {http://arxiv.org/abs/1802.08654} {arXiv:1802.08654 [astro-ph.HE]}
  \BibitemShut {NoStop}%
\bibitem [{\citenamefont {Samsing}\ \emph
  {et~al.}(2018{\natexlab{b}})\citenamefont {Samsing}, \citenamefont {Askar},\
  and\ \citenamefont {Giersz}}]{Samsing:2017oij}%
  \BibitemOpen
  \bibfield  {author} {\bibinfo {author} {\bibfnamefont {J.}~\bibnamefont
  {Samsing}}, \bibinfo {author} {\bibfnamefont {A.}~\bibnamefont {Askar}}, \
  and\ \bibinfo {author} {\bibfnamefont {M.}~\bibnamefont {Giersz}},\ }\href
  {\doibase 10.3847/1538-4357/aaab52} {\bibfield  {journal} {\bibinfo
  {journal} {Astrophys. J.}\ }\textbf {\bibinfo {volume} {855}},\ \bibinfo
  {pages} {124} (\bibinfo {year} {2018}{\natexlab{b}})},\ \Eprint
  {http://arxiv.org/abs/1712.06186} {arXiv:1712.06186 [astro-ph.HE]}
  \BibitemShut {NoStop}%
\bibitem [{\citenamefont {Leigh}\ \emph {et~al.}(2018)\citenamefont {Leigh}
  \emph {et~al.}}]{Leigh:2017wff}%
  \BibitemOpen
  \bibfield  {author} {\bibinfo {author} {\bibfnamefont {N.~W.~C.}\
  \bibnamefont {Leigh}} \emph {et~al.},\ }\href {\doibase
  10.1093/mnras/stx3134} {\bibfield  {journal} {\bibinfo  {journal} {Mon. Not.
  Roy. Astron. Soc.}\ }\textbf {\bibinfo {volume} {474}},\ \bibinfo {pages}
  {5672} (\bibinfo {year} {2018})},\ \Eprint {http://arxiv.org/abs/1711.10494}
  {arXiv:1711.10494 [astro-ph.GA]} \BibitemShut {NoStop}%
\bibitem [{\citenamefont {Rodriguez}\ \emph
  {et~al.}(2018{\natexlab{a}})\citenamefont {Rodriguez}, \citenamefont
  {Amaro-Seoane}, \citenamefont {Chatterjee},\ and\ \citenamefont
  {Rasio}}]{Rodriguez:2017pec}%
  \BibitemOpen
  \bibfield  {author} {\bibinfo {author} {\bibfnamefont {C.~L.}\ \bibnamefont
  {Rodriguez}}, \bibinfo {author} {\bibfnamefont {P.}~\bibnamefont
  {Amaro-Seoane}}, \bibinfo {author} {\bibfnamefont {S.}~\bibnamefont
  {Chatterjee}}, \ and\ \bibinfo {author} {\bibfnamefont {F.~A.}\ \bibnamefont
  {Rasio}},\ }\href {\doibase 10.1103/PhysRevLett.120.151101} {\bibfield
  {journal} {\bibinfo  {journal} {Phys. Rev. Lett.}\ }\textbf {\bibinfo
  {volume} {120}},\ \bibinfo {pages} {151101} (\bibinfo {year}
  {2018}{\natexlab{a}})},\ \Eprint {http://arxiv.org/abs/1712.04937}
  {arXiv:1712.04937 [astro-ph.HE]} \BibitemShut {NoStop}%
\bibitem [{\citenamefont {Antonini}\ \emph {et~al.}(2016)\citenamefont
  {Antonini}, \citenamefont {Chatterjee}, \citenamefont {Rodriguez},
  \citenamefont {Morscher}, \citenamefont {Pattabiraman}, \citenamefont
  {Kalogera},\ and\ \citenamefont {Rasio}}]{Antonini:2015zsa}%
  \BibitemOpen
  \bibfield  {author} {\bibinfo {author} {\bibfnamefont {F.}~\bibnamefont
  {Antonini}}, \bibinfo {author} {\bibfnamefont {S.}~\bibnamefont
  {Chatterjee}}, \bibinfo {author} {\bibfnamefont {C.~L.}\ \bibnamefont
  {Rodriguez}}, \bibinfo {author} {\bibfnamefont {M.}~\bibnamefont {Morscher}},
  \bibinfo {author} {\bibfnamefont {B.}~\bibnamefont {Pattabiraman}}, \bibinfo
  {author} {\bibfnamefont {V.}~\bibnamefont {Kalogera}}, \ and\ \bibinfo
  {author} {\bibfnamefont {F.~A.}\ \bibnamefont {Rasio}},\ }\href {\doibase
  10.3847/0004-637X/816/2/65} {\bibfield  {journal} {\bibinfo  {journal}
  {Astrophys. J.}\ }\textbf {\bibinfo {volume} {816}},\ \bibinfo {pages} {65}
  (\bibinfo {year} {2016})},\ \Eprint {http://arxiv.org/abs/1509.05080}
  {arXiv:1509.05080 [astro-ph.GA]} \BibitemShut {NoStop}%
\bibitem [{\citenamefont {Samsing}\ and\ \citenamefont
  {D'Orazio}(2018)}]{Samsing:2018isx}%
  \BibitemOpen
  \bibfield  {author} {\bibinfo {author} {\bibfnamefont {J.}~\bibnamefont
  {Samsing}}\ and\ \bibinfo {author} {\bibfnamefont {D.~J.}\ \bibnamefont
  {D'Orazio}},\ }\href {\doibase 10.1093/mnras/sty2334} {\bibfield  {journal}
  {\bibinfo  {journal} {Mon. Not. Roy. Astron. Soc.}\ }\textbf {\bibinfo
  {volume} {481}},\ \bibinfo {pages} {5445} (\bibinfo {year} {2018})},\ \Eprint
  {http://arxiv.org/abs/1804.06519} {arXiv:1804.06519 [astro-ph.HE]}
  \BibitemShut {NoStop}%
\bibitem [{\citenamefont {D'Orazio}\ and\ \citenamefont
  {Samsing}(2018)}]{DOrazio:2018jnv}%
  \BibitemOpen
  \bibfield  {author} {\bibinfo {author} {\bibfnamefont {D.~J.}\ \bibnamefont
  {D'Orazio}}\ and\ \bibinfo {author} {\bibfnamefont {J.}~\bibnamefont
  {Samsing}},\ }\href {\doibase 10.1093/mnras/sty2568} {\bibfield  {journal}
  {\bibinfo  {journal} {Mon. Not. Roy. Astron. Soc.}\ }\textbf {\bibinfo
  {volume} {481}},\ \bibinfo {pages} {4775} (\bibinfo {year} {2018})},\ \Eprint
  {http://arxiv.org/abs/1805.06194} {arXiv:1805.06194 [astro-ph.HE]}
  \BibitemShut {NoStop}%
\bibitem [{\citenamefont {Samsing}\ and\ \citenamefont
  {D'Orazio}(2019)}]{Samsing:2018nxk}%
  \BibitemOpen
  \bibfield  {author} {\bibinfo {author} {\bibfnamefont {J.}~\bibnamefont
  {Samsing}}\ and\ \bibinfo {author} {\bibfnamefont {D.~J.}\ \bibnamefont
  {D'Orazio}},\ }\href {\doibase 10.1103/PhysRevD.99.063006} {\bibfield
  {journal} {\bibinfo  {journal} {Phys. Rev.}\ }\textbf {\bibinfo {volume}
  {D99}},\ \bibinfo {pages} {063006} (\bibinfo {year} {2019})},\ \Eprint
  {http://arxiv.org/abs/1807.08864} {arXiv:1807.08864 [astro-ph.HE]}
  \BibitemShut {NoStop}%
\bibitem [{\citenamefont {{Prince}}(2003)}]{Prince:2003aa}%
  \BibitemOpen
  \bibfield  {author} {\bibinfo {author} {\bibfnamefont {T.}~\bibnamefont
  {{Prince}}},\ }\href@noop {} {\bibfield  {journal} {\bibinfo  {journal}
  {American Astronomical Society Meeting}\ }\textbf {\bibinfo {volume} {202}},\
  \bibinfo {pages} {3701} (\bibinfo {year} {2003})}\BibitemShut {NoStop}%
\bibitem [{\citenamefont {Kawamura}\ \emph {et~al.}(2011)\citenamefont
  {Kawamura} \emph {et~al.}}]{Kawamura:2011zz}%
  \BibitemOpen
  \bibfield  {author} {\bibinfo {author} {\bibfnamefont {S.}~\bibnamefont
  {Kawamura}} \emph {et~al.},\ }\bibfield  {booktitle} {\emph {\bibinfo
  {booktitle} {{Laser interferometer space antenna. Proceedings, 8th
  International LISA Symposium, Stanford, USA, June 28-July 2, 2010}}},\ }\href
  {\doibase 10.1088/0264-9381/28/9/094011} {\bibfield  {journal} {\bibinfo
  {journal} {Class. Quant. Grav.}\ }\textbf {\bibinfo {volume} {28}},\ \bibinfo
  {pages} {094011} (\bibinfo {year} {2011})}\BibitemShut {NoStop}%
\bibitem [{\citenamefont {Abramovici}\ \emph {et~al.}(1992)\citenamefont
  {Abramovici}, \citenamefont {Althouse}, \citenamefont {Drever}, \citenamefont
  {Gursel}, \citenamefont {Kawamura} \emph {et~al.}}]{Abramovici:1992ah}%
  \BibitemOpen
  \bibfield  {author} {\bibinfo {author} {\bibfnamefont {A.}~\bibnamefont
  {Abramovici}}, \bibinfo {author} {\bibfnamefont {W.~E.}\ \bibnamefont
  {Althouse}}, \bibinfo {author} {\bibfnamefont {R.~W.}\ \bibnamefont
  {Drever}}, \bibinfo {author} {\bibfnamefont {Y.}~\bibnamefont {Gursel}},
  \bibinfo {author} {\bibfnamefont {S.}~\bibnamefont {Kawamura}},  \emph
  {et~al.},\ }\href {\doibase 10.1126/science.256.5055.325} {\bibfield
  {journal} {\bibinfo  {journal} {Science}\ }\textbf {\bibinfo {volume}
  {256}},\ \bibinfo {pages} {325} (\bibinfo {year} {1992})}\BibitemShut
  {NoStop}%
\bibitem [{\citenamefont {Harry}(2010)}]{Harry:2010zz}%
  \BibitemOpen
  \bibfield  {author} {\bibinfo {author} {\bibfnamefont {G.~M.}\ \bibnamefont
  {Harry}} (\bibinfo {collaboration} {LIGO Scientific Collaboration}),\ }\href
  {\doibase 10.1088/0264-9381/27/8/084006} {\bibfield  {journal} {\bibinfo
  {journal} {Class.Quant.Grav.}\ }\textbf {\bibinfo {volume} {27}},\ \bibinfo
  {pages} {084006} (\bibinfo {year} {2010})}\BibitemShut {NoStop}%
\bibitem [{\citenamefont {Aasi}\ \emph {et~al.}(2015)\citenamefont {Aasi} \emph
  {et~al.}}]{TheLIGOScientific:2014jea}%
  \BibitemOpen
  \bibfield  {author} {\bibinfo {author} {\bibfnamefont {J.}~\bibnamefont
  {Aasi}} \emph {et~al.} (\bibinfo {collaboration} {LIGO Scientific}),\ }\href
  {\doibase 10.1088/0264-9381/32/7/074001} {\bibfield  {journal} {\bibinfo
  {journal} {Class. Quant. Grav.}\ }\textbf {\bibinfo {volume} {32}},\ \bibinfo
  {pages} {074001} (\bibinfo {year} {2015})},\ \Eprint
  {http://arxiv.org/abs/1411.4547} {arXiv:1411.4547 [gr-qc]} \BibitemShut
  {NoStop}%
\bibitem [{\citenamefont {Rodriguez}\ \emph
  {et~al.}(2018{\natexlab{b}})\citenamefont {Rodriguez}, \citenamefont
  {Amaro-Seoane}, \citenamefont {Chatterjee}, \citenamefont {Kremer},
  \citenamefont {Rasio}, \citenamefont {Samsing}, \citenamefont {Ye},\ and\
  \citenamefont {Zevin}}]{Rodriguez:2018pss}%
  \BibitemOpen
  \bibfield  {author} {\bibinfo {author} {\bibfnamefont {C.~L.}\ \bibnamefont
  {Rodriguez}}, \bibinfo {author} {\bibfnamefont {P.}~\bibnamefont
  {Amaro-Seoane}}, \bibinfo {author} {\bibfnamefont {S.}~\bibnamefont
  {Chatterjee}}, \bibinfo {author} {\bibfnamefont {K.}~\bibnamefont {Kremer}},
  \bibinfo {author} {\bibfnamefont {F.~A.}\ \bibnamefont {Rasio}}, \bibinfo
  {author} {\bibfnamefont {J.}~\bibnamefont {Samsing}}, \bibinfo {author}
  {\bibfnamefont {C.~S.}\ \bibnamefont {Ye}}, \ and\ \bibinfo {author}
  {\bibfnamefont {M.}~\bibnamefont {Zevin}},\ }\href {\doibase
  10.1103/PhysRevD.98.123005} {\bibfield  {journal} {\bibinfo  {journal} {Phys.
  Rev.}\ }\textbf {\bibinfo {volume} {D98}},\ \bibinfo {pages} {123005}
  (\bibinfo {year} {2018}{\natexlab{b}})},\ \Eprint
  {http://arxiv.org/abs/1811.04926} {arXiv:1811.04926 [astro-ph.HE]}
  \BibitemShut {NoStop}%
\bibitem [{\citenamefont {Yunes}\ and\ \citenamefont {Pretorius}(2009)}]{PPE}%
  \BibitemOpen
  \bibfield  {author} {\bibinfo {author} {\bibfnamefont {N.}~\bibnamefont
  {Yunes}}\ and\ \bibinfo {author} {\bibfnamefont {F.}~\bibnamefont
  {Pretorius}},\ }\href {\doibase 10.1103/PhysRevD.80.122003} {\bibfield
  {journal} {\bibinfo  {journal} {Phys.Rev.}\ }\textbf {\bibinfo {volume}
  {D80}},\ \bibinfo {pages} {122003} (\bibinfo {year} {2009})},\ \Eprint
  {http://arxiv.org/abs/0909.3328} {arXiv:0909.3328 [gr-qc]} \BibitemShut
  {NoStop}%
\bibitem [{\citenamefont {Abbott}\ \emph {et~al.}(2019)\citenamefont {Abbott}
  \emph {et~al.}}]{LIGOScientific:2019fpa}%
  \BibitemOpen
  \bibfield  {author} {\bibinfo {author} {\bibfnamefont {B.~P.}\ \bibnamefont
  {Abbott}} \emph {et~al.} (\bibinfo {collaboration} {LIGO Scientific,
  Virgo}),\ }\href@noop {} {\  (\bibinfo {year} {2019})},\ \Eprint
  {http://arxiv.org/abs/1903.04467} {arXiv:1903.04467 [gr-qc]} \BibitemShut
  {NoStop}%
\bibitem [{\citenamefont {Loutrel}\ \emph {et~al.}(2014)\citenamefont
  {Loutrel}, \citenamefont {Yunes},\ and\ \citenamefont
  {Pretorius}}]{Loutrel:2014vja}%
  \BibitemOpen
  \bibfield  {author} {\bibinfo {author} {\bibfnamefont {N.}~\bibnamefont
  {Loutrel}}, \bibinfo {author} {\bibfnamefont {N.}~\bibnamefont {Yunes}}, \
  and\ \bibinfo {author} {\bibfnamefont {F.}~\bibnamefont {Pretorius}},\ }\href
  {\doibase 10.1103/PhysRevD.90.104010} {\bibfield  {journal} {\bibinfo
  {journal} {Phys.Rev.}\ }\textbf {\bibinfo {volume} {D90}},\ \bibinfo {pages}
  {104010} (\bibinfo {year} {2014})},\ \Eprint {http://arxiv.org/abs/1404.0092}
  {arXiv:1404.0092 [gr-qc]} \BibitemShut {NoStop}%
\bibitem [{\citenamefont {Yang}(2019)}]{Yang:2019kmf}%
  \BibitemOpen
  \bibfield  {author} {\bibinfo {author} {\bibfnamefont {H.}~\bibnamefont
  {Yang}},\ }\href@noop {} {\  (\bibinfo {year} {2019})},\ \Eprint
  {http://arxiv.org/abs/1904.11089} {arXiv:1904.11089 [gr-qc]} \BibitemShut
  {NoStop}%
\bibitem [{\citenamefont {Yang}\ \emph {et~al.}(2018)\citenamefont {Yang},
  \citenamefont {East}, \citenamefont {Paschalidis}, \citenamefont
  {Pretorius},\ and\ \citenamefont {Mendes}}]{Yang:2018bzx}%
  \BibitemOpen
  \bibfield  {author} {\bibinfo {author} {\bibfnamefont {H.}~\bibnamefont
  {Yang}}, \bibinfo {author} {\bibfnamefont {W.~E.}\ \bibnamefont {East}},
  \bibinfo {author} {\bibfnamefont {V.}~\bibnamefont {Paschalidis}}, \bibinfo
  {author} {\bibfnamefont {F.}~\bibnamefont {Pretorius}}, \ and\ \bibinfo
  {author} {\bibfnamefont {R.~F.~P.}\ \bibnamefont {Mendes}},\ }\href {\doibase
  10.1103/PhysRevD.98.044007} {\bibfield  {journal} {\bibinfo  {journal} {Phys.
  Rev.}\ }\textbf {\bibinfo {volume} {D98}},\ \bibinfo {pages} {044007}
  (\bibinfo {year} {2018})},\ \Eprint {http://arxiv.org/abs/1806.00158}
  {arXiv:1806.00158 [gr-qc]} \BibitemShut {NoStop}%
\bibitem [{\citenamefont {Stephens}\ \emph {et~al.}(2011)\citenamefont
  {Stephens}, \citenamefont {East},\ and\ \citenamefont
  {Pretorius}}]{Stephens:2011as}%
  \BibitemOpen
  \bibfield  {author} {\bibinfo {author} {\bibfnamefont {B.~C.}\ \bibnamefont
  {Stephens}}, \bibinfo {author} {\bibfnamefont {W.~E.}\ \bibnamefont {East}},
  \ and\ \bibinfo {author} {\bibfnamefont {F.}~\bibnamefont {Pretorius}},\
  }\href {\doibase 10.1088/2041-8205/737/1/L5} {\bibfield  {journal} {\bibinfo
  {journal} {Astrophys. J.}\ }\textbf {\bibinfo {volume} {737}},\ \bibinfo
  {pages} {L5} (\bibinfo {year} {2011})},\ \Eprint
  {http://arxiv.org/abs/1105.3175} {arXiv:1105.3175 [astro-ph.HE]} \BibitemShut
  {NoStop}%
\bibitem [{\citenamefont {Chaurasia}\ \emph {et~al.}(2018)\citenamefont
  {Chaurasia}, \citenamefont {Dietrich}, \citenamefont {Johnson-McDaniel},
  \citenamefont {Ujevic}, \citenamefont {Tichy},\ and\ \citenamefont
  {Brügmann}}]{Chaurasia:2018zhg}%
  \BibitemOpen
  \bibfield  {author} {\bibinfo {author} {\bibfnamefont {S.~V.}\ \bibnamefont
  {Chaurasia}}, \bibinfo {author} {\bibfnamefont {T.}~\bibnamefont {Dietrich}},
  \bibinfo {author} {\bibfnamefont {N.~K.}\ \bibnamefont {Johnson-McDaniel}},
  \bibinfo {author} {\bibfnamefont {M.}~\bibnamefont {Ujevic}}, \bibinfo
  {author} {\bibfnamefont {W.}~\bibnamefont {Tichy}}, \ and\ \bibinfo {author}
  {\bibfnamefont {B.}~\bibnamefont {Brügmann}},\ }\href {\doibase
  10.1103/PhysRevD.98.104005} {\bibfield  {journal} {\bibinfo  {journal} {Phys.
  Rev.}\ }\textbf {\bibinfo {volume} {D98}},\ \bibinfo {pages} {104005}
  (\bibinfo {year} {2018})},\ \Eprint {http://arxiv.org/abs/1807.06857}
  {arXiv:1807.06857 [gr-qc]} \BibitemShut {NoStop}%
\bibitem [{\citenamefont {Papenfort}\ \emph {et~al.}(2018)\citenamefont
  {Papenfort}, \citenamefont {Gold},\ and\ \citenamefont
  {Rezzolla}}]{Papenfort:2018bjk}%
  \BibitemOpen
  \bibfield  {author} {\bibinfo {author} {\bibfnamefont {L.~J.}\ \bibnamefont
  {Papenfort}}, \bibinfo {author} {\bibfnamefont {R.}~\bibnamefont {Gold}}, \
  and\ \bibinfo {author} {\bibfnamefont {L.}~\bibnamefont {Rezzolla}},\ }\href
  {\doibase 10.1103/PhysRevD.98.104028} {\bibfield  {journal} {\bibinfo
  {journal} {Phys. Rev.}\ }\textbf {\bibinfo {volume} {D98}},\ \bibinfo {pages}
  {104028} (\bibinfo {year} {2018})},\ \Eprint
  {http://arxiv.org/abs/1807.03795} {arXiv:1807.03795 [gr-qc]} \BibitemShut
  {NoStop}%
\bibitem [{\citenamefont {Chirenti}\ \emph {et~al.}(2017)\citenamefont
  {Chirenti}, \citenamefont {Gold},\ and\ \citenamefont
  {Miller}}]{Chirenti:2016xys}%
  \BibitemOpen
  \bibfield  {author} {\bibinfo {author} {\bibfnamefont {C.}~\bibnamefont
  {Chirenti}}, \bibinfo {author} {\bibfnamefont {R.}~\bibnamefont {Gold}}, \
  and\ \bibinfo {author} {\bibfnamefont {M.~C.}\ \bibnamefont {Miller}},\
  }\href {\doibase 10.3847/1538-4357/aa5ebb} {\bibfield  {journal} {\bibinfo
  {journal} {Astrophys. J.}\ }\textbf {\bibinfo {volume} {837}},\ \bibinfo
  {pages} {67} (\bibinfo {year} {2017})},\ \Eprint
  {http://arxiv.org/abs/1612.07097} {arXiv:1612.07097 [astro-ph.HE]}
  \BibitemShut {NoStop}%
\bibitem [{\citenamefont {East}\ \emph {et~al.}(2016)\citenamefont {East},
  \citenamefont {Paschalidis}, \citenamefont {Pretorius},\ and\ \citenamefont
  {Shapiro}}]{East:2015vix}%
  \BibitemOpen
  \bibfield  {author} {\bibinfo {author} {\bibfnamefont {W.~E.}\ \bibnamefont
  {East}}, \bibinfo {author} {\bibfnamefont {V.}~\bibnamefont {Paschalidis}},
  \bibinfo {author} {\bibfnamefont {F.}~\bibnamefont {Pretorius}}, \ and\
  \bibinfo {author} {\bibfnamefont {S.~L.}\ \bibnamefont {Shapiro}},\ }\href
  {\doibase 10.1103/PhysRevD.93.024011} {\bibfield  {journal} {\bibinfo
  {journal} {Phys. Rev.}\ }\textbf {\bibinfo {volume} {D93}},\ \bibinfo {pages}
  {024011} (\bibinfo {year} {2016})},\ \Eprint
  {http://arxiv.org/abs/1511.01093} {arXiv:1511.01093 [astro-ph.HE]}
  \BibitemShut {NoStop}%
\bibitem [{\citenamefont {Choptuik}\ \emph {et~al.}(2015)\citenamefont
  {Choptuik}, \citenamefont {Lehner},\ and\ \citenamefont
  {Pretorius}}]{Choptuik:2015mma}%
  \BibitemOpen
  \bibfield  {author} {\bibinfo {author} {\bibfnamefont {M.~W.}\ \bibnamefont
  {Choptuik}}, \bibinfo {author} {\bibfnamefont {L.}~\bibnamefont {Lehner}}, \
  and\ \bibinfo {author} {\bibfnamefont {F.}~\bibnamefont {Pretorius}},\
  }\href@noop {} {\  (\bibinfo {year} {2015})},\ \Eprint
  {http://arxiv.org/abs/1502.06853} {arXiv:1502.06853 [gr-qc]} \BibitemShut
  {NoStop}%
\bibitem [{\citenamefont {Rosofsky}\ \emph {et~al.}(2019)\citenamefont
  {Rosofsky}, \citenamefont {Gold}, \citenamefont {Chirenti}, \citenamefont
  {Huerta},\ and\ \citenamefont {Miller}}]{Rosofsky:2018vyg}%
  \BibitemOpen
  \bibfield  {author} {\bibinfo {author} {\bibfnamefont {S.}~\bibnamefont
  {Rosofsky}}, \bibinfo {author} {\bibfnamefont {R.}~\bibnamefont {Gold}},
  \bibinfo {author} {\bibfnamefont {C.}~\bibnamefont {Chirenti}}, \bibinfo
  {author} {\bibfnamefont {E.~A.}\ \bibnamefont {Huerta}}, \ and\ \bibinfo
  {author} {\bibfnamefont {M.~C.}\ \bibnamefont {Miller}},\ }\href {\doibase
  10.1103/PhysRevD.99.084024} {\bibfield  {journal} {\bibinfo  {journal} {Phys.
  Rev.}\ }\textbf {\bibinfo {volume} {D99}},\ \bibinfo {pages} {084024}
  (\bibinfo {year} {2019})},\ \Eprint {http://arxiv.org/abs/1812.06126}
  {arXiv:1812.06126 [gr-qc]} \BibitemShut {NoStop}%
\bibitem [{\citenamefont {Poisson}(2014)}]{PW}%
  \BibitemOpen
  \bibfield  {author} {\bibinfo {author} {\bibfnamefont {C.~M.}\ \bibnamefont
  {Poisson}, \bibfnamefont {E.~and~Will}},\ }\href@noop {} {\emph {\bibinfo
  {title} {Gravity: Newtonian, Post-Newtonian, Relativistic}}}\ (\bibinfo
  {publisher} {Cambridge University Press},\ \bibinfo {address} {Cambridge},\
  \bibinfo {year} {2014})\BibitemShut {NoStop}%
\bibitem [{\citenamefont {Blanchet}(2014)}]{Blanchet:2013haa}%
  \BibitemOpen
  \bibfield  {author} {\bibinfo {author} {\bibfnamefont {L.}~\bibnamefont
  {Blanchet}},\ }\href {\doibase 10.12942/lrr-2014-2} {\bibfield  {journal}
  {\bibinfo  {journal} {Living Rev. Rel.}\ }\textbf {\bibinfo {volume} {17}},\
  \bibinfo {pages} {2} (\bibinfo {year} {2014})},\ \Eprint
  {http://arxiv.org/abs/1310.1528} {arXiv:1310.1528 [gr-qc]} \BibitemShut
  {NoStop}%
\bibitem [{\citenamefont {{East}}\ \emph {et~al.}(2012)\citenamefont {{East}},
  \citenamefont {{Pretorius}},\ and\ \citenamefont
  {{Stephens}}}]{2012PhRvD..85l4009E}%
  \BibitemOpen
  \bibfield  {author} {\bibinfo {author} {\bibfnamefont {W.~E.}\ \bibnamefont
  {{East}}}, \bibinfo {author} {\bibfnamefont {F.}~\bibnamefont {{Pretorius}}},
  \ and\ \bibinfo {author} {\bibfnamefont {B.~C.}\ \bibnamefont {{Stephens}}},\
  }\href {\doibase 10.1103/PhysRevD.85.124009} {\bibfield  {journal} {\bibinfo
  {journal} {{Phys. Rev. D.}}\ }\textbf {\bibinfo {volume} {85}},\ \bibinfo
  {eid} {124009} (\bibinfo {year} {2012})},\ \Eprint
  {http://arxiv.org/abs/1111.3055} {arXiv:1111.3055 [astro-ph.HE]} \BibitemShut
  {NoStop}%
\bibitem [{\citenamefont {{East}}\ \emph {et~al.}(2013)\citenamefont {{East}},
  \citenamefont {{McWilliams}}, \citenamefont {{Levin}},\ and\ \citenamefont
  {{Pretorius}}}]{2013PhRvD..87d3004E}%
  \BibitemOpen
  \bibfield  {author} {\bibinfo {author} {\bibfnamefont {W.~E.}\ \bibnamefont
  {{East}}}, \bibinfo {author} {\bibfnamefont {S.~T.}\ \bibnamefont
  {{McWilliams}}}, \bibinfo {author} {\bibfnamefont {J.}~\bibnamefont
  {{Levin}}}, \ and\ \bibinfo {author} {\bibfnamefont {F.}~\bibnamefont
  {{Pretorius}}},\ }\href {\doibase 10.1103/PhysRevD.87.043004} {\bibfield
  {journal} {\bibinfo  {journal} {{Phys. Rev. D.}}\ }\textbf {\bibinfo {volume}
  {87}},\ \bibinfo {eid} {043004} (\bibinfo {year} {2013})},\ \Eprint
  {http://arxiv.org/abs/1212.0837} {arXiv:1212.0837 [gr-qc]} \BibitemShut
  {NoStop}%
\bibitem [{\citenamefont {Salemi}(2019)}]{Salemi:2019owp}%
  \BibitemOpen
  \bibfield  {author} {\bibinfo {author} {\bibfnamefont {F.}~\bibnamefont
  {Salemi}} (\bibinfo {collaboration} {LIGO Scientific, Virgo}),\ }\href@noop
  {} {\  (\bibinfo {year} {2019})},\ \Eprint {http://arxiv.org/abs/1907.09384}
  {arXiv:1907.09384 [astro-ph.HE]} \BibitemShut {NoStop}%
\bibitem [{\citenamefont {Tai}\ \emph {et~al.}(2014)\citenamefont {Tai},
  \citenamefont {McWilliams},\ and\ \citenamefont {Pretorius}}]{Tai:2014bfa}%
  \BibitemOpen
  \bibfield  {author} {\bibinfo {author} {\bibfnamefont {K.~S.}\ \bibnamefont
  {Tai}}, \bibinfo {author} {\bibfnamefont {S.~T.}\ \bibnamefont {McWilliams}},
  \ and\ \bibinfo {author} {\bibfnamefont {F.}~\bibnamefont {Pretorius}},\
  }\href {\doibase 10.1103/PhysRevD.90.103001} {\bibfield  {journal} {\bibinfo
  {journal} {Phys. Rev.}\ }\textbf {\bibinfo {volume} {D90}},\ \bibinfo {pages}
  {103001} (\bibinfo {year} {2014})},\ \Eprint {http://arxiv.org/abs/1403.7754}
  {arXiv:1403.7754 [gr-qc]} \BibitemShut {NoStop}%
\bibitem [{\citenamefont {Cabero}\ \emph {et~al.}(2019)\citenamefont {Cabero}
  \emph {et~al.}}]{Cabero:2019orq}%
  \BibitemOpen
  \bibfield  {author} {\bibinfo {author} {\bibfnamefont {M.}~\bibnamefont
  {Cabero}} \emph {et~al.},\ }\href {\doibase 10.1088/1361-6382/ab2e14}
  {\bibfield  {journal} {\bibinfo  {journal} {Class. Quant. Grav.}\ }\textbf
  {\bibinfo {volume} {36}},\ \bibinfo {pages} {155010} (\bibinfo {year}
  {2019})},\ \Eprint {http://arxiv.org/abs/1901.05093} {arXiv:1901.05093
  [physics.ins-det]} \BibitemShut {NoStop}%
\bibitem [{\citenamefont {Abbott}\ \emph {et~al.}(2016)\citenamefont {Abbott}
  \emph {et~al.}}]{TheLIGOScientific:2016zmo}%
  \BibitemOpen
  \bibfield  {author} {\bibinfo {author} {\bibfnamefont {B.~P.}\ \bibnamefont
  {Abbott}} \emph {et~al.} (\bibinfo {collaboration} {LIGO Scientific,
  Virgo}),\ }\href {\doibase 10.1088/0264-9381/33/13/134001} {\bibfield
  {journal} {\bibinfo  {journal} {Class. Quant. Grav.}\ }\textbf {\bibinfo
  {volume} {33}},\ \bibinfo {pages} {134001} (\bibinfo {year} {2016})},\
  \Eprint {http://arxiv.org/abs/1602.03844} {arXiv:1602.03844 [gr-qc]}
  \BibitemShut {NoStop}%
\bibitem [{\citenamefont {Abbott}\ \emph {et~al.}(2018)\citenamefont {Abbott}
  \emph {et~al.}}]{LIGOScientific:2018mvr}%
  \BibitemOpen
  \bibfield  {author} {\bibinfo {author} {\bibfnamefont {B.~P.}\ \bibnamefont
  {Abbott}} \emph {et~al.} (\bibinfo {collaboration} {LIGO Scientific,
  Virgo}),\ }\href@noop {} {\  (\bibinfo {year} {2018})},\ \Eprint
  {http://arxiv.org/abs/1811.12907} {arXiv:1811.12907 [astro-ph.HE]}
  \BibitemShut {NoStop}%
\bibitem [{\citenamefont {Loutrel}\ and\ \citenamefont
  {Yunes}(2017{\natexlab{a}})}]{Loutrel:2017fgu}%
  \BibitemOpen
  \bibfield  {author} {\bibinfo {author} {\bibfnamefont {N.}~\bibnamefont
  {Loutrel}}\ and\ \bibinfo {author} {\bibfnamefont {N.}~\bibnamefont
  {Yunes}},\ }\href {\doibase 10.1088/1361-6382/aa7449} {\bibfield  {journal}
  {\bibinfo  {journal} {Class. Quant. Grav.}\ }\textbf {\bibinfo {volume}
  {34}},\ \bibinfo {pages} {135011} (\bibinfo {year} {2017}{\natexlab{a}})},\
  \Eprint {http://arxiv.org/abs/1702.01818} {arXiv:1702.01818 [gr-qc]}
  \BibitemShut {NoStop}%
\bibitem [{\citenamefont {Wei}\ and\ \citenamefont
  {Huerta}(2019)}]{Wei:2019zlc}%
  \BibitemOpen
  \bibfield  {author} {\bibinfo {author} {\bibfnamefont {W.}~\bibnamefont
  {Wei}}\ and\ \bibinfo {author} {\bibfnamefont {E.~A.}\ \bibnamefont
  {Huerta}},\ }\href@noop {} {\  (\bibinfo {year} {2019})},\ \Eprint
  {http://arxiv.org/abs/1901.00869} {arXiv:1901.00869 [gr-qc]} \BibitemShut
  {NoStop}%
\bibitem [{\citenamefont {Rebei}\ \emph {et~al.}(2019)\citenamefont {Rebei},
  \citenamefont {Huerta}, \citenamefont {Wang}, \citenamefont {Habib},
  \citenamefont {Haas}, \citenamefont {Johnson},\ and\ \citenamefont
  {George}}]{Rebei:2018lzh}%
  \BibitemOpen
  \bibfield  {author} {\bibinfo {author} {\bibfnamefont {A.}~\bibnamefont
  {Rebei}}, \bibinfo {author} {\bibfnamefont {E.~A.}\ \bibnamefont {Huerta}},
  \bibinfo {author} {\bibfnamefont {S.}~\bibnamefont {Wang}}, \bibinfo {author}
  {\bibfnamefont {S.}~\bibnamefont {Habib}}, \bibinfo {author} {\bibfnamefont
  {R.}~\bibnamefont {Haas}}, \bibinfo {author} {\bibfnamefont {D.}~\bibnamefont
  {Johnson}}, \ and\ \bibinfo {author} {\bibfnamefont {D.}~\bibnamefont
  {George}},\ }\href {\doibase 10.1103/PhysRevD.100.044025} {\bibfield
  {journal} {\bibinfo  {journal} {Phys. Rev.}\ }\textbf {\bibinfo {volume}
  {D100}},\ \bibinfo {pages} {044025} (\bibinfo {year} {2019})},\ \Eprint
  {http://arxiv.org/abs/1807.09787} {arXiv:1807.09787 [gr-qc]} \BibitemShut
  {NoStop}%
\bibitem [{\citenamefont {George}\ and\ \citenamefont
  {Huerta}(2018{\natexlab{a}})}]{George:2017pmj}%
  \BibitemOpen
  \bibfield  {author} {\bibinfo {author} {\bibfnamefont {D.}~\bibnamefont
  {George}}\ and\ \bibinfo {author} {\bibfnamefont {E.~A.}\ \bibnamefont
  {Huerta}},\ }\href {\doibase 10.1016/j.physletb.2017.12.053} {\bibfield
  {journal} {\bibinfo  {journal} {Phys. Lett.}\ }\textbf {\bibinfo {volume}
  {B778}},\ \bibinfo {pages} {64} (\bibinfo {year} {2018}{\natexlab{a}})},\
  \Eprint {http://arxiv.org/abs/1711.03121} {arXiv:1711.03121 [gr-qc]}
  \BibitemShut {NoStop}%
\bibitem [{\citenamefont {George}\ and\ \citenamefont
  {Huerta}(2017)}]{George:2017vlv}%
  \BibitemOpen
  \bibfield  {author} {\bibinfo {author} {\bibfnamefont {D.}~\bibnamefont
  {George}}\ and\ \bibinfo {author} {\bibfnamefont {E.~A.}\ \bibnamefont
  {Huerta}},\ }in\ \href@noop {} {\emph {\bibinfo {booktitle} {{NiPS Summer
  School 2017 Gubbio, Perugia, Italy, June 30-July 3, 2017}}}}\ (\bibinfo
  {year} {2017})\ \Eprint {http://arxiv.org/abs/1711.07966} {arXiv:1711.07966
  [gr-qc]} \BibitemShut {NoStop}%
\bibitem [{\citenamefont {Shen}\ \emph
  {et~al.}(2019{\natexlab{a}})\citenamefont {Shen}, \citenamefont {George},
  \citenamefont {Huerta},\ and\ \citenamefont {Zhao}}]{Shen:2019ohi}%
  \BibitemOpen
  \bibfield  {author} {\bibinfo {author} {\bibfnamefont {H.}~\bibnamefont
  {Shen}}, \bibinfo {author} {\bibfnamefont {D.}~\bibnamefont {George}},
  \bibinfo {author} {\bibfnamefont {E.~A.}\ \bibnamefont {Huerta}}, \ and\
  \bibinfo {author} {\bibfnamefont {Z.}~\bibnamefont {Zhao}},\ }\href {\doibase
  10.1109/ICASSP.2019.8683061} {\  (\bibinfo {year} {2019}{\natexlab{a}}),\
  10.1109/ICASSP.2019.8683061},\ \Eprint {http://arxiv.org/abs/1903.03105}
  {arXiv:1903.03105 [astro-ph.CO]} \BibitemShut {NoStop}%
\bibitem [{\citenamefont {Luo}\ \emph {et~al.}(2020)\citenamefont {Luo},
  \citenamefont {Lin}, \citenamefont {Chen},\ and\ \citenamefont
  {Huang}}]{Luo:2019hvt}%
  \BibitemOpen
  \bibfield  {author} {\bibinfo {author} {\bibfnamefont {H.-M.}\ \bibnamefont
  {Luo}}, \bibinfo {author} {\bibfnamefont {W.}~\bibnamefont {Lin}}, \bibinfo
  {author} {\bibfnamefont {Z.-C.}\ \bibnamefont {Chen}}, \ and\ \bibinfo
  {author} {\bibfnamefont {Q.-G.}\ \bibnamefont {Huang}},\ }\href {\doibase
  10.1007/s11467-019-0936-x} {\bibfield  {journal} {\bibinfo  {journal} {Front.
  Phys.(Beijing)}\ }\textbf {\bibinfo {volume} {15}},\ \bibinfo {pages} {14601}
  (\bibinfo {year} {2020})}\BibitemShut {NoStop}%
\bibitem [{\citenamefont {Kim}\ \emph {et~al.}(2019)\citenamefont {Kim},
  \citenamefont {Li}, \citenamefont {Lo}, \citenamefont {Sachdev},\ and\
  \citenamefont {Yuen}}]{Kim:2019ktw}%
  \BibitemOpen
  \bibfield  {author} {\bibinfo {author} {\bibfnamefont {K.}~\bibnamefont
  {Kim}}, \bibinfo {author} {\bibfnamefont {T.~G.~F.}\ \bibnamefont {Li}},
  \bibinfo {author} {\bibfnamefont {R.~K.~L.}\ \bibnamefont {Lo}}, \bibinfo
  {author} {\bibfnamefont {S.}~\bibnamefont {Sachdev}}, \ and\ \bibinfo
  {author} {\bibfnamefont {R.~S.~H.}\ \bibnamefont {Yuen}},\ }\href@noop {} {\
  (\bibinfo {year} {2019})},\ \Eprint {http://arxiv.org/abs/1912.07740}
  {arXiv:1912.07740 [astro-ph.IM]} \BibitemShut {NoStop}%
\bibitem [{\citenamefont {George}\ and\ \citenamefont
  {Huerta}(2018{\natexlab{b}})}]{George:2016hay}%
  \BibitemOpen
  \bibfield  {author} {\bibinfo {author} {\bibfnamefont {D.}~\bibnamefont
  {George}}\ and\ \bibinfo {author} {\bibfnamefont {E.~A.}\ \bibnamefont
  {Huerta}},\ }\href {\doibase 10.1103/PhysRevD.97.044039} {\bibfield
  {journal} {\bibinfo  {journal} {Phys. Rev.}\ }\textbf {\bibinfo {volume}
  {D97}},\ \bibinfo {pages} {044039} (\bibinfo {year} {2018}{\natexlab{b}})},\
  \Eprint {http://arxiv.org/abs/1701.00008} {arXiv:1701.00008 [astro-ph.IM]}
  \BibitemShut {NoStop}%
\bibitem [{\citenamefont {Gabbard}\ \emph {et~al.}(2018)\citenamefont
  {Gabbard}, \citenamefont {Williams}, \citenamefont {Hayes},\ and\
  \citenamefont {Messenger}}]{Gabbard:2017lja}%
  \BibitemOpen
  \bibfield  {author} {\bibinfo {author} {\bibfnamefont {H.}~\bibnamefont
  {Gabbard}}, \bibinfo {author} {\bibfnamefont {M.}~\bibnamefont {Williams}},
  \bibinfo {author} {\bibfnamefont {F.}~\bibnamefont {Hayes}}, \ and\ \bibinfo
  {author} {\bibfnamefont {C.}~\bibnamefont {Messenger}},\ }\href {\doibase
  10.1103/PhysRevLett.120.141103} {\bibfield  {journal} {\bibinfo  {journal}
  {Phys. Rev. Lett.}\ }\textbf {\bibinfo {volume} {120}},\ \bibinfo {pages}
  {141103} (\bibinfo {year} {2018})},\ \Eprint
  {http://arxiv.org/abs/1712.06041} {arXiv:1712.06041 [astro-ph.IM]}
  \BibitemShut {NoStop}%
\bibitem [{\citenamefont {Shen}\ \emph
  {et~al.}(2019{\natexlab{b}})\citenamefont {Shen}, \citenamefont {Huerta},\
  and\ \citenamefont {Zhao}}]{Shen:2019vep}%
  \BibitemOpen
  \bibfield  {author} {\bibinfo {author} {\bibfnamefont {H.}~\bibnamefont
  {Shen}}, \bibinfo {author} {\bibfnamefont {E.~A.}\ \bibnamefont {Huerta}}, \
  and\ \bibinfo {author} {\bibfnamefont {Z.}~\bibnamefont {Zhao}},\ }\href@noop
  {} {\  (\bibinfo {year} {2019}{\natexlab{b}})},\ \Eprint
  {http://arxiv.org/abs/1903.01998} {arXiv:1903.01998 [gr-qc]} \BibitemShut
  {NoStop}%
\bibitem [{\citenamefont {Chang}\ \emph {et~al.}(2019)\citenamefont {Chang}
  \emph {et~al.}}]{Chang:2019edd}%
  \BibitemOpen
  \bibfield  {author} {\bibinfo {author} {\bibfnamefont {P.}~\bibnamefont
  {Chang}} \emph {et~al.},\ }\href@noop {} {\  (\bibinfo {year} {2019})},\
  \Eprint {http://arxiv.org/abs/1903.04590} {arXiv:1903.04590 [astro-ph.IM]}
  \BibitemShut {NoStop}%
\bibitem [{\citenamefont {Allen}\ \emph {et~al.}(2019)\citenamefont {Allen}
  \emph {et~al.}}]{Allen:2019dkq}%
  \BibitemOpen
  \bibfield  {author} {\bibinfo {author} {\bibfnamefont {G.}~\bibnamefont
  {Allen}} \emph {et~al.}\ }(\bibinfo {year} {2019})\ \Eprint
  {http://arxiv.org/abs/1902.00522} {arXiv:1902.00522 [astro-ph.IM]}
  \BibitemShut {NoStop}%
\bibitem [{\citenamefont {Gebhard}\ \emph {et~al.}(2019)\citenamefont
  {Gebhard}, \citenamefont {Kilbertus}, \citenamefont {Harry},\ and\
  \citenamefont {Schölkopf}}]{Gebhard:2019ldz}%
  \BibitemOpen
  \bibfield  {author} {\bibinfo {author} {\bibfnamefont {T.~D.}\ \bibnamefont
  {Gebhard}}, \bibinfo {author} {\bibfnamefont {N.}~\bibnamefont {Kilbertus}},
  \bibinfo {author} {\bibfnamefont {I.}~\bibnamefont {Harry}}, \ and\ \bibinfo
  {author} {\bibfnamefont {B.}~\bibnamefont {Schölkopf}},\ }\bibfield
  {booktitle} {\emph {\bibinfo {booktitle} {{Phys. Rev. D 100, 063015
  (2019)}}},\ }\href {\doibase 10.1103/PhysRevD.100.063015} {\bibfield
  {journal} {\bibinfo  {journal} {Phys. Rev.}\ }\textbf {\bibinfo {volume}
  {D100}},\ \bibinfo {pages} {063015} (\bibinfo {year} {2019})},\ \Eprint
  {http://arxiv.org/abs/1904.08693} {arXiv:1904.08693 [astro-ph.IM]}
  \BibitemShut {NoStop}%
\bibitem [{\citenamefont {Huerta}\ \emph {et~al.}(2020)\citenamefont {Huerta},
  \citenamefont {Shen}, \citenamefont {Loutrel}, \citenamefont {Pretorius},\
  and\ \citenamefont {Samsing}}]{Huerta:2019ecc}%
  \BibitemOpen
  \bibfield  {author} {\bibinfo {author} {\bibfnamefont {E.}~\bibnamefont
  {Huerta}}, \bibinfo {author} {\bibfnamefont {H.}~\bibnamefont {Shen}},
  \bibinfo {author} {\bibfnamefont {N.}~\bibnamefont {Loutrel}}, \bibinfo
  {author} {\bibfnamefont {F.}~\bibnamefont {Pretorius}}, \ and\ \bibinfo
  {author} {\bibfnamefont {J.}~\bibnamefont {Samsing}},\ }\href@noop {}
  {\bibfield  {journal} {\bibinfo  {journal} {{In preparation}}\ } (\bibinfo
  {year} {2020})}\BibitemShut {NoStop}%
\bibitem [{\citenamefont {Krolak}\ \emph {et~al.}(1991)\citenamefont {Krolak},
  \citenamefont {Lobo},\ and\ \citenamefont {Meers}}]{PhysRevD.43.2470}%
  \BibitemOpen
  \bibfield  {author} {\bibinfo {author} {\bibfnamefont {A.}~\bibnamefont
  {Krolak}}, \bibinfo {author} {\bibfnamefont {J.~A.}\ \bibnamefont {Lobo}}, \
  and\ \bibinfo {author} {\bibfnamefont {B.~J.}\ \bibnamefont {Meers}},\ }\href
  {\doibase 10.1103/PhysRevD.43.2470} {\bibfield  {journal} {\bibinfo
  {journal} {Phys. Rev. D}\ }\textbf {\bibinfo {volume} {43}},\ \bibinfo
  {pages} {2470} (\bibinfo {year} {1991})}\BibitemShut {NoStop}%
\bibitem [{\citenamefont {{Blair}}(2005)}]{2005dgw..book.....B}%
  \BibitemOpen
  \bibfield  {author} {\bibinfo {author} {\bibfnamefont {D.~G.}\ \bibnamefont
  {{Blair}}},\ }\href@noop {} {\emph {\bibinfo {title} {The Detection of
  Gravitational Waves, by Edited by David G.~Blair, Cambridge, UK: Cambridge
  University Press, 2005}}}\ (\bibinfo {year} {2005})\BibitemShut {NoStop}%
\bibitem [{\citenamefont {Martel}\ and\ \citenamefont
  {Poisson}(1999)}]{Martel:1999tm}%
  \BibitemOpen
  \bibfield  {author} {\bibinfo {author} {\bibfnamefont {K.}~\bibnamefont
  {Martel}}\ and\ \bibinfo {author} {\bibfnamefont {E.}~\bibnamefont
  {Poisson}},\ }\href {\doibase 10.1103/PhysRevD.60.124008} {\bibfield
  {journal} {\bibinfo  {journal} {Phys. Rev.}\ }\textbf {\bibinfo {volume}
  {D60}},\ \bibinfo {pages} {124008} (\bibinfo {year} {1999})},\ \Eprint
  {http://arxiv.org/abs/gr-qc/9907006} {arXiv:gr-qc/9907006 [gr-qc]}
  \BibitemShut {NoStop}%
\bibitem [{\citenamefont {Usman}\ \emph {et~al.}(2016)\citenamefont {Usman}
  \emph {et~al.}}]{Usman:2015kfa}%
  \BibitemOpen
  \bibfield  {author} {\bibinfo {author} {\bibfnamefont {S.~A.}\ \bibnamefont
  {Usman}} \emph {et~al.},\ }\href {\doibase 10.1088/0264-9381/33/21/215004}
  {\bibfield  {journal} {\bibinfo  {journal} {Class. Quant. Grav.}\ }\textbf
  {\bibinfo {volume} {33}},\ \bibinfo {pages} {215004} (\bibinfo {year}
  {2016})},\ \Eprint {http://arxiv.org/abs/1508.02357} {arXiv:1508.02357
  [gr-qc]} \BibitemShut {NoStop}%
\bibitem [{\citenamefont {Sachdev}\ \emph {et~al.}(2019)\citenamefont {Sachdev}
  \emph {et~al.}}]{Sachdev:2019vvd}%
  \BibitemOpen
  \bibfield  {author} {\bibinfo {author} {\bibfnamefont {S.}~\bibnamefont
  {Sachdev}} \emph {et~al.},\ }\href@noop {} {\  (\bibinfo {year} {2019})},\
  \Eprint {http://arxiv.org/abs/1901.08580} {arXiv:1901.08580 [gr-qc]}
  \BibitemShut {NoStop}%
\bibitem [{\citenamefont {Taracchini}\ \emph {et~al.}(2014)\citenamefont
  {Taracchini} \emph {et~al.}}]{Taracchini:2013rva}%
  \BibitemOpen
  \bibfield  {author} {\bibinfo {author} {\bibfnamefont {A.}~\bibnamefont
  {Taracchini}} \emph {et~al.},\ }\href {\doibase 10.1103/PhysRevD.89.061502}
  {\bibfield  {journal} {\bibinfo  {journal} {Phys. Rev.}\ }\textbf {\bibinfo
  {volume} {D89}},\ \bibinfo {pages} {061502} (\bibinfo {year} {2014})},\
  \Eprint {http://arxiv.org/abs/1311.2544} {arXiv:1311.2544 [gr-qc]}
  \BibitemShut {NoStop}%
\bibitem [{\citenamefont {Loutrel}\ and\ \citenamefont
  {Yunes}(2017{\natexlab{b}})}]{Loutrel:2016cdw}%
  \BibitemOpen
  \bibfield  {author} {\bibinfo {author} {\bibfnamefont {N.}~\bibnamefont
  {Loutrel}}\ and\ \bibinfo {author} {\bibfnamefont {N.}~\bibnamefont
  {Yunes}},\ }\href {\doibase 10.1088/1361-6382/aa59c3} {\bibfield  {journal}
  {\bibinfo  {journal} {Class. Quant. Grav.}\ }\textbf {\bibinfo {volume}
  {34}},\ \bibinfo {pages} {044003} (\bibinfo {year} {2017}{\natexlab{b}})},\
  \Eprint {http://arxiv.org/abs/1607.05409} {arXiv:1607.05409 [gr-qc]}
  \BibitemShut {NoStop}%
\bibitem [{\citenamefont {Forseth}\ \emph {et~al.}(2016)\citenamefont
  {Forseth}, \citenamefont {Evans},\ and\ \citenamefont
  {Hopper}}]{Forseth:2015oua}%
  \BibitemOpen
  \bibfield  {author} {\bibinfo {author} {\bibfnamefont {E.}~\bibnamefont
  {Forseth}}, \bibinfo {author} {\bibfnamefont {C.~R.}\ \bibnamefont {Evans}},
  \ and\ \bibinfo {author} {\bibfnamefont {S.}~\bibnamefont {Hopper}},\ }\href
  {\doibase 10.1103/PhysRevD.93.064058} {\bibfield  {journal} {\bibinfo
  {journal} {Phys. Rev.}\ }\textbf {\bibinfo {volume} {D93}},\ \bibinfo {pages}
  {064058} (\bibinfo {year} {2016})},\ \Eprint
  {http://arxiv.org/abs/1512.03051} {arXiv:1512.03051 [gr-qc]} \BibitemShut
  {NoStop}%
\bibitem [{\citenamefont {Yunes}\ \emph {et~al.}(2009)\citenamefont {Yunes},
  \citenamefont {Arun}, \citenamefont {Berti},\ and\ \citenamefont
  {Will}}]{PhysRevD.80.084001}%
  \BibitemOpen
  \bibfield  {author} {\bibinfo {author} {\bibfnamefont {N.}~\bibnamefont
  {Yunes}}, \bibinfo {author} {\bibfnamefont {K.~G.}\ \bibnamefont {Arun}},
  \bibinfo {author} {\bibfnamefont {E.}~\bibnamefont {Berti}}, \ and\ \bibinfo
  {author} {\bibfnamefont {C.~M.}\ \bibnamefont {Will}},\ }\href {\doibase
  10.1103/PhysRevD.80.084001} {\bibfield  {journal} {\bibinfo  {journal} {Phys.
  Rev. D}\ }\textbf {\bibinfo {volume} {80}},\ \bibinfo {pages} {084001}
  (\bibinfo {year} {2009})}\BibitemShut {NoStop}%
\bibitem [{\citenamefont {Moreno-Garrido}\ \emph {et~al.}(1995)\citenamefont
  {Moreno-Garrido}, \citenamefont {Mediavilla},\ and\ \citenamefont
  {Buitrago}}]{10.1093/mnras/274.1.115}%
  \BibitemOpen
  \bibfield  {author} {\bibinfo {author} {\bibfnamefont {C.}~\bibnamefont
  {Moreno-Garrido}}, \bibinfo {author} {\bibfnamefont {E.}~\bibnamefont
  {Mediavilla}}, \ and\ \bibinfo {author} {\bibfnamefont {J.}~\bibnamefont
  {Buitrago}},\ }\href {\doibase 10.1093/mnras/274.1.115} {\bibfield  {journal}
  {\bibinfo  {journal} {Monthly Notices of the Royal Astronomical Society}\
  }\textbf {\bibinfo {volume} {274}},\ \bibinfo {pages} {115} (\bibinfo {year}
  {1995})},\ \Eprint
  {http://arxiv.org/abs/http://oup.prod.sis.lan/mnras/article-pdf/274/1/115/18539844/mnras274-0115.pdf}
  {http://oup.prod.sis.lan/mnras/article-pdf/274/1/115/18539844/mnras274-0115.pdf}
  \BibitemShut {NoStop}%
\bibitem [{\citenamefont {Moore}\ \emph {et~al.}(2018)\citenamefont {Moore},
  \citenamefont {Robson}, \citenamefont {Loutrel},\ and\ \citenamefont
  {Yunes}}]{Moore:2018kvz}%
  \BibitemOpen
  \bibfield  {author} {\bibinfo {author} {\bibfnamefont {B.}~\bibnamefont
  {Moore}}, \bibinfo {author} {\bibfnamefont {T.}~\bibnamefont {Robson}},
  \bibinfo {author} {\bibfnamefont {N.}~\bibnamefont {Loutrel}}, \ and\
  \bibinfo {author} {\bibfnamefont {N.}~\bibnamefont {Yunes}},\ }\href@noop {}
  {\  (\bibinfo {year} {2018})},\ \Eprint {http://arxiv.org/abs/1807.07163}
  {arXiv:1807.07163 [gr-qc]} \BibitemShut {NoStop}%
\bibitem [{\citenamefont {Abramowitz}\ and\ \citenamefont {Stegun}(1972)}]{AS}%
  \BibitemOpen
  \bibfield  {author} {\bibinfo {author} {\bibfnamefont {M.}~\bibnamefont
  {Abramowitz}}\ and\ \bibinfo {author} {\bibfnamefont {I.~A.}\ \bibnamefont
  {Stegun}},\ }\href@noop {} {\emph {\bibinfo {title} {Handbook of Mathematical
  Functions with Formulas, Graphs, and Mathematical Tables}}}\ (\bibinfo
  {publisher} {Dover Publications},\ \bibinfo {address} {Mineola, New York},\
  \bibinfo {year} {1972})\BibitemShut {NoStop}%
\bibitem [{\citenamefont {Peters}\ and\ \citenamefont
  {Mathews}(1963)}]{PetersMathews}%
  \BibitemOpen
  \bibfield  {author} {\bibinfo {author} {\bibfnamefont {P.~C.}\ \bibnamefont
  {Peters}}\ and\ \bibinfo {author} {\bibfnamefont {J.}~\bibnamefont
  {Mathews}},\ }\href@noop {} {\bibfield  {journal} {\bibinfo  {journal}
  {Physical Review}\ }\textbf {\bibinfo {volume} {131}},\ \bibinfo {pages}
  {435} (\bibinfo {year} {1963})}\BibitemShut {NoStop}%
\bibitem [{\citenamefont {Boyd}()}]{Boyd}%
  \BibitemOpen
  \bibfield  {author} {\bibinfo {author} {\bibfnamefont {J.~P.}\ \bibnamefont
  {Boyd}},\ }\href {\doibase 10.1023/A:1006145903624} {\bibfield  {journal}
  {\bibinfo  {journal} {Acta Applicandae Mathematica}\ }\textbf {\bibinfo
  {volume} {56}},\ \bibinfo {pages} {1}}\BibitemShut {NoStop}%
\bibitem [{\citenamefont {Johansson}\ \emph {et~al.}(2013)\citenamefont
  {Johansson} \emph {et~al.}}]{mpmath}%
  \BibitemOpen
  \bibfield  {author} {\bibinfo {author} {\bibfnamefont {F.}~\bibnamefont
  {Johansson}} \emph {et~al.},\ }\href@noop {} {\emph {\bibinfo {title}
  {mpmath: a {P}ython library for arbitrary-precision floating-point arithmetic
  (version 0.19)}}} (\bibinfo {year} {2013}),\ \bibinfo {note} {{\tt
  http://mpmath.org/}}\BibitemShut {NoStop}%
\bibitem [{\citenamefont {Barsotti}\ \emph {et~al.}(2018)\citenamefont
  {Barsotti}, \citenamefont {Fritschel}, \citenamefont {Evans},\ and\
  \citenamefont {Gras}}]{LIGOsn}%
  \BibitemOpen
  \bibfield  {author} {\bibinfo {author} {\bibfnamefont {L.}~\bibnamefont
  {Barsotti}}, \bibinfo {author} {\bibfnamefont {P.}~\bibnamefont {Fritschel}},
  \bibinfo {author} {\bibfnamefont {M.}~\bibnamefont {Evans}}, \ and\ \bibinfo
  {author} {\bibfnamefont {S.}~\bibnamefont {Gras}},\ }\href@noop {} {\bibfield
   {journal} {\bibinfo  {journal} {LIGO Document T1800044-v5}\ } (\bibinfo
  {year} {2018})}\BibitemShut {NoStop}%
\bibitem [{\citenamefont {Buonanno}\ \emph {et~al.}(2009)\citenamefont
  {Buonanno}, \citenamefont {Iyer}, \citenamefont {Ochsner}, \citenamefont
  {Pan},\ and\ \citenamefont {Sathyaprakash}}]{Buonanno:2009zt}%
  \BibitemOpen
  \bibfield  {author} {\bibinfo {author} {\bibfnamefont {A.}~\bibnamefont
  {Buonanno}}, \bibinfo {author} {\bibfnamefont {B.}~\bibnamefont {Iyer}},
  \bibinfo {author} {\bibfnamefont {E.}~\bibnamefont {Ochsner}}, \bibinfo
  {author} {\bibfnamefont {Y.}~\bibnamefont {Pan}}, \ and\ \bibinfo {author}
  {\bibfnamefont {B.~S.}\ \bibnamefont {Sathyaprakash}},\ }\href {\doibase
  10.1103/PhysRevD.80.084043} {\bibfield  {journal} {\bibinfo  {journal} {Phys.
  Rev.}\ }\textbf {\bibinfo {volume} {D80}},\ \bibinfo {pages} {084043}
  (\bibinfo {year} {2009})},\ \Eprint {http://arxiv.org/abs/0907.0700}
  {arXiv:0907.0700 [gr-qc]} \BibitemShut {NoStop}%
\bibitem [{\citenamefont {Boetzel}\ \emph {et~al.}(2017)\citenamefont
  {Boetzel}, \citenamefont {Susobhanan}, \citenamefont {Gopakumar},
  \citenamefont {Klein},\ and\ \citenamefont {Jetzer}}]{Boetzel:2017zza}%
  \BibitemOpen
  \bibfield  {author} {\bibinfo {author} {\bibfnamefont {Y.}~\bibnamefont
  {Boetzel}}, \bibinfo {author} {\bibfnamefont {A.}~\bibnamefont {Susobhanan}},
  \bibinfo {author} {\bibfnamefont {A.}~\bibnamefont {Gopakumar}}, \bibinfo
  {author} {\bibfnamefont {A.}~\bibnamefont {Klein}}, \ and\ \bibinfo {author}
  {\bibfnamefont {P.}~\bibnamefont {Jetzer}},\ }\href {\doibase
  10.1103/PhysRevD.96.044011} {\bibfield  {journal} {\bibinfo  {journal} {Phys.
  Rev.}\ }\textbf {\bibinfo {volume} {D96}},\ \bibinfo {pages} {044011}
  (\bibinfo {year} {2017})},\ \Eprint {http://arxiv.org/abs/1707.02088}
  {arXiv:1707.02088 [gr-qc]} \BibitemShut {NoStop}%
\bibitem [{\citenamefont {Kremer}\ \emph {et~al.}(2018)\citenamefont {Kremer},
  \citenamefont {Chatterjee}, \citenamefont {Breivik}, \citenamefont
  {Rodriguez}, \citenamefont {Larson},\ and\ \citenamefont
  {Rasio}}]{Kremer:2018tzm}%
  \BibitemOpen
  \bibfield  {author} {\bibinfo {author} {\bibfnamefont {K.}~\bibnamefont
  {Kremer}}, \bibinfo {author} {\bibfnamefont {S.}~\bibnamefont {Chatterjee}},
  \bibinfo {author} {\bibfnamefont {K.}~\bibnamefont {Breivik}}, \bibinfo
  {author} {\bibfnamefont {C.~L.}\ \bibnamefont {Rodriguez}}, \bibinfo {author}
  {\bibfnamefont {S.~L.}\ \bibnamefont {Larson}}, \ and\ \bibinfo {author}
  {\bibfnamefont {F.~A.}\ \bibnamefont {Rasio}},\ }\href {\doibase
  10.1103/PhysRevLett.120.191103} {\bibfield  {journal} {\bibinfo  {journal}
  {Phys. Rev. Lett.}\ }\textbf {\bibinfo {volume} {120}},\ \bibinfo {pages}
  {191103} (\bibinfo {year} {2018})},\ \Eprint
  {http://arxiv.org/abs/1802.05661} {arXiv:1802.05661 [astro-ph.HE]}
  \BibitemShut {NoStop}%
\end{thebibliography}%
\end{document}